\documentclass[11pt]{article}

\usepackage[margin=1in]{geometry}
\usepackage{amsmath,amssymb,amsfonts,mathtools}
\usepackage{amsthm}
\usepackage{graphicx}
\usepackage{booktabs}
\usepackage{enumitem}
\usepackage{microtype}
\usepackage{natbib}
\usepackage{xcolor}
\definecolor{linknavy}{HTML}{1A5FB4}
\usepackage[colorlinks=true,citecolor=linknavy,linkcolor=linknavy,urlcolor=linknavy]{hyperref}

\newtheorem{theorem}{Theorem}
\newtheorem{proposition}[theorem]{Proposition}
\newtheorem{lemma}[theorem]{Lemma}
\newtheorem{corollary}[theorem]{Corollary}
\newtheorem{definition}[theorem]{Definition}

\theoremstyle{remark}

\newcommand{\claimtag}[1]{}

\graphicspath{{figures/}}

\begin{document}

\title{Uncertainty Quantification for Topological Treatment Effects}
\author{%
  Hugo G.\,Souto\thanks{Dell Research. Correspondence: hugo.souto@dell.com}%
  \and
  Ioannis Diamantis\thanks{Department of Data Analytics and Digitalisation, Maastricht University, School of Business and Economics, P.O.Box 616, 6200 MD, Maastricht, The Netherlands. Email: i.diamantis@maastrichtuniversity.nl}%
}
\date{}

\maketitle

\begin{abstract}
We present a unified framework for uncertainty quantification over filtration scale, extending the topological average treatment effect (TATE) analysis of \citet{kim-lee} to conditional mean confidence and individual effect prediction. For the conditional topological average treatment effect (CTATE), we derive multiplier confidence bands at a fixed covariate value, simultaneous over filtration scale, and prove validity in a correctly specified finite-dimensional regression regime, with explicit control of smoothing bias and nuisance-estimation error. For the individual topological treatment effect (ITTE), we integrate existing weighted and functional split-conformal methods to obtain finite-sample marginal prediction coverage under known target-to-calibration likelihood ratios, with a total-variation bound for estimated ratios. We also extend the current TATE silhouette stability analysis to the diagram-independent bound $\|\Phi_r(D)-\Phi_r(D')\|_\infty \le (1+2r)W_1(D,D')$ for $r\geq1$, recovering the known $r=1$ constant, and propagate diagram approximation error to prediction bands. The installable Python library \texttt{tcda\_uq} integrates mean confidence, individual prediction, and topological preprocessing, with tutorials and reproduction materials. Simulations and a PACCE lesion-measurement application illustrate the distinct coverage targets and the importance of declaring the target population.
\end{abstract}

\section{Introduction}
\label{sec:intro}

Treatment effects are often summarized through changes in scalar outcomes, such as survival time, blood pressure, or total tumor burden. In many applications, however, the outcome contains additional structure that may itself be scientifically relevant.
For example, longitudinal cancer studies may record multiple lesion measurements across visits. An intervention can then affect not only the overall level of these measurements, but also the geometric pattern they form, potentially revealing early structural changes in treatment response before changes in conventional summary measures become apparent. Similar questions arise when
outcomes are naturally represented as point clouds, images, networks, or other structured objects. Reducing such data to a single scalar may therefore discard features of the treatment response that are of substantive interest.

Persistent homology provides one way to summarize this structure by describing
features that persist across multiple filtration scales. This makes it possible to ask
causal questions about structural aspects of an outcome rather than only about a
conventional scalar endpoint. Building on the mean-effect framework of
\citet{kim-lee}, we represent each potential outcome by the power-weighted silhouette
of its persistence diagram and study treatment-induced differences between these
functional summaries. The resulting objects are continuous functions of filtration
scale, allowing both the magnitude and the scale at which structural differences
appear to be retained. Formally, we work in the separable Banach space $C(T)$ with
the supremum norm, where $T$ is a compact filtration interval.

Defining a topological treatment effect, however, is only part of the inferential
problem. For such effects to be used empirically, one must also quantify how much
uncertainty surrounds them. This distinction is particularly important because
different scientific questions require fundamentally different uncertainty
statements. At the population level, the topological average treatment effect (TATE) \citep{kim-lee}
describes the average treatment-induced difference between potential topological
summaries. Conditioning this contrast on baseline characteristics gives the
conditional topological average treatment effect (CTATE). These are
deterministic mean curves, so the relevant inferential objects are simultaneous
confidence bands that quantify estimation uncertainty over the entire filtration
interval. \citet[Sections 3--5]{kim-lee} already identify TATE, derive its
efficient augmented inverse-probability-weighted (AIPW) estimator, establish
functional weak convergence, and discuss
simultaneous confidence bands. Their Remark 2 describes cross-fitting, and their
Corollary 5.4 uses multiplier calibration for a supremum test. These results
provide the mean-effect foundations of our framework. Fixed-covariate CTATE
inference additionally uses established pseudo-outcome regression ideas
\citep{semenova2021debiased}.

A different uncertainty problem arises when the question concerns a new individual.
We take the individual topological treatment effect (ITTE), the contrast between
that unit's two potential summary curves, as an explicit prediction target in this
topological framework. Even if the population-level quantities were
known without estimation error, the effect for a new unit would remain random because
treatment responses can vary across individuals. A confidence band for a mean effect
therefore cannot generally be interpreted as a prediction band for an individual
effect. Individual prediction must instead account for the distribution of
unit-to-unit variation in addition to uncertainty introduced by estimation. The
central aim of this paper is to develop uncertainty-quantification procedures
appropriate to these distinct topological causal targets: simultaneous confidence
bands for TATE and CTATE, and marginal prediction bands for the ITTE of a new unit.

For mean effects, our construction uses the functional influence process of the
AIPW estimator of \citet[Equations (8)--(9)]{kim-lee}, implemented with cross-fitting. Cross-fitting
and orthogonality address nuisance-estimation error under suitable rate conditions
\citep{dml}. Simultaneous inference over the filtration interval,
however, requires more than pointwise asymptotic normality or double robustness:
it requires process-level convergence and valid calibration of the supremum of the
estimation error. Following their functional weak-convergence analysis, we state
these requirements explicitly and verify a bounded, uniformly Lipschitz
sufficient regime that also checks multiplier validity and variance consistency.
For CTATE, we distinguish a general
linear-smoother formulation from a fully specified fixed-dimensional regression
case, making explicit the role of second-stage approximation bias. We also clarify
the relationship with existing functional-data procedures: the simultaneous bands
of \citet{liebl} require their own process regularity conditions, while the
interval-wise procedure of \citet{pini} is a testing procedure rather than
a confidence band unless an appropriate family of tests is inverted.

Individual prediction requires a different construction. Functional conformal
prediction supplies prediction sets for curve-valued responses
\citep{diquigiovanni}, while weighted conformal prediction accommodates covariate
shift when the relevant likelihood ratio is known \citep{tibshirani}. In causal
applications this shift arises naturally: observed outcomes within one treatment arm
and potential outcomes for the target population generally have different covariate
distributions \citep{lei-candes}. We therefore construct a weighted functional
split-conformal band separately for each potential topological outcome and combine
the two arm-specific bands by a union bound. The resulting band gives finite-sample
marginal coverage for the ITTE when the target-to-calibration likelihood ratios are
known. The construction does not require independence between the two potential
outcomes and does not identify their joint law. Related arm-differencing logic
appears in causal conformal inference, including \citet[Appendix A.1, Lemma A.1]{schroeder}; our
measurable-transformation argument establishes the applicability of this existing
machinery to the topological summaries considered here rather than a new general
conformal principle.

The weighting step also exposes an important practical limitation. Under weak
overlap, a target unit can receive sufficiently large weight that the full-population
conformal radius becomes infinite. Overlap weighting instead targets a population
concentrated where both treatment arms are represented \citep{li}, yielding bounded
weights but changing the population to which the prediction statement applies.
Bounded weights alone do not guarantee a finite radius at every sample size, since
sufficient calibration mass is still required. Moreover, estimated likelihood
ratios do not inherit the exact finite-sample guarantee merely because they are
estimated on independent data. We quantify this loss explicitly through a
total-variation discrepancy between the intended and induced target distributions.

The main contribution is a unified framework for uncertainty quantification
that develops both CTATE confidence and ITTE prediction alongside population
TATE confidence. The TATE estimand, identification, AIPW representation, and
functional weak-convergence foundation follow \citet{kim-lee}. We restate these
foundations to make the framework self-contained and extend the TATE inference
with explicit studentized calibration and sufficient conditions that verify
variance consistency and multiplier validity under cross-fitting
(Theorems~\ref{thm:tate-band}--\ref{thm:tate-sufficient}).

The conditional-mean contribution extends their marginal TATE analysis to
simultaneous confidence for CTATE at a prespecified covariate value.
Proposition~\ref{prop:ctate-band} establishes coverage under explicit conditions
on the smoothing bias, first-stage remainder, convergence rate, and multiplier
process. Corollary~\ref{cor:ols} then verifies a concrete sufficient regime for
correctly specified finite-dimensional regression of cross-fitted AIPW
pseudo-outcomes. Thus the conditional contribution includes both a general
coverage statement under process assumptions and a proved regression case.
It uses established pseudo-outcome regression tools \citep{semenova2021debiased};
the arm-outcome regressions in \citet[Example 5.1]{kim-lee} serve instead as
nuisance functions for marginal TATE estimation.

The individual-effect contribution takes ITTE as an explicit prediction target
and combines weighted functional conformal calibration and arm differencing
to obtain finite-sample marginal coverage under known likelihood ratios, with
an explicit total-variation correction for estimated ratios. The conformal
principles are inherited from the literature cited above; their integration
and coverage analysis here concern the topological individual contrast.
Uniformity throughout is over the filtration interval, with CTATE confidence
at a fixed covariate value and ITTE prediction marginal over the declared
target population.

We also provide the installable, reusable Python library \texttt{tcda\_uq},
which brings topological preprocessing, mean confidence, individual prediction,
and target-specific evaluation into a common interface with executable tutorials
and reproduction materials (Section~\ref{subsec:software}). Supporting this
framework, we also extend the silhouette stability analysis of
\citet[Theorem 5.3]{kim-lee} by proving, for every fixed $r\geq1$, the
diagram-independent bound
\[
    \|\Phi_r(D)-\Phi_r(D')\|_\infty
    \leq (1+2r)W_1(D,D').
\]
Appendix C.5 of \citet{kim-lee} already gives the same constant $3$ at $r=1$; for $r>1$,
the bound above removes the dependence on diagram lifetime bounds in their
constant. We use it to propagate known persistence-diagram approximation errors
into prediction bands. We further provide simultaneous prediction over prespecified
collections of homology degrees and silhouette exponents and a deterministic
extension from gridwise bands to the full filtration interval.

We examine these guarantees and their assumptions through synthetic benchmarks and
an application to longitudinal lesion measurements from the PACCE trial
\citep{hecht2009}. It illustrates
how topological summaries and their uncertainty bands can be brought to an applied
setting, while also demonstrating why statistical validity cannot be separated from
the definition of the target population.

Several limitations are important for interpreting the resulting uncertainty
statements. The guarantees concern treatment effects on a prespecified topological
summary rather than on the complete persistence diagram or the underlying outcome:
silhouettes are not injective, so a zero silhouette effect does not imply equality
of diagrams. The ITTE prediction guarantee is marginal over a new target unit and
does not provide conditional coverage at every covariate value or identify the joint
law of the two potential outcomes. Finally, confidence and prediction should not be
interchanged. In particular, we show that a shrinking confidence band for the mean
effect has asymptotically zero individual prediction coverage when the
individual-effect distribution has no atom at the mean curve. Thus increasingly
precise estimation of an average topological effect does not, by itself, resolve
uncertainty about the treatment effect experienced by a new individual.

The remainder of the paper is organized as follows.
Section~\ref{sec:setup} defines the topological representation, causal assumptions,
and inferential targets. Section~\ref{sec:confidence} develops simultaneous
confidence bands for mean effects, and Section~\ref{sec:conformal} develops
individual prediction bands. Section~\ref{sec:extensions} considers overlap
populations, propagation of persistence-diagram approximation error, joint
prediction, and extension from discrete grids to continuous filtration intervals.
The numerical and application sections evaluate the procedures while distinguishing
the sampling population, inferential target, and coverage event. Section~\ref{sec:conclusion} concludes with a synthesis of the framework and outlines open research directions, while the appendices provide
notation, proofs, additional experiments, and a register of the mathematical claims.

\section{Problem setup and estimands}
\label{sec:setup}

We begin by connecting the three ingredients of the framework: the potential-outcomes formulation of the causal problem, the topological representation of structured outcomes, and the causal quantities defined from that representation. The basic idea is straightforward. Each unit has two potentially structured outcomes, one under each treatment condition. We map each of these outcomes to a continuous curve that summarizes its topological structure and then define treatment effects as contrasts between the resulting curves. This section makes each step precise and establishes the notation used throughout the paper.

Throughout, \(0<\alpha<1\); armwise errors also lie in \((0,1)\).

\subsection{Observed data and causal assumptions}

Let \((X_i,A_i,Y_i^0,Y_i^1), \ i=1,\ldots,n\), be independent and identically distributed full-data units. Here, \(X_i\) denotes baseline covariates, \(A_i\in\{0,1\}\) the treatment assignment, and \(Y_i^0,Y_i^1\) the potential outcomes under control and treatment, respectively. Unlike in conventional causal analyses, \(Y_i^a\) need not be scalar or even finite-dimensional: it may be a point cloud, image, network, or another structured object. Only one of the two potential outcomes is observed, so the observed data for unit \(i\) are \(O_i=(X_i,A_i,Y_i), \ Y_i=Y_i^{A_i}\).

We work under the standard assumptions that connect these observed data to the potential-outcome distributions of interest. Specifically, we assume consistency, \(Y_i=Y_i^{A_i}\), conditional exchangeability, \((Y^0,Y^1)\perp A\mid X\), and strict positivity,
\(0<\pi(X)<1\) almost surely, where \(\pi(x)=P(A=1\mid X=x)\) is the propensity score. Regular conditional distributions are assumed to exist; for example, the underlying spaces may be taken to be standard Borel spaces.

Together, these assumptions identify the conditional marginal distribution of each potential outcome from the corresponding observed treatment arm. They do not, however, identify the dependence between \(Y^0\) and \(Y^1\). This distinction will matter later: mean treatment effects can be identified from the two arm-specific marginal laws, whereas the joint distribution of the two potential outcomes for the same individual generally cannot.

Some of our inferential results require a stronger form of positivity. Where explicitly stated, we assume that for some \(\epsilon>0\),
\begin{equation}
\label{eq:strong-overlap}
\epsilon\leq\pi(X)\leq1-\epsilon
\qquad\text{almost surely}.
\end{equation}

We refer to \eqref{eq:strong-overlap} as
\emph{strong overlap}. Strict positivity, together with consistency and conditional exchangeability, identifies the arm-specific potential-outcome distributions. The stronger condition
\eqref{eq:strong-overlap} is imposed where required for
the asymptotic inference results of Section~\ref{sec:confidence}.
For the overlap-weighted target considered in
Section~\ref{subsec:overlap}, positivity is required
only on the support of the target population.

The causal assumptions above tell us when the observed treatment arms can be used to learn about the potential outcomes. They do not yet specify \emph{what aspect} of a structured potential outcome will serve as the response in our causal analysis. We turn to that representation next.

\subsection{The topological representation}

When an outcome is structured, a treatment may alter more than its overall level. It may change the geometry, connectivity, or organization of the object itself. A point cloud, for example, may develop or lose clusters and loops; analogous structural changes can occur in images, networks, and other complex data. Our goal is therefore to convert each structured potential outcome into a common mathematical object that preserves information about such structure while remaining amenable to statistical analysis.

Persistent homology provides the first step in this conversion. Starting from an outcome, we construct a \emph{filtration}, a nested family of objects indexed by a scale parameter \(t\). As \(t\) varies, topological features such as connected components, loops, and higher-dimensional cavities may appear and disappear. A \emph{persistence diagram} records each such feature through its birth and death scales. Features that survive across a wide range of scales have large persistence and represent comparatively persistent topological structure, whereas short-lived features appear closer to the diagonal of the diagram.

A persistence diagram is already a useful summary of the original structured outcome, but it remains a collection of points whose cardinality may vary from one unit to another. For the functional inferential framework developed below, it is convenient to take one further step and represent each diagram by a continuous curve on a common domain. We use the \emph{power-weighted persistence silhouette} for this purpose. The complete representation can therefore be viewed as the prespecified pipeline
\[
\text{structured outcome}
\;\longrightarrow\;
\text{filtration}
\;\longrightarrow\;
\text{persistence diagram}
\;\longrightarrow\;
\text{silhouette curve in }C(T).
\]
This final step is important for what follows: after the transformation, every unit is represented by a function on the same filtration interval, so causal contrasts can be defined and studied pointwise in filtration scale while retaining their functional nature.

Readers seeking further mathematical background on topological representations and their role in causal analysis may consult \citet[Section~2]{kim-lee} and \citet[Sections~2.2--3]{soutodiamantis2026tcda}. We give here the definitions needed for the present framework.

Fix a nondegenerate compact filtration interval \(T=[t_-,t_+]\), a homology degree \(d\), and an exponent \(r\geq1\). The degree \(d\) specifies the type of topological feature under consideration (for example, connected components when \(d=0\) and loops when \(d=1\)), while \(r\) determines how strongly the representation emphasizes features with greater persistence.

Let \(D=\{(b_j,e_j)\}_{j=1}^{N_D}, \ b_j<e_j<\infty\), be a finite multiset of off-diagonal birth--death pairs from the degree-\(d\) persistence diagram. Associate with each pair the triangular tent function
\[
\lambda_j(t) = \max\{0,\min(t-b_j,e_j-t)\}.
\]
The tent is zero outside the lifetime of the feature, increases after its birth \(b_j\), and decreases toward zero as \(t\) approaches its death \(e_j\). Its contribution therefore records both where in the filtration the feature exists and how long it persists.

The power-weighted silhouette combines these individual contributions into the single curve
\begin{equation}
\Phi_r(D)(t)
=
\frac{\sum_j(e_j-b_j)^r\lambda_j(t)}
     {\sum_j(e_j-b_j)^r}.
\label{eq:silhouette}
\end{equation}
Thus, \(\Phi_r(D)\) is a persistence-weighted average of the individual tents. The exponent \(r\) controls their relative importance: increasing \(r\) gives greater weight to long-lived features relative to short-lived ones. We set \(\Phi_r(\varnothing)=0\), discard zero-persistence points, and treat the removal or finite truncation of essential classes as part of the prespecified representation pipeline.

The silhouette construction follows \citet{chazal} and \citet[Equation~(2)]{kim-lee}. Each silhouette is nonnegative and \(1\)-Lipschitz, as recorded in their Lemma~2.1. Moreover, if every persistence is bounded above by \(c\), then \(\|\Phi_r(D)\|_\infty\leq c/2\). These properties are useful because they place the topological summaries in a well-behaved common function space and later provide regularity for simultaneous inference over filtration scale.

Formally, let \(G_{d,r}:\mathcal Y\longrightarrow C(T)\) denote the resulting prespecified Borel map from the original outcome space to the space of continuous functions on \(T\). Applying the same transformation to the two potential outcomes gives
\[
\phi_{i,d}^a=G_{d,r}(Y_i^a),
\qquad a\in\{0,1\},
\]
while consistency implies that the observed topological response is
\[
\phi_{i,d}
=
G_{d,r}(Y_i)
=
\phi_{i,d}^{A_i}.
\]
The original causal problem has therefore been transformed into a functional one: unit \(i\) has two potential topological curves, \(\phi_{i,d}^0\) and \(\phi_{i,d}^1\), of which only \(\phi_{i,d}^{A_i}\) is observed.

This transformation also makes clear the scope of the causal quantities studied below. They describe treatment-induced changes in the \emph{prespecified topological summary}, not treatment effects on the complete original outcome or on the persistence diagram itself; see \citet{soutodiamantis2026tcda} for further discussion. We assume
\[
E\|\phi_d^a\|_\infty<\infty,
\qquad a=0,1,
\]
so that population mean curves are well defined as Bochner expectations in \(C(T)\). Stronger moment and regularity conditions are introduced only when required by particular inferential results. Likewise, alternative vectorizations of persistence diagrams would require their own state spaces and regularity arguments and are not obtained by a purely formal substitution in \eqref{eq:silhouette}.

With the potential outcomes now represented by the pair of curves
\[
(\phi_{i,d}^0,\phi_{i,d}^1)\in C(T)\times C(T),
\]
the remaining question is \emph{which contrast between these curves is scientifically relevant}. Just as ordinary causal inference distinguishes population-average, conditional-average, and individual treatment effects, the topological setting leads naturally to three corresponding functional targets. The next subsection defines these quantities and clarifies the different uncertainty statements associated with each.

\subsection{Three targets}

Once each potential outcome has been represented by a curve in \(C(T)\), treatment effects can be defined by comparing the two potential topological summaries. The appropriate comparison depends on the scientific question. We distinguish three targets: the average structural effect in the target population, the average structural effect among units with specified baseline characteristics, and the treatment-induced structural change for an individual unit. These correspond, respectively, to TATE, CTATE, and ITTE. The first two are population-level summaries: they average the contrast between the potential topological curves either marginally or conditionally on covariates. The third is fundamentally different. It is the unit-specific contrast between the two potential curves and therefore remains a random object when considering a new individual.

\begin{definition}[Mean and individual topological effects]\label{def:targets}
For the fixed pair $(d,r)$, define
\begin{align}
 \psi_d(t)&=E\{\phi_d^1(t)-\phi_d^0(t)\},\label{eq:tate}\\
 \tau_d(t,x)&=E\{\phi_d^1(t)-\phi_d^0(t)\mid X=x\},\label{eq:ctate}\\
 \delta_{i,d}(t)&=\phi_{i,d}^1(t)-\phi_{i,d}^0(t).\label{eq:itte}
\end{align}
\end{definition}

Equation~\eqref{eq:tate} restates the TATE introduced by \citet[Equation~(3)]{kim-lee}; \eqref{eq:ctate} gives its conditional-mean analogue, and \eqref{eq:itte} defines the individual contrast studied here. In each case, the effect is itself a curve indexed by the filtration scale (t): its value at a given (t) describes the treatment-induced difference in the topological summary at that scale. Consequently, these targets are naturally paired with different statistical procedures: population-mean curves require simultaneous confidence bands, whereas individual contrasts require predictive intervals.

The identification of conditional means follows the standard argument of \citet[Equations (4)--(5)]{kim-lee}. Choosing measurable conditional-mean versions $\mu_{a,d}(t,x)=E\{\phi_d^a(t)\mid X=x\}$ with continuous paths in $t$, standard causal assumptions identify these quantities from the observed data: for $P_X$-almost every $x$,
\begin{equation}\label{eq:identification}
 \mu_{a,d}(t,x) = E\{\phi_d(t)\mid X=x,A=a\}.
\end{equation}

The identification in \eqref{eq:identification} holds $P_X$-almost everywhere. Inference at a prespecified covariate value $x$ requires an identified version of the conditional mean, justified by suitable continuity or regression-model assumptions. In particular, inference at points outside the covariate support is necessarily model-dependent. Because individual silhouettes are $1$-Lipschitz, their conditional means $\mu_{a,d}(\cdot, x)$ can be chosen $1$-Lipschitz in $t$, implying that $\psi_d$, $\tau_d(\cdot,x)$, and $\delta_{i,d}$ are $2$-Lipschitz. 

Crucially, the joint distribution of the individual contrast $\delta_{i,d}$ is not point-identified from the arm-specific marginals alone. However, the prediction bands developed in Section~\ref{sec:conformal} remain valid across every valid coupling of these marginal distributions that satisfies the stated assumptions.

\subsection{AIPW estimation and independent conformal calibration}
We use the AIPW score and estimator of
\citet[Equations (8)--(9)]{kim-lee}, with the standard cross-fitting
extension described in their Remark 2 and in \citet{dml}.
Write $\eta=(\pi,\mu_{0,d},\mu_{1,d})$ and define

\begin{equation}\label{eq:aipw-score}
\rho_d(t,O;\eta) =\mu_{1,d}(t,X)-\mu_{0,d}(t,X)
 +\frac{A}{\pi(X)}\{\phi_d(t)-\mu_{1,d}(t,X)\}
 -\frac{1-A}{1-\pi(X)}
       \{\phi_d(t)-\mu_{0,d}(t,X)\}.
\end{equation}

At the true nuisance functions, $E\{\rho_d(t,O;\eta)\mid X=x\}=\tau_d(t,x)$ and $E\rho_d(t,O;\eta)=\psi_d(t)$. At distributions where it is square integrable, $\xi_d(t,O)=\rho_d(t,O;\eta)-\psi_d(t)$ is the pointwise efficient
influence function in the nonparametric observed-data model.
The score is doubly robust at the population level: either a correct
propensity or two correct outcome regressions suffice for the mean
identity, subject to integrability and nonzero denominators
\citep{bangrobins2005doubly,kim-lee}. These mean-score properties
are inherited foundations of the present uncertainty framework.

Use a fixed number $K$ of cross-fitting folds $I_k$, with the partition fixed in advance or randomized independently of the observations, and each nuisance fit $\widehat\eta^{(-k)}$ obtained without fold $I_k$. For $i\in I_k$, set

\begin{equation}\label{eq:estimators}
 \widehat\rho_{i,d}(t)=\rho_d(t,O_i;\widehat\eta^{(-k)}),\qquad
 \widehat\psi_d(t)=\frac1n\sum_{i=1}^n\widehat\rho_{i,d}(t),\qquad
 \widehat\xi_{i,d}(t)=\widehat\rho_{i,d}(t)-\widehat\psi_d(t).
\end{equation}

Section~\ref{sec:confidence} states sufficient conditions for
root-$n$ inference. Double robustness of a mean identity alone is not a bootstrap validity theorem.

Conformal prediction instead uses a training sample independent of the calibration sample and the new unit. All fitted centers, scales, and estimated weights are fixed using the training sample before calibration. The AIPW cross-fitting scheme is not substituted for this split-conformal construction without a separate justification.

\subsection{Target distributions and covariate shift}

To apply conformal prediction methods to functional topological summaries in order to perform proper uncertainty quantification for ITTE, we must first verify that applying the transformation $G_{d,r}$ preserves the underlying statistical structure of the data, and we must account for the distribution shift inherent in counterfactual prediction.

First, we establish that the transformation to the function space $C(T)$ retains the necessary independence and exchangeability properties required for conformal inference.

\begin{lemma}[Transformation of potential outcomes]\label{lem:transform}
The fixed Borel map $G_{d,r}$ preserves the i.i.d. property of the units, consistency, and conditional exchangeability. Conditional on independent training data, a fixed measurable scoring rule applied separately to calibration and test pairs from the same law produces exchangeable scores.
\end{lemma}

\begin{proof}
Coordinatewise measurable maps preserve independence and identical
distribution. Consistency is preserved by applying the same map to the observed and potential outcomes. Conditional independence is preserved under measurable functions of the potential outcomes. Conditioning on independent training data fixes the scoring rule and leaves the calibration and test pairs identically distributed.
\end{proof}

Second, we must account for covariate shift between the observational data used for calibration and the target population for prediction. In an observational study, calibration pairs in treatment arm $a$ follow the observed distribution $P_a = \mathcal{L}(X, \phi_d \mid A = a)$, whereas a new potential outcome in the full target population follows $Q_a = \mathcal{L}(X, \phi_d^a)$. While their conditional response laws given covariates $X$ coincide, their covariate distributions generally differ. 

To correct for this discrepancy in our prediction framework, we employ likelihood-ratio weights. Under standard positivity assumptions, the exact density ratio relating the target covariate distribution to the arm-specific covariate distribution is given by:
\begin{equation}\label{eq:weight}
 w_a(x)=\frac{dP_X}{dP_{X\mid A=a}}(x)
       =\frac{P(A=a)}{e_a(x)},\qquad
 e_1(x)=\pi(x),\quad e_0(x)=1-\pi(x),
\end{equation}
where any arm-specific positive multiplicative constants cancel out in the weighted conformal quantiles. When these weights must be estimated, the resulting approximation error is explicitly bounded in Section~\ref{subsec:estimated}.

Finally, to maintain strict statistical validity throughout the subsequent procedures, all structural choices (including the homology degree $d$, exponent $r$, filtration interval $T$, essential-class rules, and evaluation grids) are held fixed prior to calibration. Any hyperparameter selection performed using independent training data is treated conditionally on that training sample, with the corresponding target population declared explicitly.
\section{Confidence bands for mean topological effects}
\label{sec:confidence}

We now develop simultaneous confidence bands for the two mean topological treatment effects defined in Section~\ref{sec:setup}. Unlike individual prediction bands, which target the random treatment effect of a new unit, confidence bands quantify estimation
uncertainty for deterministic mean curves. The population TATE band has root-$n$ width under the conditions below. For CTATE at a fixed covariate value, the width depends on the second-stage estimation procedure: it is of order $v_n(x)$ for a general linear smoother and of order $n^{-1/2}$ in the correctly specified fixed-dimensional
regression case.

We organize this section into three parts. First, Section~\ref{subsec:tate} constructs simultaneous confidence bands for the population-average curve $\psi_d$ (TATE), stating high-level empirical process requirements (Theorem~\ref{thm:tate-band}), verifying them under a concrete non-parametric cross-fitting regime (Theorem~\ref{thm:tate-sufficient}), and examining studentized versus unstudentized choices. Second, Section~\ref{subsec:ctate} addresses treatment effect heterogeneity across subgroups by constructing simultaneous bands for the conditional mean curve $\tau_d(\cdot,x)$ at a prespecified covariate profile $x$ (CTATE) through linear smoothing of AIPW pseudo-outcomes (Proposition~\ref{prop:ctate-band}) and fixed-dimensional ordinary least squares (Corollary~\ref{cor:ols}). Finally, Section~\ref{subsec:benchmarks} evaluates existing functional data analysis benchmarks, explaining why their regularity requirements are often unsuited for topological summaries.

\subsection{TATE: process assumptions and a multiplier band}\label{subsec:tate}

To conduct simultaneous inference on the population-average curve $\psi_d$, we analyze the asymptotic distribution of the cross-fitted AIPW estimator $\widehat\psi_d$. Let $\rho_d(t,O;\eta)$ denote the uncentered AIPW score from Section~\ref{sec:setup}, and let
\[
 \xi_d(t,O)=\rho_d(t,O;\eta)-\psi_d(t)
\]
denote the corresponding centered influence curve, with pointwise variance $\sigma_d(t)^2=E\xi_d(t,O)^2$. To calibrate simultaneous coverage over the filtration domain $T$ without repeatedly refitting the first-stage machine-learning nuisance models, we employ a Gaussian multiplier bootstrap. Let $g_1,\ldots,g_n$ be independent standard Gaussian variables, independent of the observed data, and define

\begin{equation}\label{eq:tate-bootstrap}
 \widehat G_{n,d}^{\#}(t)=\frac1{\sqrt n}\sum_{i=1}^n
           g_i\widehat\xi_{i,d}(t),\qquad
 \widehat\sigma_d(t)^2=\frac1n\sum_{i=1}^n\widehat\xi_{i,d}(t)^2.
\end{equation}

The following assumptions isolate the requirements of the band.

\begin{enumerate}[label=(T\arabic*)]
\item\label{ass:t1} Uniform asymptotic linearity holds:
\[
 \left\|\sqrt n(\widehat\psi_d-\psi_d)
       -\frac1{\sqrt n}\sum_i\xi_d(\cdot,O_i)\right\|_\infty=o_p(1).
\]
\item\label{ass:t2} The oracle process converges weakly in $C(T)$:
$n^{-1/2}\sum_i\xi_d(\cdot,O_i)\Rightarrow G_d$, for a tight
centered Gaussian process $G_d$.
\item\label{ass:t3} Conditional on the data,
$\widehat G_{n,d}^{\#}\Rightarrow G_d$ in probability in $C(T)$.
\item\label{ass:t4} The variance estimate is uniformly consistent,
$\|\widehat\sigma_d-\sigma_d\|_\infty=o_p(1)$, and
$0<\inf_T\sigma_d\leq\sup_T\sigma_d<\infty$. The distribution
function of $\|G_d/\sigma_d\|_\infty$ is continuous and strictly
increasing in a neighborhood of its $(1-\alpha)$ quantile.
\end{enumerate}

On a sample for which $\inf_T\widehat\sigma_d=0$, return the
whole real-line band instead of dividing by zero. Under (T4) the
probability of this exceptional event tends to zero, so this convention
does not change the asymptotic statement.

\begin{theorem}[Studentized TATE band]\label{thm:tate-band}
Under \textup{(T1)--(T4)}, let
\[
 \widehat c_{d,1-\alpha}
 =\operatorname{Quantile}_{1-\alpha}
 \left(\left\|\widehat G_{n,d}^{\#}/\widehat\sigma_d\right\|_\infty
       \,\middle|\,O_1,\ldots,O_n\right).
\]
Then
\begin{equation}\label{eq:tate-band}
 C_{n,d}(t)=
 \left[\widehat\psi_d(t)\ \pm\
       \frac{\widehat c_{d,1-\alpha}\widehat\sigma_d(t)}{\sqrt n}\right]
\end{equation}

satisfies $P\{\psi_d(t)\in C_{n,d}(t)\ \forall t\in T\}\to1-\alpha$.
Its supremum full width is $O_p(n^{-1/2})$.
\end{theorem}

The proof is deferred to Appendix~\ref{app:proofs}.
While Theorem~\ref{thm:tate-band} establishes simultaneous asymptotic validity under high-level process conditions \textup{(T1)--(T4)}, empirical applications require primitive, checkable conditions on the underlying data-generating process and first-stage learners. The next result supplies a self-contained sufficient regime for the cross-fitted AIPW estimator and its Gaussian multiplier process, following the functional AIPW analysis of \citet{kim-lee}.

\begin{theorem}[A sufficient regime for cross-fitted TATE inference]
\label{thm:tate-sufficient}
Assume the causal conditions of Section~\ref{sec:setup} and the
following conditions, for the fixed degree and exponent.

\begin{enumerate}[label=(S\arabic*)]
\item The potential curves are $1$-Lipschitz and uniformly bounded:
$0\leq\phi_d^a(t)\leq M$ almost surely for a finite deterministic $M$.
Strong overlap holds for some $\epsilon>0$.
\item The number $K$ of folds is fixed, their partition is independent of the observations, and $n_k/n\to\lambda_k>0$.
For every fold, the fitted propensity lies in
$[\epsilon,1-\epsilon]$. The fitted outcome regressions are jointly
measurable, uniformly bounded in absolute value by a finite constant, and uniformly Lipschitz in $t$ with a finite deterministic bound, all uniformly over $x$ and folds.
\item For each fold, with $\|\cdot\|_{P_X,2}$ denoting the
$L^2(P_X)$ norm,
\begin{align*}
 p_k&=\|\widehat\pi^{(-k)}-\pi\|_{P_X,2}=o_p(1),\\
 b_{a,k}&=\sup_{t\in T}
 \|\widehat\mu_{a,d}^{(-k)}(t,\cdot)-\mu_{a,d}(t,\cdot)\|_{P_X,2}
 =o_p(1),\\
 p_k(b_{0,k}+b_{1,k})&=o_p(n^{-1/2}).
\end{align*}
\end{enumerate}

Then \textup{(T1)--(T3)} and uniform variance consistency hold for the Gaussian multiplier process in \eqref{eq:tate-bootstrap}. If the nondegeneracy and quantile conditions in \textup{(T4)} also hold, Theorem~\ref{thm:tate-band} applies.
\end{theorem}

The proof is deferred to Appendix~\ref{app:proofs}.
These are sufficient, not necessary, conditions. The uniform bounds may also be imposed on events whose probability tends to one by localization. In particular, a finite diagram for every unit does not by itself imply the uniform bound in (S1). A deterministic lifetime bound is one way to obtain it. Conditions on a fitted regression must be checked for that learner; they do not follow from the continuity of the true summary. The proof uses the entropy of the filtration-scale index and a foldwise nuisance remainder, in the spirit of \citet{dml} and \citet{vdv-wellner}.

In persistent homology, summary curves such as silhouettes can vanish identically outside the filtration range of persistent features, meaning that the true variance $\sigma_d(t)^2$ may equal zero on portions of $T$. At points where the variance vanishes, division by
$\widehat\sigma_d(t)$ is not justified by this theorem. An
unstudentized alternative uses the conditional quantile
$\widehat q_{1-\alpha}$ of $\|\widehat G_{n,d}^{\#}\|_\infty$ and
the band $[\widehat\psi_d(t)\pm\widehat q_{1-\alpha}/\sqrt n]$.
It has the same asymptotic conclusion under (T1)--(T3) and continuity and local strict increase of the distribution of $\|G_d\|_\infty$ at the relevant quantile. Indeed, (T1)--(T2) give convergence of the raw supremum to $\|G_d\|_\infty$, while (T3) gives conditional convergence of the raw bootstrap supremum; the stated quantile condition then yields coverage by the argument of Theorem~\ref{thm:tate-band}. Alternatively, a studentized analysis can prespecify a subinterval on which variance is bounded away from zero. An arbitrary numerical variance floor is not covered without an argument for its effect on the statistic and its calibration.

\subsection{CTATE at a fixed covariate value}\label{subsec:ctate}

While the TATE band characterizes the overall population-average effect, clinical and scientific studies often require understanding treatment effect heterogeneity: what is the expected topological response for a subgroup with a specific covariate profile $X=x$?

To address this question, we construct simultaneous confidence bands for the conditional topological average treatment effect $\tau_d(\cdot,x)$. By double robustness of the AIPW construction, the conditional expectation of the oracle pseudo-outcome satisfies $E\{\rho_d(t,O;\eta)\mid X=x\}=\tau_d(t,x)$. Estimating CTATE therefore reduces to a second-stage regression of the cross-fitted pseudo-outcomes $\widehat\rho_{i,d}(t)$ on covariates $X_i$ \citep{semenova2021debiased}. The functional linear smoother in \citet[Example 5.1]{kim-lee} concerned the first-stage arm regressions $\mu_a$; here, the linear smoother operates on the second-stage pseudo-outcomes to target $\tau_d(\cdot,x)$ directly, subject to its own bias and simultaneous-coverage requirements over $T$. For smoother weights $a_{i,n}(x)$ independent of $t$, define

\begin{equation}\label{eq:ctate-estimate}
 \widehat\tau_d(t,x)=\sum_i a_{i,n}(x)\widehat\rho_{i,d}(t),\qquad
 v_n(x)^2=\sum_i a_{i,n}(x)^2.
\end{equation}

Put $\varepsilon_{i,d}(t)=\rho_d(t,O_i;\eta)-\tau_d(t,X_i)$.
Let $\widehat r_{i,d}$ be residual curves whose variance and multiplier properties satisfy (C4) below, and define

\begin{equation}\label{eq:ctate-se}
 \widehat s_{d,x}(t)^2=
       \sum_i a_{i,n}(x)^2\widehat r_{i,d}(t)^2.
\end{equation}

The general linear-smoother conditions are:

\begin{enumerate}[label=(C\arabic*)]
\item $\sum_i a_{i,n}(x)=1$, $v_n(x)\to0$, and
$\max_i|a_{i,n}(x)|/v_n(x)\to0$ in probability.
\item Both the covariate-smoothing bias and the complete first-stage
remainder are negligible:
\begin{align*}
 \left\|\sum_i a_{i,n}(x)\tau_d(\cdot,X_i)-\tau_d(\cdot,x)\right\|_\infty
 &=o_p\{v_n(x)\},\\
 \left\|\sum_i a_{i,n}(x)
       \{\widehat\rho_{i,d}-\rho_d(\cdot,O_i;\eta)\}\right\|_\infty
 &=o_p\{v_n(x)\}.
\end{align*}
\item $v_n(x)^{-1}\sum_i a_{i,n}(x)\varepsilon_{i,d}
\Rightarrow H_{d,x}$ in $C(T)$, where $H_{d,x}$ is a tight centered
Gaussian process with continuous standard deviation $\sigma_{d,x}$
bounded away from zero and infinity.
\item Uniformly in $t$,
$\widehat s_{d,x}(t)/v_n(x)\to\sigma_{d,x}(t)$ in probability.
Conditionally on the data, the signed process
\[
 M_{n,d,x}^{\#}(t)=
 \frac{\sum_i g_i a_{i,n}(x)\widehat r_{i,d}(t)}{\widehat s_{d,x}(t)}
\]
converges weakly in probability to $H_{d,x}/\sigma_{d,x}$ in $C(T)$.
The distribution of $\|H_{d,x}/\sigma_{d,x}\|_\infty$ is continuous
and strictly increasing near its $(1-\alpha)$ quantile.
\end{enumerate}

The same whole real-line convention applies when
$\inf_T\widehat s_{d,x}=0$; under (C4) its probability tends to zero.

\begin{proposition}[Fixed-$x$ CTATE band]\label{prop:ctate-band}
Under \textup{(C1)--(C4)}, define the critical value using the absolute supremum,

\begin{equation}\label{eq:ctate-critical}
 \widehat c_{d,x,1-\alpha}=
 \operatorname{Quantile}_{1-\alpha}
 \left(\sup_{t\in T}\left|
 \frac{\sum_i g_i a_{i,n}(x)\widehat r_{i,d}(t)}{\widehat s_{d,x}(t)}
 \right|\,\middle|\,O_1,\ldots,O_n\right).
\end{equation}

Then
\begin{equation}\label{eq:ctate-band}
 C_{n,d,x}(t)=
 [\widehat\tau_d(t,x)\pm\widehat c_{d,x,1-\alpha}\widehat s_{d,x}(t)]
\end{equation}
has simultaneous coverage tending to $1-\alpha$ over $T$, at this
fixed $x$, and supremum width $O_p\{v_n(x)\}$.
\end{proposition}

The proof is deferred to Appendix~\ref{app:proofs}.
The critical value in \eqref{eq:ctate-critical} is a quantile of
$\sup_t|M_{n,d,x}^{\#}(t)|$, not of $\sup_tM_{n,d,x}^{\#}(t)$.

An unstudentized construction is useful when the limiting variance can
vanish at some filtration scales. Retain (C1)--(C2) and the weak
convergence in (C3), allowing a tight centered Gaussian limit $H_{d,x}$
with zero variance at some points. Replace (C4) by conditional weak convergence in probability in $C(T)$ of
$v_n(x)^{-1}\sum_i g_i a_{i,n}(x)\widehat r_{i,d}$ to $H_{d,x}$,
and assume that the distribution of $\|H_{d,x}\|_\infty$ is continuous
and strictly increasing near its $(1-\alpha)$ quantile. If
\[
 \widehat q_{d,x,1-\alpha}
 =\operatorname{Quantile}_{1-\alpha}
 \left(\left\|\sum_i g_i a_{i,n}(x)\widehat r_{i,d}\right\|_\infty
       \,\middle|\,O_1,\ldots,O_n\right),
\]
then $[\widehat\tau_d(t,x)\pm\widehat q_{d,x,1-\alpha}]$ has the
same asymptotic coverage and width order. Indeed, the error divided by
$v_n(x)$ converges to $H_{d,x}$, and the raw bootstrap quantile divided
by $v_n(x)$ consistently estimates its supremum quantile. This avoids
division by a numerical variance floor. It does not remove smoothing
or projection bias.

\begin{corollary}[A verified fixed-dimensional case]\label{cor:ols}
Assume (S1)--(S3). Let $b(X)\in\mathbb R^q$ be a bounded,
fixed-dimensional basis containing an intercept, with
$Q=E\{b(X)b(X)^\top\}$ positive definite. Suppose the selected
conditional-mean version is correctly specified:
\[
 \tau_d(t,z)=b(z)^\top\beta_d(t)
\]
for $P_X$-almost every $z$ and at the prescribed $x$.
Use ordinary least squares on the cross-fitted pseudo-outcomes, with
\begin{align*}
 Q_n&=\frac1n\sum_i b(X_i)b(X_i)^\top,\\
 a_{i,n}(x)&=\frac1n b(x)^\top Q_n^{-1}b(X_i),\\
 \widehat r_{i,d}(t)&=\widehat\rho_{i,d}(t)
                     -b(X_i)^\top\widehat\beta_d(t).
\end{align*}

Define \(\zeta_{d,x}(t,O)=b(x)^\top Q^{-1}b(X) \{\rho_d(t,O;\eta)-\tau_d(t,X)\}\). If $\operatorname{Var}\{\zeta_{d,x}(t,O)\}$ is bounded away from zero and the corresponding standardized Gaussian supremum satisfies the quantile condition in (C4), Proposition~\ref{prop:ctate-band} applies with $v_n(x)^2=n^{-1}b(x)^\top Q_n^{-1}b(x)$ and root-$n$ width. Ordinary fitted residuals suffice in this fixed-dimensional case.
\end{corollary}

The proof is deferred to Appendix~\ref{app:proofs}.
For the OLS construction, a singular $Q_n$ also triggers a whole
real-line band. Positive definiteness of $Q$ makes this an event whose
probability tends to zero. The inverse formulas and their asymptotic
expansions are evaluated on its complement.

Correct specification, or a negligible approximation bias under a
separately justified approximation scheme, is essential. The identity $E(\rho_d\mid X)=\tau_d$ does not make an arbitrary regression model correct. For local smoothing in $p$ covariate dimensions, the typical scale is $(nh^p)^{-1/2}$, and bias must be negligible at that scale. A constant true CTATE does not force fitted covariate-regression weights to equal $1/n$. An intercept-only fit has those weights; it targets a CTATE only when the corresponding constant model is valid. For example, if $X$ is uniform on $[-1,1]$ and $\tau_d(t,x)=x^2 g(t)$ for a nonzero curve $g$, regression on $(1,X)$ converges to $g(t)/3$ at $x=0$, whereas the true CTATE there is zero. Even oracle pseudo-outcomes do not remove this approximation bias, and a shrinking band around that regression limit misses the target. This example can be realized with an arm-one diagram that contains one fixed bar with probability $X^2$, and an empty arm-zero diagram, taking $g$ to be the bar's tent.

These statements do not give uniform coverage over $x$. A finite
prespecified collection of covariate values can be addressed by an
additional simultaneous calibration or error allocation. General
uniform inference over a covariate region requires further conditions. If a response projection $P$ is applied, the same operation must be included in the bootstrap and variance construction. For fixed $P$, the natural target is $P\tau_d$ or $P\psi_d$. Inference for the unprojected target additionally requires negligible projection bias; with a changing projection, its effect on stochastic convergence must also be controlled.

\subsection{Two functional-data benchmarks}\label{subsec:benchmarks}

To benchmark our multiplier bootstrap bands against existing tools in functional data analysis, we examine two prominent methodologies: the fast-and-fair simultaneous confidence bands of \citet{liebl} and the interval-wise testing procedure of \citet{pini}. While both approaches are widely applied in functional statistics, each encounters fundamental limitations when applied to causal topological effects. We review their core mechanics below to clarify why a multiplier-based construction is appropriate under our assumptions.

The simultaneous bands of \citet{liebl} require a sufficiently
smooth standardized limiting process and consistent estimates of the
covariance and its derivatives along the diagonal. These assumptions
are stronger than continuity or Lipschitz continuity of the observed
silhouettes. The expected-excursion construction yields an upper bound on the noncoverage probability. Its general guarantee is asymptotic coverage at least $1-\alpha$, not necessarily equality. A fair local allocation over a partition $t_-=a_0<\cdots<a_J=t_+$ assigns error budgets $\alpha(a_j-a_{j-1})/(t_+-t_-)$. Equivalently, the domain can first be rescaled to $[0,1]$.

The interval-wise procedure of \citet{pini} is a hypothesis-testing procedure. Suppose valid interval-null $p$-values $p_I$ are available on a prespecified finite grid and set
\[
 \widetilde p(t_j)=\max_{I\ni t_j}p_I.
\]
For a true-null interval $I$,
\begin{equation}\label{eq:iwt-control}
 P\{\exists t_j\in I:\widetilde p(t_j)\leq\alpha\}
 \leq P\{p_I\leq\alpha\}\leq\alpha+o(1).
\end{equation}
Thus the error event is any false rejection within $I$, not rejection at every point of $I$. This is interval-wise familywise control; control over arbitrary disconnected true-null sets is a separate property \citep{abramowicz}. In particular, a true global null obtains control over the whole domain when the domain is one of the tested intervals. A confidence set requires an explicit inversion over candidate mean curves. Without that inversion we report adjusted $p$-values, type-I error, and power, not confidence-band width or coverage. Permutation validity for an AIPW-based procedure also needs its own justification; it is not implied merely by using a published functional test.

\section{Conformal prediction for the individual topological effect}
\label{sec:conformal}

We now construct a marginal prediction band for ITTE of a new unit. Since only one potential outcome is observed for each sampled unit, we first construct a weighted split-conformal band for each potential topological outcome and then take their Minkowski difference. This construction uses only the two arm-specific marginal laws and
does not require identification of the joint law of $(\phi_d^0,\phi_d^1)$.
The component methods are inherited from functional conformal prediction
\citep{diquigiovanni}, weighted conformal prediction under covariate shift
\citep{tibshirani}, and causal counterfactual prediction and arm differencing
\citep{lei-candes,schroeder}. Their integration for simultaneous prediction
of the topological individual contrast is the focus here; the weighted-rank
principle and union bound are established tools.
We organize the development into four parts. First, Section~\ref{subsec:arm-calibration} introduces arm-specific calibration under covariate shift. Second, Section~\ref{subsec:arm-differencing} forms individual prediction bands by interval subtraction and establishes exact finite-sample coverage under oracle weights (Theorem~\ref{thm:conformal}). Third, Section~\ref{subsec:estimated} addresses the practical setting where propensity scores are unknown, quantifying coverage stability under estimated likelihood ratios. Finally, Section~\ref{subsec:shrinking} demonstrates why shrinking mean confidence bands cannot serve as individual prediction bands.

\subsection{Arm-specific calibration}\label{subsec:arm-calibration}
Let the training sample, calibration sample, and new full-data unit be independent. Write $I_a$ for calibration indices with $A_i=a$ and
$m_a=|I_a|$. Conditional on their counts, arm-$a$ calibration pairs
are i.i.d. from $P_a$ in Section~\ref{sec:setup}. Fit centers
$\widehat\mu_{a,d}(t,x)$ and scales $\widehat s_{a,d}(t,x)$ using
training data only. These functions are jointly measurable, continuous in $t$ for each $x$, and the scales are strictly positive on $T$. The unscaled choice is $\widehat s_{a,d}\equiv1$. Compactness then ensures a finite supremum score for each continuous response:

\begin{equation}\label{eq:conformal-score}
 S_{i,a}=\sup_{t\in T}
   \frac{|\phi_{i,d}(t)-\widehat\mu_{a,d}(t,X_i)|}
        {\widehat s_{a,d}(t,X_i)},\qquad i\in I_a.
\end{equation}

The score is measurable because its supremum can be taken over a fixed countable dense subset of $T$. Regression accuracy affects efficiency, but is not an assumption for the true-weight coverage statement.

For weights $w_a$ proportional to \eqref{eq:weight}, define

\begin{align}
 W_a(x)&=\sum_{i\in I_a}w_a(X_i)+w_a(x),\nonumber\\
 p_{i,a}(x)&=w_a(X_i)/W_a(x),\qquad
 p_{\infty,a}(x)=w_a(x)/W_a(x),\label{eq:conformal-mass}\\
 \widehat q_a(x)&=\inf\left\{q\in\mathbb R\cup\{+\infty\}:
   \sum_{i\in I_a}p_{i,a}(x)\mathbf1\{S_{i,a}\leq q\}
       +p_{\infty,a}(x)\mathbf1\{q=+\infty\}\geq1-\alpha_a\right\}.
 \label{eq:conformal-quantile}
\end{align}

Thus the quantile includes a mass at $+\infty$ with the weight of the target point. The arm band is

\begin{equation}\label{eq:arm-band}
 \widehat C_{a,d}(x)(t)=
 [\widehat\mu_{a,d}(t,x)\pm\widehat q_a(x)\widehat s_{a,d}(t,x)].
\end{equation}

The definition permits $m_a=0$, when the radius is infinite. Under the full-population assumptions the denominator is positive almost surely. For a more general nonnegative tilt, any zero denominator is assigned the full real-line band; it occurs only off the target support or when no usable mass is available.

\subsection{Individual coverage by arm differencing}\label{subsec:arm-differencing}

Take the pointwise Minkowski difference of the intervals:

\begin{equation}\label{eq:effect-band}
 \widetilde C_d(x)(t)=
 \left[\widehat\mu_{1,d}(t,x)-\widehat\mu_{0,d}(t,x)
 \ \pm\ \{\widehat q_1(x)\widehat s_{1,d}(t,x)
          +\widehat q_0(x)\widehat s_{0,d}(t,x)\}\right].
\end{equation}

\begin{theorem}[True-weight marginal ITTE coverage]\label{thm:conformal}
Under the identification and measurability assumptions of
Section~\ref{sec:setup}, the independent splitting and score
conditions above, and true likelihood-ratio weights, for each arm
\[
 P\{\phi_{n+1,d}^a(t)\in\widehat C_{a,d}(X_{n+1})(t)
          \ \forall t\in T\mid\mathcal T,m_0,m_1\}\geq1-\alpha_a,
\]
where $\mathcal T$ denotes the training data. Consequently, if
$\alpha_0+\alpha_1\leq\alpha$,

\begin{equation}\label{eq:conformal-coverage}
 P\{\delta_{n+1,d}(t)\in\widetilde C_d(X_{n+1})(t)
       \ \forall t\in T\mid\mathcal T,m_0,m_1\}
 \geq1-\alpha_0-\alpha_1\geq1-\alpha.
\end{equation}

Probability averages over calibration observations and the independent new unit, and is marginal over the new covariates.
\end{theorem}

The proof is deferred to Appendix~\ref{app:proofs}.
The symmetric allocation is $\alpha_0=\alpha_1=\alpha/2$.
Separate calibration samples do not imply independence of the coverage events for the two potential outcomes of the same new unit. No \v{S}id\'ak product is used. The coverage bound is a lower bound; ties, discreteness, and the difference construction can make it conservative. A nontrivial distribution-free conditional guarantee at every fixed $x$ is not asserted \citep{barber}.
While Theorem~\ref{thm:conformal} yields valid finite-sample coverage when the true likelihood-ratio weights $w_a$ are known, observational studies typically require estimating propensity scores from finite samples. We address this practical setting next.

\subsection{Estimated likelihood ratios}\label{subsec:estimated}

Theorem~\ref{thm:conformal} guarantees nominal finite-sample coverage under the idealized setting where the true likelihood-ratio weights $w_a(x) = P(A=a)/e_a(x)$ in \eqref{eq:weight} are known exactly. In observational applications, however, the propensity score $e_a(x)$, and consequently the true weight $w_a(x)$, is unknown and must be estimated from data. Substituting estimated weights $\widehat w_a$ into the weighted quantile \eqref{eq:conformal-quantile} perturbs exact weighted exchangeability. To evaluate the sensitivity of the prediction band to estimation error, we bound the resulting coverage discrepancy in terms of the total variation distance between the target distribution and the law induced by the estimated weights.

Let $\widehat w_a$ be positive, training-measurable weights satisfying $0<E_{P_a}\widehat w_a(X)<\infty$ almost surely. Define
\[
 \frac{d\widehat Q_a}{dP_a}(x,\phi)
  =\frac{\widehat w_a(x)}{E_{P_a}\widehat w_a(X)},\qquad
 \Delta_a(\mathcal T)=\operatorname{TV}(Q_a,\widehat Q_a),
\]
with $\operatorname{TV}(P,Q)=\sup_B|P(B)-Q(B)|$.

Conditional on training data, the weighted procedure has a finite-sample coverage lower bound for $\widehat Q_a$. For the intended full-population target, the same procedure therefore satisfies

\begin{align}
 P\{\delta_{n+1,d}\in\widetilde C_d(X_{n+1})\mid\mathcal T,m_0,m_1\}
 &\geq1-\alpha_0-\alpha_1-\Delta_0(\mathcal T)-\Delta_1(\mathcal T),
 \label{eq:tv-conditional}\\
 P\{\delta_{n+1,d}\in\widetilde C_d(X_{n+1})\}
 &\geq1-\alpha_0-\alpha_1-E(\Delta_0+\Delta_1).
 \label{eq:tv-marginal}
\end{align}

For exact weights $w_a$ normalized by $E_{P_a}w_a=1$,
\[
 \Delta_a=\frac12E_{P_a}\left|
 w_a(X)-\frac{\widehat w_a(X)}{E_{P_a}\widehat w_a(X)}\right|.
\]
Thus consistency of normalized ratios in $L^1(P_a)$ is a sufficient
condition for vanishing coverage error. If $\Delta_a=o_p(1)$, boundedness of total variation implies $E\Delta_a\to0$. Neither independent fitting nor an AIPW product-rate condition alone proves this conclusion. Other causal conformal constructions can have different approximation or doubly robust properties \citep{lei-candes,alaa}; such properties are not attributed to \eqref{eq:conformal-score} without a separate theorem.

\subsection{Shrinking mean bands used for individual prediction}\label{subsec:shrinking}

A confidence band for the deterministic mean curve $\psi_d$ does not generally provide an individual prediction band. For a consistent mean estimator, a band whose radius tends to zero can asymptotically cover only individual effects equal to the mean curve. Proposition~\ref{prop:collapse} makes this atom-at-the-mean qualification explicit; without such an atom, prediction coverage tends to zero.

\begin{proposition}[The role of an atom at the mean]\label{prop:collapse}
Let $B_n(t)=[\widehat\psi_d(t)\pm r_n(t)]$, with $r_n(t)\geq0$,
be based on data independent of the new effect $\delta_{n+1,d}$.
If $\|\widehat\psi_d-\psi_d\|_\infty+\|r_n\|_\infty\to0$
in probability, then

\begin{equation}\label{eq:collapse}
 \limsup_n P\{\delta_{n+1,d}\in B_n\}
 \leq P\{\delta_{n+1,d}=\psi_d\text{ in }C(T)\}.
\end{equation}

In particular, the prediction coverage tends to zero when the latter
probability is zero. If additionally $P\{\psi_d\in B_n\}\to1-\alpha$
and the atom has mass $p$, then the prediction coverage tends to
$p(1-\alpha)$.
\end{proposition}

Nonzero individual variance does not exclude an atom at the mean.
Prediction-band width also need not have a positive limit in every
model: a positive limiting residual quantile, together with quantile
consistency and nondegenerate scaling, is a sufficient condition for
nonvanishing width. Degenerate conditional outcomes can permit
shrinking prediction bands. The two-arm construction may remain
conservative even if the effect itself is nearly deterministic, because it does not exploit the unknown dependence between potential outcomes.
\section{Overlap targets, topological stability, and implementation}
\label{sec:extensions}

The statistical methods developed in Sections~\ref{sec:confidence} and~\ref{sec:conformal} treat summary curves largely as generic continuous functions. In this section, we develop the machinery required to connect those inferential guarantees to the specific geometric, topological, and computational properties of persistent homology.

We address four central challenges arising in topological causal inference. First, Section~\ref{subsec:overlap} addresses large full-population conformal weights by targeting an overlap population with bounded balancing weights (Theorem~\ref{thm:overlap}), establishing exact conditions for finite prediction radii (Lemma~\ref{lem:finite}). Second, Section~\ref{subsec:stability} bridges persistence diagram space to silhouette function space through a uniform Wasserstein stability theorem (Theorem~\ref{thm:stability}) and translates diagram approximation error into rigorous prediction band inflation (Theorem~\ref{thm:inflation}). Third, Section~\ref{subsec:joint} establishes simultaneous prediction coverage across multiple homology degrees and silhouette power exponents using one calibrated maximum score (Theorem~\ref{thm:joint}). Finally, Section~\ref{subsec:grid} provides a deterministic interpolation lemma bridging discrete computational grids to continuous interval bands (Lemma~\ref{lem:grid}), and Section~\ref{subsec:software} introduces the companion software package \texttt{tcda\_uq}.

\subsection{Prediction for an overlap population}\label{subsec:overlap}

In observational studies, full-population likelihood-ratio weights $w_1(x) = P(A=1)/\pi(x)$ and $w_0(x) = P(A=0)/(1-\pi(x))$ explode when the propensity score $\pi(x)$ approaches zero or one. In split conformal prediction, when a new test unit has an extreme propensity, its normalized conformal mass $p_{\infty,a}(x) = w_a(x)/W_a(x)$ in \eqref{eq:conformal-mass} can exceed the error budget $\alpha_a$. The weighted quantile \eqref{eq:conformal-quantile} then evaluates to $+\infty$, returning the uninformative real-line band $[-\infty, +\infty]$. To reduce weight instability, we consider a declared overlap target; finite radii still require the calibration-mass condition below.

For a nonnegative measurable tilt $h(X)$ with $0<Eh(X)<\infty$,
define $dQ_{h,X}/dP_X=h/Eh$. Suppose $h(X)=0$ whenever either
treatment has probability zero. Relative to the observed arm law, the target likelihood ratio is proportional to \(u_{h,a}(x)=h(x)/e_a(x)\) on the arm support. The target preserves the conditional potential-outcome law given $X$ and changes the covariate distribution. For the overlap tilt $h(x)=\pi(x)\{1-\pi(x)\}$, the ratios can be chosen as

\begin{equation}\label{eq:overlap-weights}
 u_{\mathrm{ov},1}(x)=1-\pi(x),\qquad
 u_{\mathrm{ov},0}(x)=\pi(x).
\end{equation}

These are the balancing weights of \citet{li}, here used for
weighted conformal calibration.

\begin{theorem}[Overlap-population prediction]\label{thm:overlap}
Assume i.i.d. units, consistency, conditional exchangeability, and
$E[\pi(X)\{1-\pi(X)\}]>0$. Under the independent training and score
conditions of Section~\ref{sec:conformal}, use the true weights in \eqref{eq:overlap-weights}, including the target-point atom at infinity. For a new full-data unit drawn from the overlap population,
\[
 P_{Q_{\mathrm{ov}}}\{\delta_{n+1,d}(t)\in
       \widetilde C_d^{\mathrm{ov}}(X_{n+1})(t)\ \forall t\in T\}
 \geq1-\alpha_0-\alpha_1.
\]
The same statement holds conditionally on training data and calibration arm counts. Global strict positivity is unnecessary because the target has zero mass where $\pi$ is zero or one.
\end{theorem}

The proof is deferred to Appendix~\ref{app:proofs}.
This is a guarantee for a different target population. Replacing an
unbounded full-population ratio by a clipped or overlap weight does not retain exact full-population validity. With estimated ratios, the total-variation bound in Section~\ref{subsec:estimated} applies, with $Q_a$ replaced by the declared target arm law.

\begin{lemma}[Finite weighted quantiles]\label{lem:finite}
Suppose all calibration scores are finite. Put
$U_a=\sum_{i\in I_a}u_a(X_i)$ and assume $U_a+u_a(x)>0$.
The weighted $(1-\alpha_a)$ quantile is finite if and only if

\begin{equation}\label{eq:finite-condition}
 \frac{u_a(x)}{U_a+u_a(x)}\leq\alpha_a.
\end{equation}

If $0\leq u_a\leq B<\infty$, a sufficient condition uniform in $x$
is $U_a\geq B(1-\alpha_a)/\alpha_a$. Under this boundedness
condition, if $E_{P_a}u_a(X)>0$, the sufficient calibration-mass
condition holds with probability tending to one as $m_a\to\infty$.
\end{lemma}

The proof is deferred to Appendix~\ref{app:proofs}.
Bounded weights therefore do not guarantee finite radii at every sample size. For overlap weights take $B=1$. A formal coverage guarantee can hold when the returned band is unbounded, so coverage, frequency of unbounded bands, and width conditional on finiteness are different performance measures. If a known deterministic support bound is available, intersecting a band with that support preserves coverage; this is a separate way to obtain finite, potentially uninformative, intervals. Overlap weighting guarantees bounded conformal weights and may reduce the frequency of unbounded prediction bands. Nevertheless, bounded weights alone do not guarantee finite prediction radii at finite sample sizes. As established by Lemma 10, sufficient calibration mass relative to the test-point weight is also required. In practice, persistence diagrams are estimated from finite, noisy point clouds rather than observed exactly. We examine how diagram approximation error propagates to silhouette prediction bands next.

\subsection{A uniform Wasserstein stability bound}\label{subsec:stability}

In empirical topological data analysis, persistence diagrams are typically constructed from sample point clouds, triangulations, or digitized biomedical images subject to measurement error or finite-sample discretization. Consequently, the analyst observes an approximate diagram $\widehat D$ rather than the true underlying diagram $D$. To establish that inference conducted on approximate diagrams remains statistically valid for the true topological outcomes, we must control how perturbation in diagram space distorts the resulting silhouette curve $\Phi_r(D)$.

Let $W_1$ be the persistence-diagram $1$-Wasserstein distance with
$\ell^\infty$ matching cost and arbitrary diagonal matches. Diagrams
are finite multisets, with the essential-class convention of
Section~\ref{sec:setup}. Whereas calibration mass determines whether a radius is finite, the next result controls error from approximating the topological summary defined in \eqref{eq:silhouette}.

\citet[Theorem 5.3]{kim-lee} already establish a $W_1$ stability
bound with constant $1+2Lr c^{r-1}$, under their persistence-ratio
bound and with $c$ determined by diagram lifetimes. Their Appendix C.5
gives the diagram-independent constant $3$ when $r=1$. The next result
recovers that special case and extends a diagram-independent constant
to all $r\geq1$, including empty diagrams under our convention.

\begin{theorem}[Stability of the normalized power silhouette]
\label{thm:stability}
For every $r\geq1$ and every such pair of diagrams,
\begin{equation}\label{eq:stability}
 \|\Phi_r(D)-\Phi_r(E)\|_\infty
       \leq K_r W_1(D,E),\qquad K_r=1+2r.
\end{equation}
No positive lower bound on persistence is required.
\end{theorem}

The proof is deferred to Appendix~\ref{app:proofs}.
The norm can be taken on $\mathbb R$ or on the specified interval $T$. The proof bounds the change in the normalized weights directly. In particular, the finite-diagram bound does not deteriorate merely because a short-lived feature approaches the diagonal. This statement concerns $W_1$, not the bottleneck distance: an unbounded number of small bars can have substantial total weight while their bottleneck distance to the diagonal is small.

The direction of \eqref{eq:stability} cannot be reversed. For
$D=\{(0,2)\}$ and $E=\{(0,2),(0,2)\}$, the silhouettes agree for
every $r$, whereas $W_1(D,E)=1$. Consequently, a confidence or prediction band for a silhouette does not itself give a diagram-distance confidence set. If a hard persistence threshold is used, apply the theorem to the post-threshold diagrams; the threshold map is not automatically Lipschitz in the original diagrams. Thresholding also changes the target.

\begin{theorem}[Propagation of diagram approximation]\label{thm:inflation}
Suppose a fixed measurable approximation pipeline is applied in the
same way to calibration and target units. Let
$\widehat\phi_{n+1,d}^a=\Phi_r(\widehat D_{n+1,d}^a)$, and let
$[L_a(x,t),U_a(x,t)]$ cover this approximate target curve with
probability at least $1-\alpha_a$. If, under the target arm law,
\[
 P\{W_1(D_{n+1,d}^a,\widehat D_{n+1,d}^a)\leq\varepsilon_a\}
       \geq1-\gamma_a,
\]
then the inflated band
\begin{equation}\label{eq:inflation}
 [L_a(x,t)-K_r\varepsilon_a,\ U_a(x,t)+K_r\varepsilon_a]
\end{equation}
covers the exact target curve with probability at least
$1-\alpha_a-\gamma_a$. Differencing the two inflated arm bands gives
ITTE coverage at least $1-\alpha_0-\alpha_1-\gamma_0-\gamma_1$,
and adds $K_r(\varepsilon_0+\varepsilon_1)$ to the effect half-width.
\end{theorem}

The proof is deferred to Appendix~\ref{app:proofs}.
No independence of approximation failure and calibration failure is
needed. With estimated conformal weights, the marginal ITTE lower bound becomes
$1-\alpha_0-\alpha_1-\gamma_0-\gamma_1-E\Delta_0-E\Delta_1$, with the discrepancies defined in Section~\ref{subsec:estimated}. A conditional version given training data uses $\Delta_0(\mathcal T)+\Delta_1(\mathcal T)$ instead of their expectations and requires the approximation-probability bounds under the same conditioning. An observed small difference
between two silhouette curves is not an upper bound on the diagram
error required by this theorem.
Having established stability with respect to diagram approximation, we next address the discrete design parameters that define the summary curves themselves: the homology degree and the silhouette power exponent.

\subsection{Joint degrees and exponents}\label{subsec:joint}

A single filtration generates persistence diagrams across multiple homology degrees (for example, $d=0$ for connected components, $d=1$ for one-dimensional loops, and $d=2$ for two-dimensional cavities). In addition, the power exponent $r \ge 1$ in \eqref{eq:silhouette} can be adjusted to weight feature persistence. Evaluating each degree or exponent separately raises a multiplicity problem, while selecting one post hoc invalidates coverage. To resolve this, we construct a joint prediction band covering an arbitrary prespecified collection of degree-exponent pairs simultaneously.

Let $\mathcal J$ be a finite, prespecified collection of pairs $(d,r)$. Use common calibration indices within each arm, and training-only fits, to form the single scalar score
\begin{equation}\label{eq:joint-score}
 S_{i,a}^{\mathcal J}=
 \max_{(d,r)\in\mathcal J}\sup_{t\in T}
 \frac{|\phi_{i,d}^r(t)-\widehat\mu_{a,d}^r(t,X_i)|}
      {\widehat s_{a,d}^r(t,X_i)}.
\end{equation}

This is the multivariate functional max-score construction of
\citet{diquigiovanni}, combined with weighted calibration.
\begin{theorem}[Joint ITTE prediction]\label{thm:joint}
Under the true-weight conditions of Theorem~\ref{thm:conformal},
calibrate \eqref{eq:joint-score} at arm errors $\alpha_0,\alpha_1$. Use the common arm quantile for all members of $\mathcal J$ and difference the arm intervals. Then
\[
 P\{\delta_{n+1,d}^r(t)\in\widetilde C_{d,r}^{\mathcal J}
 (X_{n+1})(t)\ \forall(d,r)\in\mathcal J,\ \forall t\in T\}
 \geq1-\alpha_0-\alpha_1.
\]
\end{theorem}

The proof is deferred to Appendix~\ref{app:proofs}.

No further correction across $\mathcal J$ is needed when its maximum
is inside the score. Alternatively, construct separate effect bands at error $\alpha/|\mathcal J|$, using error $\alpha/(2|\mathcal J|)$
for each arm of each band; the union bound then gives joint coverage
at least $1-\alpha$. A \v{S}id\'ak correction needs additional
dependence assumptions and has no such general guarantee here.

The exponent is part of the estimand. For a diagram with bars $(0,2)$ and $(0,1)$,
\[
 \Phi_r(D)(3/2)=\frac{2^{r-1}}{2^r+1},
\]
which is strictly increasing in $r$. Thus different exponents can
define different causal summaries, although they need not differ for
every diagram or data-generating law. Choose the exponent in advance, using independent training data with a declared conditional target, or include it in $\mathcal J$. Width comparisons across exponents are sensitivity analyses across targets, not efficiency comparisons for a single fixed target.

\subsection{Covariate-adaptive widths}\label{subsec:adaptive}

A positive training-only scale $\widehat s_{a,d}(t,x)$ is already accommodated in the score definition \eqref{eq:conformal-score}. Scaling by $\widehat s_{a,d}(t,x)$ modulates band width according to local heteroscedasticity across covariates without modifying the target population or compromising the marginal finite-sample coverage guarantee. In contrast, attempting to produce locally adaptive bands by inserting a kernel factor $K_h(x_0,X)$ into the conformal weights alters the inferential target itself, redefining it to a localized covariate population centered at $x_0$. Such localized reweighting requires deriving the corresponding density ratio for that specific target and does not establish exact conditional coverage at every covariate value.

\subsection{From a grid to a continuous band}\label{subsec:grid}

All preceding coverage guarantees (Theorems~\ref{thm:tate-band}, \ref{thm:conformal}, \ref{thm:overlap}, and~\ref{thm:joint}) establish simultaneous validity over the continuous filtration interval $t \in T$. In computational software, however, quantiles and band boundaries are necessarily evaluated on a discrete grid of filtration points $t_0 < t_1 < \dots < t_m$. To bridge this discretization gap rigorously, we establish a deterministic extension lemma based on the Lipschitz continuity of silhouette curves.

Let $t_-=t_0<\cdots<t_m=t_+$, set $\Delta=\max_j(t_{j+1}-t_j)$, and write $I_m$ for piecewise linear interpolation on this grid.
\begin{lemma}[Deterministic grid extension]\label{lem:grid}
Let $f$ be $L$-Lipschitz. On the event $f(t_j)\in[\ell_j,u_j]$ for every grid point, the continuous band

\begin{equation}\label{eq:grid-band}
 [\,I_m\ell(t)-L\Delta/2,\ I_m u(t)+L\Delta/2\,],\qquad t\in T,
\end{equation}

contains $f$ throughout $T$. Any probability lower bound for the
gridwise event therefore also applies to this inflated continuous band.
\end{lemma}

The proof is deferred to Appendix~\ref{app:proofs}.
One can replace $L\Delta/2$ by the smaller, interval-specific quantity $2L(t-t_j)(t_{j+1}-t)/(t_{j+1}-t_j)$. For an arm silhouette take $L=1$; for TATE, a Lipschitz CTATE version, or ITTE take $L=2$. The added full width is $L\Delta$. For a root-$n$ band this inflation preserves the width order if $\Delta=O(n^{-1/2})$, and is asymptotically negligible if $\Delta=o(n^{-1/2})$. Replace $n^{-1/2}$ by the justified CTATE scale when appropriate. These statements extend a valid gridwise band; they do not by themselves prove validity of a growing-grid bootstrap. A grid maximum without this extension establishes only gridwise prediction coverage. For asymptotic confidence inference, any claim that a computed grid critical value approximates the continuous critical value also requires suitable grid refinement and process equicontinuity.

\subsection{Software and reproducibility}\label{subsec:software}

The inferential framework developed across this work is implemented in \texttt{tcda\_uq}, an open-source Python library providing a unified interface for TATE mean confidence bands (Section~\ref{subsec:tate}), fixed-covariate CTATE confidence bands (Section~\ref{subsec:ctate}), full-population ITTE prediction bands (Section~\ref{sec:conformal}), and overlap-population prediction bands (Section~\ref{subsec:overlap}), incorporating the diagram inflation (Section~\ref{subsec:stability}), joint max-score (Section~\ref{subsec:joint}), and grid extension (Section~\ref{subsec:grid}) guarantees. Version 0.1.0,
installation instructions, and source code are available at
\url{https://github.com/hugogobato/tcda_uq}. The package accepts
grid-valued summaries and includes persistence-diagram, point-cloud,
and image-to-silhouette preprocessing. Its prediction interface
declares the target population and whether the likelihood ratios are
known or estimated. Tutorials cover the three inferential targets,
ORBIT point clouds, and a CT-slice pipeline; the reproduction materials
retain the frozen experiment sources and all $11{,}200$ production
fitted-sample records.

The library is designed for reuse across these data modalities, with
the prespecified representation and statistical assumptions determining
the scope of each guarantee. Its computed bands and inclusion diagnostics
are on finite grids; continuous-interval coverage additionally needs
the extension above or another justified discretization argument.
\citet{kim-lee} already provide Python code for their mean-effect
methodology. The software contribution here is the packaged integration
of mean confidence and individual prediction, with target declarations,
tutorials, and reproduction evidence.

\section{Simulation Studies}
\label{sec:experiments}

Three designs are used for this paper's simulation studies. A bounded diagram-valued design supplies analytic targets and correctly specified regression models. A functional design generates curves directly, isolating the statistical procedures, while a point-cloud design includes persistence and silhouette computation. The latter two designs also expose failures outside the sufficient regime.

\subsection{Design scope and evaluation metrics}
\label{sec:sim-design}

All reported simulations evaluate three performance metrics across the complete evaluation grid: 1. simultaneous coverage, defined as the empirical frequency with which a band covers its functional target across all grid points at once; 2. grid-averaged width, measuring band informativeness; and 3. the unbounded share, recording the fraction of test draws at which the weighted conformal quantile is infinite due to weak overlap. Mean coverage is evaluated separately for each reported component and covariate profile. Confidence bands for mean effects (TATE and fixed-profile CTATE) target nominal $0.90$ coverage; conformal prediction bands for the individual effect (ITTE) target nominal $0.95$ coverage (this discrepancy is an artifact of independent implementation workflows rather than a deliberate methodological choice). Each primary setting aggregates between $200$ and $2{,}000$ independent Monte Carlo fitted-sample replications (depending on computational complexity of running the setting experiments), and prediction coverage evaluates $200$ freshly drawn target units per fitted sample. Monte Carlo standard errors are computed with the fitted sample as the independent replication unit. Complete mathematical formulations of all data-generating processes, exact nuisance parameter matrices, and exhaustive numerical tables are compiled in Appendix~\ref{app:experiments}. In addition to the three synthetic designs evaluated below, we also investigate the synthetic ORBIT point-cloud benchmark introduced by \citet{adams} and studied by \citet{kim-lee}, which generates chaotic dynamical-system trajectories on the unit torus. Because both potential-outcome clouds are mathematically constructed and retainable, this benchmark permits paired Monte Carlo reference calculations. The complete mathematical formulation of the linked twist map, simulation design, and empirical results for ORBIT are detailed in Appendix~\ref{app:orbit}.

\subsection{A bounded diagram-valued benchmark}
\label{sec:sim-bounded}
\label{tab:bounded-rerun}

The bounded diagram-valued design has bounded responses, correctly specified regression models, and targets known in closed form. Each unit's potential persistence diagram is either empty or contains the two bars $(0,2)$ and $(1,3)$. Over the filtration grid $\mathcal{T} = [1.1, 1.9]$, the plain power-weighted silhouette is identically zero (when empty) or $1/2$ (when non-empty). Covariates $X = (X_1, X_2, X_3)$ are sampled iid from $\operatorname{Unif}([-1,1]^3)$. Non-empty diagram probabilities under control and treatment are $q_0(X) = 0.3 + 0.1 X_1$ and $q_1(X) = 0.6 + 0.1 X_2$. Treatment assignment is generated with known propensity score $\pi(X) = 0.25 + 0.5\operatorname{expit}(X_1)$, which is bounded away from zero and one in $[0.37, 0.63]$.

Under this construction, the true causal targets are exact and linear: the marginal TATE is $\psi_d(t) \equiv 0.15$, and the conditional CTATE profile is $\tau_d(t, x) = 0.15 + 0.05(x_2 - x_1)$. Independent sample-splitting folds, bounded nuisance fits, and the linear second-stage DR-learner satisfy the sufficient conditions of Theorem~\ref{thm:tate-band} and Proposition~\ref{prop:ctate-band}. Appendix~\ref{app:bounded-rerun} details the complete mathematical formulation of this benchmark, including the nuisance parameter coefficient bounds $[-2,2]$, Fourier calibration on 50 filtration points, and target covariate tilt mechanisms.

\begin{figure}[htbp]
\centering
\includegraphics[width=\textwidth]{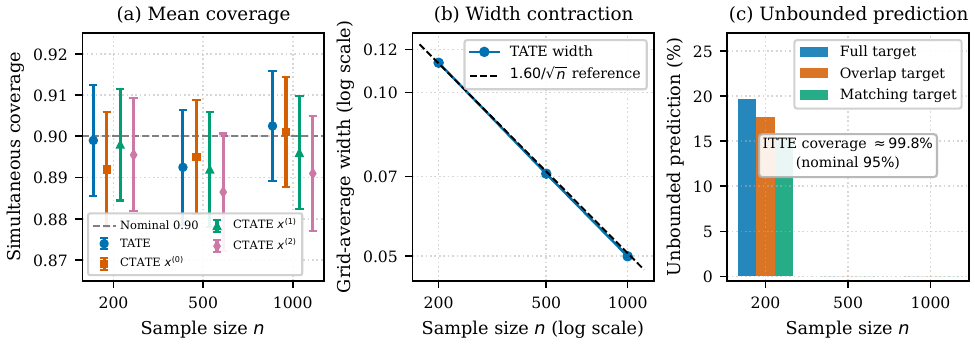}
\caption{Bounded benchmark: $2{,}000$ replications at each $n\in\{200,500,1000\}$. (a) TATE and three separately evaluated fixed-profile CTATE coverages; error bars are two Monte Carlo standard errors. (b) Mean TATE width and the $1.60/\sqrt n$ reference. (c) Unbounded prediction share under each declared target. Nominal confidence/prediction levels are $0.90/0.95$.}
\label{fig:bounded-benchmark}
\end{figure}

Figure~\ref{fig:bounded-benchmark} summarizes the results across $2{,}000$ replications for $n \in \{200, 500, 1000\}$. The empirical evidence aligns closely with asymptotic theory. In panel (a), simultaneous TATE coverage settles near the nominal $0.90$ level at all sample sizes ($0.8990 \pm 0.0067$ at $n=200$, $0.8925 \pm 0.0069$ at $n=500$, and $0.9025 \pm 0.0066$ at $n=1000$). Fixed-profile CTATE coverage across three evaluation points ($x^{(0)}, x^{(1)}, x^{(2)}$) tracks nominal coverage with equal precision, spanning $0.8865$ to $0.9010$. In panel (b), grid-averaged TATE width contracts steadily from $0.1135$ at $n=200$ to $0.0709$ at $n=500$ and $0.0499$ at $n=1000$, with unrounded $\sqrt{n} \times \text{width}$ products of $1.605$, $1.585$, and $1.579$, matching the parametric $n^{-1/2}$ rate.

In panel (c), the composed conformal ITTE prediction bands maintain empirical coverage between $0.9979$ and $0.9989$ across all sample sizes and target populations, above the nominal $0.95$ level. Armwise calibration and the Minkowski difference yield conservative composed bands; these results do not isolate the contribution of each step. At $n=200$, finite calibration sample size combined with covariate shift causes $19.6\%$ of full-population prediction bands to be unbounded. Overlap and matching tilts reduce this unbounded fraction to $17.6\%$ and $15.5\%$, respectively. At $n=500$ and $n=1000$, no unbounded bands are observed across these targets ($0.0\%$). This empirical frequency does not guarantee a finite radius in every sample; sufficient calibration mass relative to the new-point weight is still required (Lemma~\ref{lem:finite}).

\subsection{Functional design and approximation failures}
\label{sec:sim-functional}
\label{sec:sim-confidence}
\label{tab:remaining-functional}

The functional design generates summary curves directly from a functional data-generating process, bypassing homology computation to evaluate the statistical inferential procedures in isolation. Covariates $X \in \mathbb{R}^p$ are drawn from an equal two-component Gaussian mixture. Conditional outcome means follow a Fourier-in-$t$ basis model whose coefficient functions are linear in $X$, perturbed by smooth mean-zero Gaussian process noise that supplies the aleatoric spread. The complete mathematical specification of this functional generator, including the exact Gaussian subgroup means $m_0, m_1$, logistic assignment features, and explicit coefficient matrices $\Gamma_{ad}$, is detailed in Appendix~\ref{app:levelA-tate}.

Three distinct functional scenarios evaluate the limits of the asymptotic theory: 1. a Base design where the true conditional response is linear in $X$ and periodic in $t$, matching the specification of the functional linear second stage; 2. a Step design where a treatment-arm perturbation $\mathbf{1}\{X_1 > 0.7\}\sin(\pi t)$ introduces a sharp covariate discontinuity that violates the linear second-stage specification and periodic Fourier basis; and 3. a Three-component design generating three simultaneous functional curves to examine componentwise mean inference and joint prediction calibration without topological homology.

\begin{figure}[htbp]
\centering
\includegraphics[width=0.75\textwidth]{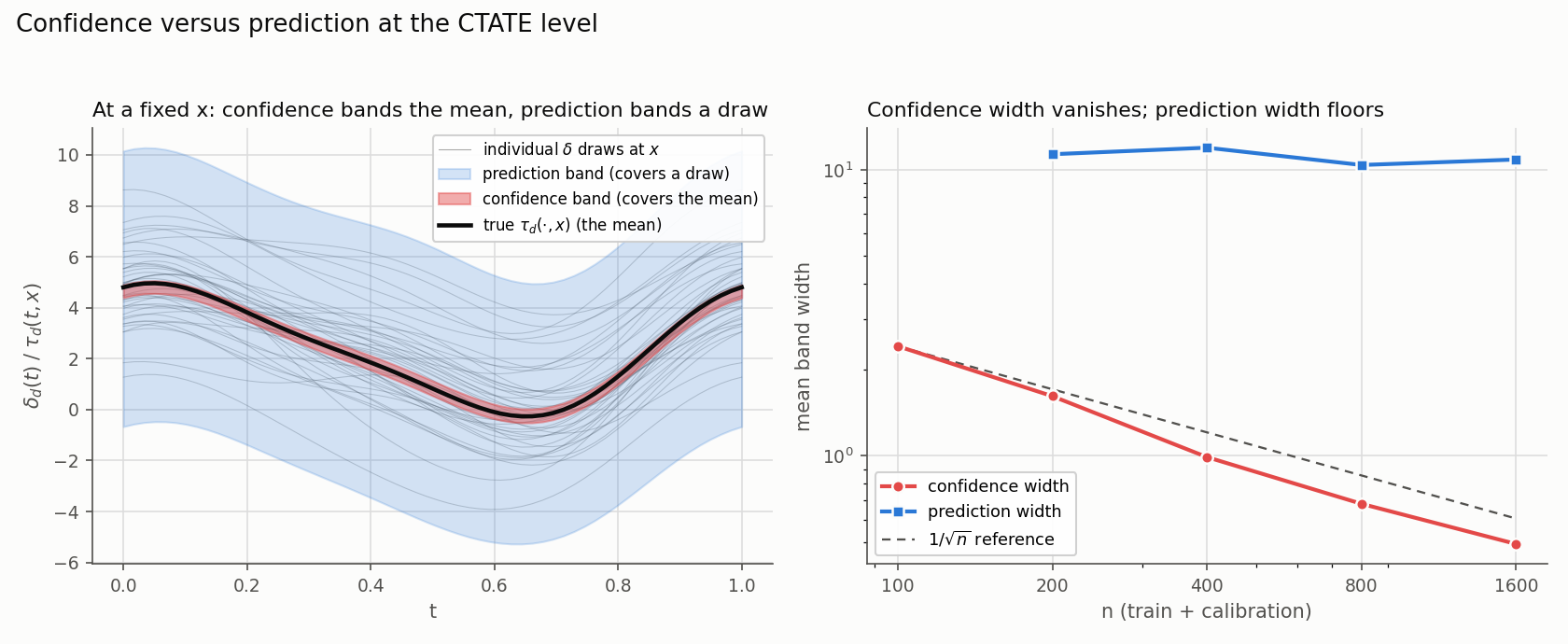}
\caption{Illustrative functional confidence versus prediction. Left: one fixed-profile fit with individual effects. Right: widths averaged over three fits per sample size. Prediction width retains individual variation in this design. These archived adaptive fits use estimated weights and scales; they do not establish exact or conditional prediction coverage.}
\label{fig:cvp}
\end{figure}

\begin{figure}[!ht]
\centering
\includegraphics[width=0.95\textwidth]{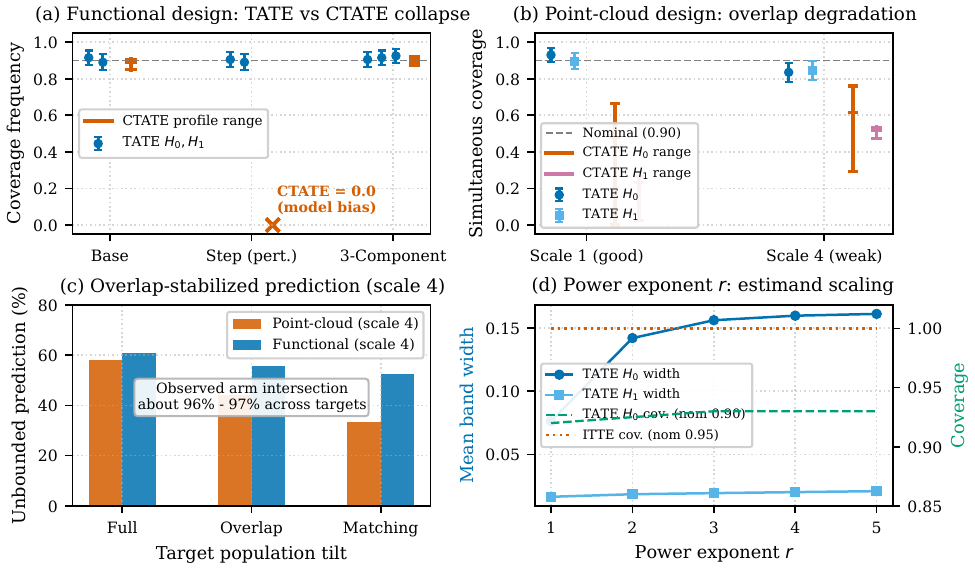}
\caption{Finite-grid stress diagnostics at $n=500$, $200$ replications per setting. (a) Component/profile-specific functional TATE and CTATE coverage under Base, Step, and Three-component scenarios. (b) Point-cloud mean coverage under logit scales $\kappa=1,4$. (c) Unbounded prediction shares under weak overlap ($\kappa=4$) using known target-to-calibration ratios in both designs, with full, overlap, and matching targets each evaluated under its own covariate law. (d) Mean confidence widths and empirical coverage across distinct exponent-specific targets $r\in\{1,2,3,4,5\}$. These frequencies do not assert exact coverage equality or a general monotonicity of width in $r$.}
\label{fig:stress-diagnostics}
\end{figure}

The contrast between the Base and Step scenarios isolates where the asymptotic guarantees hold and where they fail outside the sufficient regime. As shown in Figure~\ref{fig:stress-diagnostics}(a) ($n=500$, $200$ replications), marginal TATE coverage remains near nominal in both the Base design ($0.915 \pm 0.020$ for $H_0$, $0.890 \pm 0.022$ for $H_1$) and the Step design ($0.905 \pm 0.021$ for $H_0$, $0.890 \pm 0.022$ for $H_1$). Marginal AIPW estimation uses known propensity scores and does not depend on the misspecified CTATE second stage.

Fixed-profile CTATE coverage instead shows an approximation failure: while CTATE profile coverage spans $0.850$ to $0.905$ in the Base design, it falls to an empirical $0.000$ across all three evaluation profiles in the Step design. Because the discontinuous step perturbation cannot be represented by the linear DR-learner, point evaluation at specific covariate values suffers uncorrected approximation bias. This failure illustrates the need to control second-stage approximation bias at the queried covariate profile, even when propensity scores are known. In the Three-component scenario, whole-grid coverage evaluated separately by component and profile ranges from $0.905$ to $0.925$ for TATE and $0.875$ to $0.920$ for CTATE. These ranges are not the coverage probability of a joint event across all three components. The stepped perturbation, three-component setup, and comprehensive numerical diagnostics across sample sizes $n \in \{50, 100, 200, 500, 1000\}$ are documented in Appendices~\ref{app:functional-iid} and~\ref{app:ctate-simulation} (Tables~\ref{tab:appA-tate} and~\ref{tab:ctate-simulation}).

Figure~\ref{fig:cvp} contrasts a fitted conditional mean band with a wider individual prediction band in the illustrated functional design. The individual draws vary around the conditional mean. Its sample-size panel summarizes three fits per size, with learned weights and adaptive scales; these illustrations do not establish nominal coverage or a general limiting prediction width. The confidence and prediction bands address different targets.

\subsection{Point-cloud design and topological stress testing}
\label{sec:sim-pointcloud}
\label{sec:sim-levelb}
\label{sec:sim-prediction}
\label{sec:sim-topology}
\label{tab:remaining-full-weights}

The point-cloud design executes the full topological causal pipeline from raw data to persistent homology to power-weighted silhouettes. Each unit's potential outcome is a two-dimensional point cloud whose number of loops ($H_1$ features) is generated by a rounded step function of covariates and treatment: $L_i^a = \operatorname{clip}(\operatorname{round}(\text{base} + a \cdot \text{effect} + \text{slope} \cdot (X_i \cdot w)), 0, \text{max\_loops})$. Points are sampled on separate radius-one circles with Gaussian jitter and background noise; circle centers occupy a $3\times3$ grid with spacing three. Persistence diagrams are extracted via alpha complexes. The primary target-matched study and exponent sweep use $50$ filtration points on $[0,1]$, in standard tent units, with $r=3$ and $r\in\{1,\dots,5\}$, respectively (Appendix~\ref{app:topology-rerun}). Because the loop count is an integer step function, the conditional mean summary is nonlinear and need not be represented by the working regression. Appendix~\ref{app:levelB} gives the shared circle geometry and the historical stress configuration, which instead uses $45$ filtration points on $[0,1]$ in rescaled $\sqrt{2}\phi_d$ units. These configurations are distinct; the results below concern the primary target-matched study and its exponent sensitivities.

Figure~\ref{fig:stress-diagnostics}(b) illustrates the performance of asymptotic mean bands under good overlap (logit scale $\kappa=1$) and weak overlap ($\kappa=4$). Under good overlap, marginal TATE coverage remains near nominal ($0.930 \pm 0.018$ for $H_0$, $0.895 \pm 0.022$ for $H_1$). However, CTATE profile coverage is severely degraded by loop-count rounding, covering only $0.000$ to $0.665$ across profiles in $H_0$ and $0.025$ to $0.265$ in $H_1$. Under weak overlap ($\kappa=4$), marginal TATE coverage itself deteriorates, dropping to $0.835 \pm 0.026$ in $H_0$ and $0.845 \pm 0.026$ in $H_1$, as large propensity weights inflate variance and boundary error.

In the target-matched point-cloud runs, empirical ITTE coverage is $1.0000$ and arm-intersection coverage ranges from $0.9598$ to $0.9704$. Zero observed failures do not establish a failure probability of zero. Under weak overlap ($\kappa=4$), the full-target unbounded share is $57.9\%$; overlap and matching targets reduce it to $44.3\%$ and $33.3\%$, respectively, using true ratios. Their finite widths are approximately $2.15$--$2.19$. In the functional design with known ratios, the corresponding full/overlap/matching shares are $60.8\%$, $55.6\%$, and $52.3\%$ (Figure~\ref{fig:stress-diagnostics}c). The learned-ratio functional procedure instead gives $54.6\%$, $37.8\%$, and $24.0\%$; those results do not have the exact known-ratio guarantee. These target rows describe different populations. Complete comparisons across exact, learned, constant, clipped, and shrunk weights appear in Appendix~\ref{app:stabilization} (Table~\ref{tab:remaining-weights} and Figure~\ref{fig:remaining-weights}), with target-matched point-cloud results in Table~\ref{tab:topology-rerun}.

Figure~\ref{fig:stress-diagnostics}(d) reports the exponent study for $r\in\{1,2,3,4,5\}$. Empirical TATE coverage ranges from $0.920$ to $0.930$ for $H_0$ and $0.890$ to $0.905$ for $H_1$; empirical ITTE coverage is $1.0000$ in these runs. Grid-averaged TATE confidence width for $H_0$ increases from $0.076$ at $r=1$ to $0.161$ at $r=5$. These are mean-band widths, distinct from the finite ITTE widths in Table~\ref{tab:remaining-exponents}, which are nonmonotone. Changing $r$ reweights features by persistence and changes the functional target, so this is a sensitivity comparison across estimands. The forward error bounds of Theorems~\ref{thm:stability} and~\ref{thm:inflation} do not imply monotone statistical widths in $r$, and these separate exponent runs do not test the joint calibration of Theorem~\ref{thm:joint}. Complete exponent and multi-degree diagnostics appear in Appendix~\ref{app:degree-r} (Tables~\ref{tab:remaining-degrees} and~\ref{tab:remaining-exponents}).

\subsection{When to use which: empirical decision rule}
\label{sec:sim-regime}

The simulation results suggest the following target-specific guidance, subject to the corresponding theoretical assumptions:

1. \emph{Mean effect inference under valid assumptions}: When the scientific objective is to estimate population-level mean curves (TATE or CTATE), use multiplier confidence bands under the corresponding overlap, approximation, and process conditions. TATE has width $O_p(n^{-1/2})$ under Theorem~\ref{thm:tate-band}. At a fixed covariate profile, general CTATE width is $O_p\{v_n(x)\}$ under Proposition~\ref{prop:ctate-band}, with bias negligible at that scale. The verified correctly specified fixed-dimensional regression case gives $v_n(x)\asymp n^{-1/2}$ (Corollary~\ref{cor:ols}).

2. \emph{Individual effect prediction under misspecification}: When the scientific objective is individual-effect prediction, weighted conformal bands provide finite-sample marginal coverage under known likelihood-ratio weights and independent calibration, regardless of the accuracy of the fitted outcome predictors. If the objective instead concerns TATE or CTATE, model misspecification must be addressed within the mean-inference procedure; conformal prediction for ITTE is not a substitute for valid confidence inference on population or conditional mean effects. Estimated likelihood ratios require the additional coverage qualifications of Section 4.3.

3. \emph{Stabilization under weak overlap}: Under weak overlap, researchers may consider a prespecified overlap or matching target population to reduce instability in the conformal weights. Such reweighting changes the estimand and may reduce the frequency of unbounded prediction sets, but it does not guarantee finite or informative prediction bands at every finite sample size.

4. \emph{Multiplicity and persistence weighting}: Joint-score calibration can be useful for a prespecified collection when splitting the error budget produces coarse or infinite calibration quantiles, as observed in the three-component study. It has no universal width advantage over separate Bonferroni bands; a \v{S}id\'ak correction additionally requires dependence assumptions. Choose the persistence power exponent $r$ based on domain interest in persistent versus transient features, recognizing that different values can define different causal targets.

\section{Semi-Synthetic Benchmark and Real Clinical Application}
\label{sec:experiments-realdata}

Moving beyond purely synthetic simulation environments, we evaluate our inferential methods on two empirical settings: 1. a semi-synthetic benchmark constructed from real clinical chest computed tomography (CT) scans (SARS-CoV-2), where real patient lung-slice images are paired with synthetic treatment assignments; and 2. a genuine observational clinical application from the PACCE trial evaluating metastatic colorectal cancer lesion trajectories under chemotherapy plus targeted biologic therapies. The semi-synthetic benchmark retains both generated potential outcomes and can expose errors against an analytic reference contrast, whereas the clinical application observes one treatment arm per patient and introduces observation and population-selection issues that randomization alone does not resolve.

\subsection{SARS-CoV-2 CT-slice benchmark}
\label{sec:real-sars}

The semi-synthetic benchmark follows the CT-slice design of \citet[Section 6]{kim-lee}, using the SARS-CoV-2 dataset of \citet{soares}. The image pools contain $1{,}252$ COVID-19 and $1{,}229$ non-COVID-19 grayscale chest CT slices. The benchmark independently resamples these finite slice pools, randomly pairs the images, and generates covariates and treatment assignments. Each complete image is processed with a lower-star (sublevel) cubical filtration. Degree-zero features describe connected components born at local intensity minima and their merging scales. Only degree zero is analyzed. The preprocessing applies no lung mask, so these features summarize the supplied image's intensity geometry and are not isolated measures of lung pathology.

The causal structure is semi-synthetic: the image pools are real, but the counterfactual pairing is constructed by treating the infected pool as control and a mixture containing $75\%$ non-COVID-19 and $25\%$ COVID-19 slices as treated. The resulting target is the contrast between the two generated slice distributions, with analytic form $\psi_0(t) = 0.75 (\bar\phi_{\text{non-COVID}}(t) - \bar\phi_{\text{COVID}}(t))$. Covariates and treatment propensity are simulated independently of the images, ensuring unconfoundedness by construction.

\begin{figure}[htbp]
\centering
\includegraphics[width=0.46\textwidth]{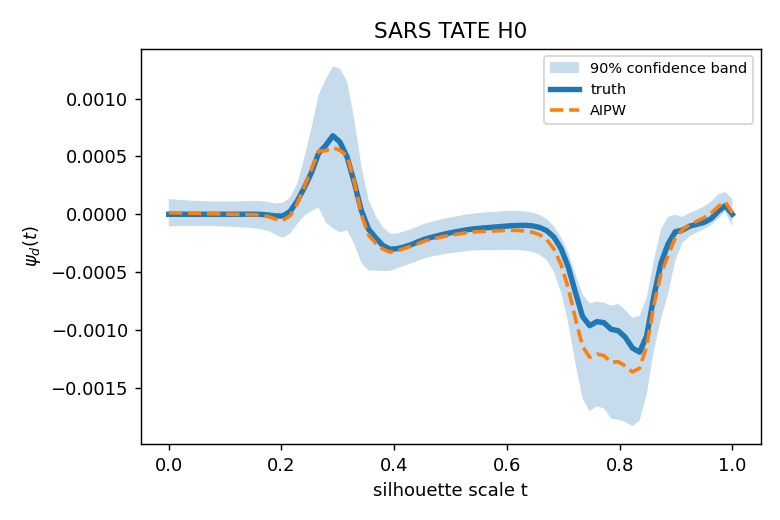}
\includegraphics[width=0.46\textwidth]{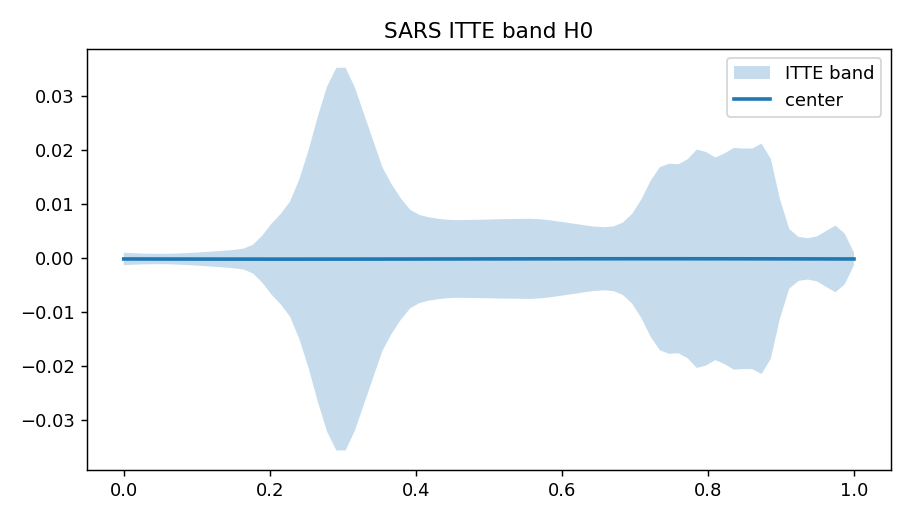}
\par\smallskip
\includegraphics[width=0.95\textwidth]{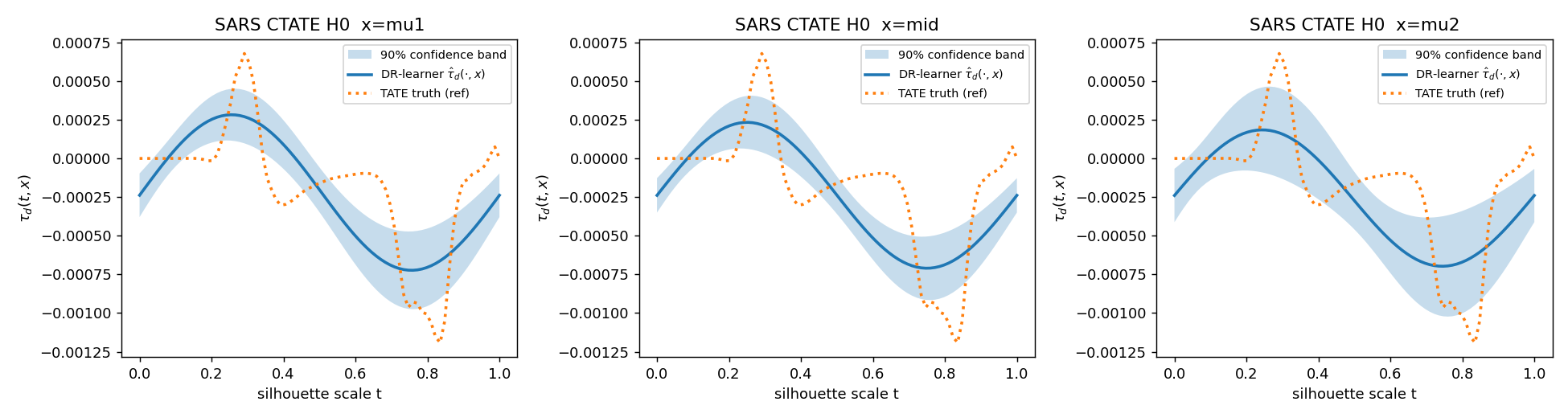}
\caption{Archived CT-slice illustration at $n=400$, $r=0.1$, in $\sqrt{2}\phi_0$ units, outside the $r\ge1$ stability regime. Top left: nominal $90\%$ multiplier band and sample-specific paired-contrast reference. Top right: qualitative nominal $90\%$ prediction band with no paired target overlaid. Bottom: working fixed-profile mean bands at three covariate locations. These panels are distinct from the repeated-sampling $r=1$ study in Appendix~\ref{app:sars-rerun}.}
\label{fig:sars}
\end{figure}

Figure~\ref{fig:sars} retains an archived single-sample illustration at $n=400$ using $r=0.1$ and the rescaled $\sqrt{2}\phi_0$ convention. This exponent is outside the $r\ge1$ stability regime. The displayed mean reference is sample-specific, and the prediction panel is qualitative: no paired individual target is overlaid or coverage frequency estimated. Appendix~\ref{app:sars-diagnostic} gives its construction and diagnostics.

A separate repeated-sampling study at $n=400$ uses $r=1$ in standard tent units (Appendix~\ref{app:sars-rerun}, Table~\ref{tab:sars-rerun}). Across $200$ independent fitted-sample replications, TATE whole-grid coverage is $0.920\pm0.019$ at nominal level $0.90$, while coverage at the three separately evaluated CTATE profiles spans $0.700$ to $0.895$. In this study, mean coverage is evaluated against the analytic empirical-pool curve rather than the sample-specific reference in Figure~\ref{fig:sars}; fixed Fourier approximation can introduce bias. Full-target ITTE coverage is $0.9999$ at nominal level $0.95$, with unbounded share $0.0512$ and finite width $0.0732$. These are repeated-sampling diagnostics for newly generated paired slice contrasts, not values read from the archived figure. Anonymous slice-pool resampling does not provide a patient-level causal interpretation.

\subsection{PACCE: an observed lesion-trajectory contrast}
\label{sec:pacce-application}

The PACCE trial compared chemotherapy plus bevacizumab with the same regimen plus panitumumab in metastatic colorectal cancer \citep{hecht2009}. We analyze the Amgen study 20040249 data supplied through Project Data Sphere.\footnote{The \href{https://data.projectdatasphere.org/projectdatasphere/html/content/84e64633-4d24-44cd-b9d6-67125010d4c2}{Project Data Sphere PACCE record} identifies Amgen as the provider and lists the protocol, case-report form, data dictionary, and both-arm SAS data. While the trial catalog lists $1{,}053$ randomized patients, the released dataset provides subject-level records for $842$ patients ($421$ per assigned arm). In accordance with the Project Data Sphere Data Use Agreement, patient-level records are restricted and cannot be redistributed; external researchers must request access directly from Project Data Sphere.} We evaluate assigned treatment, with arm one ($A=1$) denoting the panitumumab regimen and arm zero ($A=0$) denoting control, retaining chemotherapy backbone (oxaliplatin versus irinotecan) as a key baseline covariate. The outcome captures the joint geometry of multi-focal tumor burden over time constructed from central target-lesion longest diameters.

To construct a topological representation of early tumor evolution, the analysis defines an observation window encompassing baseline assessments and scheduled week-12 assessments through study day $117$. Computing persistent homology on an individual's tumor trajectory requires sufficient longitudinal geometric support. Specifically, a patient is deemed eligible if they possess: 1. at least two distinct assessment visit days, and 2. at least four total target-lesion diameter measurements within this window. Among the $842$ supplied patients, $566$ satisfy these criteria ($288$ control, $278$ panitumumab), while $276$ patients are excluded due to missing or sparse follow-up (detailed in Appendix~\ref{app:pacce-construction}). For each eligible subject, every lesion assessment generates a point in the plane with coordinates $u = \mathrm{day} / 117$ (normalized assessment day) and $z = \log\{\max(\mathrm{diameter}, 5)\}$ (logarithm of lesion diameter in millimeters, floored at $5$~mm). Alpha-complex persistent homology in degree zero is then computed for each subject's point cloud and summarized by an order-$r=3$ persistence silhouette $\widetilde\phi_0$ on an equispaced $100$-point grid over $[0, 0.83]$ (endpoint selection is detailed in Appendix~\ref{app:pacce-construction}).

Two methodological considerations govern the inferential interpretation of this contrast. First, eligibility within the $842$ supplied records depends on post-treatment observation. Letting $R \in \{0, 1\}$ denote observed geometric eligibility and $R^a$ the potential eligibility status under assignment $a \in \{0, 1\}$ (with $R = R^A$), the analysis necessarily compares observed patients with $R^1=1$ in arm one to observed patients with $R^0=1$ in arm zero. Because treatment assignment can affect disease progression, adverse events, or early withdrawal, randomization prior to this observation window does not identify the conditional average treatment effect among the observed cohort, $E\{\phi_d^1-\phi_d^0\mid R=1\}$, nor an always-observed principal-stratum effect ($R^1=R^0=1$). Second, coordinate scale parameters and the filtration grid endpoint ($0.83$) were selected using the pooled eligible cohort. Although this construction was treatment-blind, it induces data dependence between the representation and the estimation sample (examined in the split-sample sensitivity analysis in Appendix~\ref{app:pacce-pilot}). The main calculation also uses treatment-stratified folds rather than the observation-independent partition assumed by Theorem~\ref{thm:tate-sufficient}; we do not invoke that sufficient regime as an individual certificate for these exploratory clinical bands.

Under a fixed representation with treatment overlap in the observed cohort ($0<P(A=1\mid X,R=1)<1$ almost surely), we define the well-posed observable target as a covariate-standardized observed-group contrast:
\[
 \vartheta_d(t)=E_{X\mid R=1}\{\nu_{1d}(t,X)-\nu_{0d}(t,X)\},\qquad
 \nu_{ad}(t,x)=E\{\widetilde\phi_d(t)\mid A=a,X=x,R=1\},
\]
where $\widetilde\phi_d$ is the observed silhouette summary under that representation. The expectation standardizes both group-specific regression curves to the baseline covariate distribution of the observed eligible subjects, $P_{X\mid R=1}$. Interpreting $\vartheta_d(t)$ as an unconditional causal effect requires additional assumptions regarding the missing-data mechanism and the unobserved potential outcomes of excluded patients.

\begin{figure}[htbp]
\centering
\begin{minipage}{0.48\textwidth}
\centering
\includegraphics[width=\textwidth]{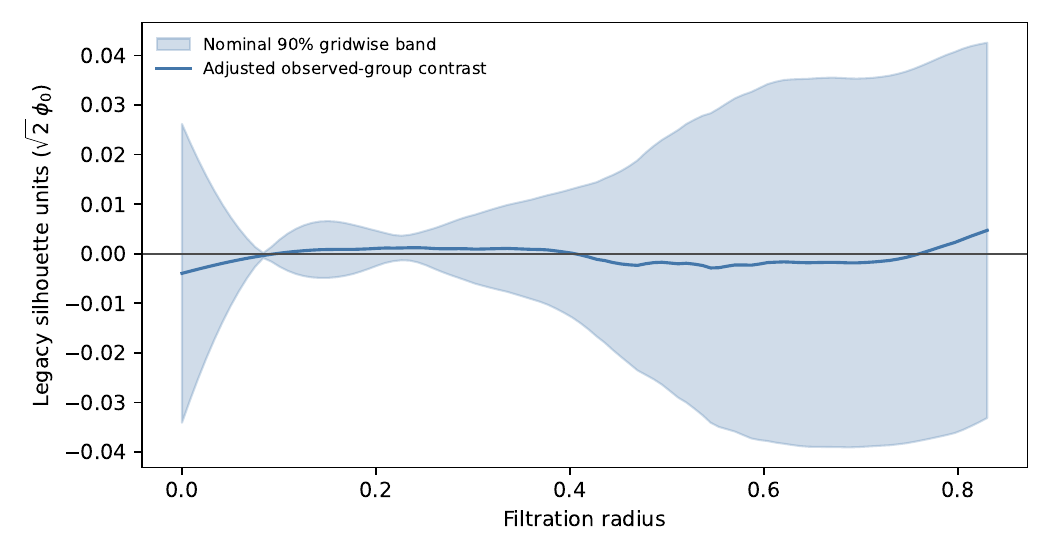}
\end{minipage}
\hfill
\begin{minipage}{0.48\textwidth}
\centering
\includegraphics[width=\textwidth]{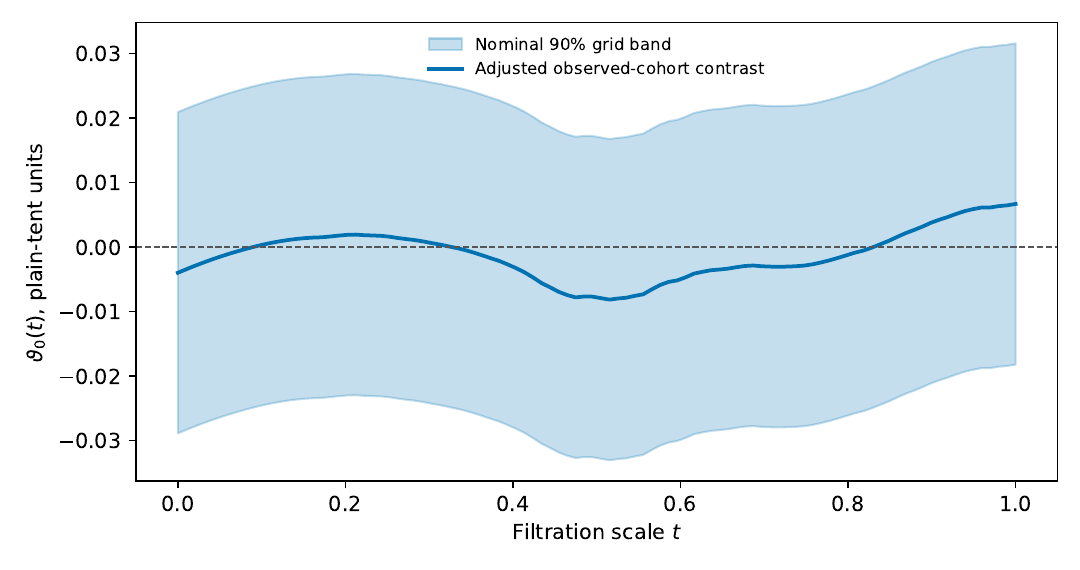}
\end{minipage}
\caption{PACCE lesion-trajectory contrasts. Left: adjusted contrast among $566$ observed eligible subjects, with nominal $90\%$ simultaneous grid multiplier band in rescaled units $\sqrt{2}\phi_0$. Right: split-sample sensitivity analysis among $419$ eligible analysis subjects with coordinate scale parameters pre-fitted on an independent held-out sample of $210$ subjects (excluded from inference), using unscaled tent units and a nominal $90\%$ Gaussian unstudentized grid band. In both specifications, the nominal confidence band contains zero across the entire filtration grid.}
\label{fig:pacce-tate}
\end{figure}

The cross-fitted AIPW calculation on the $566$ eligible subjects has peak absolute estimate $0.0047$ at $t=0.83$ and mean full width $0.0402$ in the rescaled units $\sqrt{2}\phi_0$ (Figure~\ref{fig:pacce-tate}, left panel). Its zero-curve sup test gives $p=0.982$. The constructed log-summed-diameter comparator defined in Appendix~\ref{app:pacce-comparator} gives $p=0.175$. Neither calculation rejects a zero adjusted observed-group contrast under its working models. They do not establish equivalence, exclude smaller effects, or locate the clinical mechanism of the regimen's toxicity or survival effects. In the released data, median central-review progression-free survival is $360$ versus $332$ days, and severe-or-worse adverse-event incidence is $73.6\%$ versus $88.4\%$ in the control and panitumumab arms, respectively. These descriptive benchmarks are consistent with the unfavorable clinical findings reported by \citet{hecht2009}.

A further analysis separates scale estimation from the fitted contrast by randomly reserving $210$ released subjects as a held-out sample (Figure~\ref{fig:pacce-tate}, right panel). The remaining $632$ subjects contain $419$ eligible patients, $210$ control and $209$ panitumumab. Coordinate scales are fitted on held-out records, and the analysis uses a fixed $[0,1]$ grid rather than the pooled death-quantile endpoint. With full-baseline propensity adjustment and Gaussian unstudentized calibration, the estimated contrast has peak absolute value $0.0081$ and mean full band width $0.0498$ in standard tent units. Its nominal $90\%$ grid band contains zero throughout. This split-sample sensitivity analysis isolates coordinate scale estimation from observed-group contrast estimation, addressing potential data-dependent representation leakage while maintaining the observed-cohort interpretation and its selection and model qualifications (Appendix~\ref{app:pacce-pilot}).

\begin{table}[t]
\centering\small
\caption{Exploratory PACCE fixed-profile CTATE diagnostics. Profiles cross chemotherapy backbone with baseline target-lesion burden, fixing age at $61$ years. $n_{\mathrm{exact}}$ denotes exact backbone-and-burden sample matches. Values use rescaled units $\sqrt{2}\phi_0$. Sup-test $p$-values and simultaneous confidence bands are based on a working functional DR learner; no model-free conditional coverage guarantee is claimed.}
\label{tab:pacce-ctate}
\begin{tabular}{@{}llrrrrr@{}}
\toprule
Backbone&Burden&$n_{\mathrm{exact}}$&Peak $|\widehat\tau|$&Mean half-width&Sup $p$&Bonf.\ $p$\\
\midrule
Oxaliplatin&2&99&0.011&0.039&0.834&1.000\\
&4&55&0.030&0.043&0.710&1.000\\
&6&23&0.036&0.043&0.741&1.000\\
Irinotecan&2&28&0.035&0.049&0.432&1.000\\
&4&19&0.004&0.055&0.996&1.000\\
&6&3&0.063&0.056&0.443&1.000\\
\bottomrule
\end{tabular}
\end{table}

Table~\ref{tab:pacce-ctate} reports results for six representative clinical profiles formed by crossing chemotherapy backbone (oxaliplatin versus irinotecan) with baseline target-lesion burdens of $2$, $4$, and $6$. Across all six profiles, the working CTATE bands contain zero across the entire grid; the smallest unadjusted sup-test $p$-value is $0.432$ and all six-profile Bonferroni-adjusted values are $1.000$ (Table~\ref{tab:pacce-ctate}).

We also report aggregate exploratory ITTE prediction summaries across the six age-$61$ profiles, retaining patient-level trajectories within the restricted analysis environment. With a score scale fixed at one before calibration, the composed prediction bands have mean full widths $1.639$--$1.730$ in rescaled units. Intersecting them with the deterministic support $[-\sqrt{2}t,\sqrt{2}t]$ reduces these widths to $1.053$--$1.090$ (Appendix~\ref{app:pacce-profiles}, Table~\ref{tab:pacce-prediction}). All six bands are finite and contain zero at every grid node. These summaries do not establish conditional prediction coverage, an absence of individual effects, or clinical equivalence. The broad widths reflect supremum calibration and conservative armwise composition as well as outcome variation; they are not estimates of individual-effect variability.

Among patients with sufficiently observed early lesion trajectories, we found no statistically significant covariate-adjusted difference in the selected topological summaries between treatment groups. This exploratory observed-group comparison does not establish the absence of a causal treatment effect, clinical equivalence, or the absence of meaningful structural differences. The topological comparison concerns a different outcome from the toxicity and progression-free-survival endpoints reported by \citet{hecht2009}.
\section{Conclusion}
\label{sec:conclusion}
\label{sec:discussion}

This paper has established a unified theoretical, algorithmic, and computational framework for uncertainty quantification in topological causal inference. By embedding persistent homology summaries within semiparametric causal estimation and distribution-free predictive inference, the framework enables researchers to interrogate treatment-induced structural changes across continuous filtration scales with rigorous statistical guarantees. A central contribution is bridging two fundamentally distinct inferential regimes: simultaneous confidence bands for population and conditional average treatment effects (TATE and CTATE), which quantify estimation uncertainty for deterministic mean functional curves, and marginal prediction bands for individual topological treatment effects (ITTE), which quantify unit-level counterfactual dispersion for a newly observed subject. All proposed methodologies are unified within the open-source \texttt{tcda\_uq} library, pairing mathematical theory with reproducible software.

At the population and subgroup levels, our results build upon the functional AIPW and weak-convergence foundation of \citet{kim-lee}, augmenting it with explicit studentized multiplier bootstrap calibration and sufficient cross-fitting conditions for simultaneous asymptotic confidence bands over the filtration interval. For conditional effects at a prespecified covariate profile, we develop a CTATE linear-smoother framework with explicit control of smoothing bias and first-stage nuisance-estimation error. The resulting confidence bands have width of order $O_p(v_n(x))$ in the general setting, specializing to the parametric rate $O_p(n^{-1/2})$ under correctly specified fixed-dimensional regression.

For individual prediction, the inferential challenge is fundamentally different because the two counterfactual outcomes are never jointly observed for the same unit. Rather than imposing unidentifiable assumptions on their dependence structure, we combine weighted functional split-conformal prediction with arm differencing and a union bound to construct ITTE prediction bands with finite-sample marginal coverage of at least the nominal level under known likelihood-ratio weights. When propensity scores are estimated, the resulting coverage discrepancy is controlled by an explicit total-variation bound. Our analysis also clarifies the trade-offs associated with weak overlap. Retaining the full target population preserves the intended coverage guarantee under known weights, although sparse support may lead to unbounded prediction bands. Alternatively, overlap weighting guarantees bounded calibration weights and can reduce the occurrence of infinite prediction radii, but changes the target population and hence the estimand. Importantly, bounded weights alone do not ensure finite prediction radii; sufficient calibration mass is also required, as characterized by Lemma~\ref{lem:finite}.

The topological interface of our framework hinges on the stability of power-weighted silhouettes. By establishing the diagram-independent Lipschitz bound $\|\Phi_r(D) - \Phi_r(D')\|_\infty \le (1+2r)W_1(D, D')$ for all $r \ge 1$, we generalize existing stability results and supply a mechanism to propagate persistence diagram approximation errors directly into prediction bands without requiring arbitrary persistence thresholds. Importantly, our exposition emphasizes the one-way nature of this stability: because functional summaries are non-injective projections of persistence diagrams, silhouette confidence bands cannot be inverted to obtain confidence sets in diagram Wasserstein space, and a null topological contrast on a silhouette curve does not imply identical persistent homology.

In the correctly specified benchmarks, mean confidence bands contract with sample size, whereas composed individual prediction bands remain broad. Prediction width reflects outcome variation, calibration, and armwise composition; it is not an estimate of individual-effect variability. The PACCE analysis illustrates exploratory adjusted contrasts of observed lesion trajectories. Observation selection, learned likelihood ratios, and working models prevent a causal interpretation without additional assumptions. Its topological summaries remain distinct from the toxicity and progression-free-survival endpoints studied by \citet{hecht2009}.

Several promising directions emerge for future mathematical and methodological development: 1. extending CTATE inference from fixed covariate profiles to uniform confidence bands that hold simultaneously over multidimensional continuous covariate regions, 2. developing adaptive conformal prediction methods that remain informative in the presence of weak overlap without distorting the target population or resorting to ad-hoc weight clipping, 3. formulating intrinsic causal contrasts and prediction sets directly within the metric space of persistence diagrams (such as through Wasserstein barycenters or optimal transport geometries) rather than through intermediate functional projections, and 4. generalizing the framework to continuous or time-varying treatments, which will require uniform empirical process control over both the treatment dosage index and the filtration scale. Addressing these frontiers will further expand the role of topological data analysis as a rigorous tool for causal discovery in complex, high-dimensional scientific domains.


\bibliography{manuscript}

\appendix

\section{Notation and Assumptions Ledger}
\label{app:notation}
\label{sec:notation}

\begin{center}
\begin{tabular}{p{0.24\textwidth}p{0.68\textwidth}}
\toprule
Notation & Meaning\\
\midrule
$\phi_d^a=G_{d,r}(Y^a)$ & Potential power-weighted silhouette; $d,r$ fixed unless indexed explicitly.\\
$\psi_d,\tau_d,\delta_{i,d}$ & Population mean, conditional mean, and individual topological treatment effects.\\
$P_a,Q_a$ & Observed arm law and intended target potential-outcome law.\\
$\rho_d,\xi_d$ & AIPW pseudo-outcome and its centered oracle influence function.\\
$\widehat\rho_{i,d}$ & Pseudo-outcome using nuisance fits excluding observation $i$'s fold.\\
$\widehat s_{d,x}$ & Standard error of the CTATE smoother; includes its weights.\\
$\widehat s_{a,d}$ & Positive training-only conformal score scale; not necessarily a standard error.\\
$\Delta_a$ & Total-variation discrepancy between intended and estimated-weight target laws.\\
$K_r=1+2r$ & A valid uniform finite-diagram $W_1$ stability constant.\\
$\Delta$ & Maximum spacing of the deterministic filtration grid.\\
\bottomrule
\end{tabular}
\end{center}

The causal identification and measurability conditions are stated in
Section~\ref{sec:setup}. Assumptions (T1)--(T4) state the process
requirements for the TATE band; (S1)--(S3) verify them except for the explicit nondegeneracy and quantile conditions. Conditions (C1)--(C4) govern a general fixed-$x$ CTATE smoother; Corollary~\ref{cor:ols} verifies a correctly specified finite-dimensional case.

Theorem~\ref{thm:conformal} uses independent training, finite
measurable scores, the true likelihood ratio, and the target-point
atom. It does not require the outcome-regression consistency conditions for the mean-band theorems. Theorem~\ref{thm:overlap} replaces global positivity by positive mass of the overlap population. Approximation inflation, joint scores, and grid extension impose the additional conditions in their own statements.

\section{Deferred Statements and Proofs}
\label{app:proofs}

\subsection{The TATE band}
The first proof records standard supremum calibration. The sufficient-regime
proof then spells out the foldwise remainder and Gaussian multiplier checks
for the AIPW construction of \citet{kim-lee}, using the cross-fitting
and empirical-process tools of \citet{dml} and \citet{vdv-wellner}.
These self-contained arguments document the assumptions used in this
framework rather than introduce another TATE estimator.
\begin{proof}[Proof of Theorem~\ref{thm:tate-band}]
Assumptions (T1), (T2), and (T4), followed by the continuous mapping
theorem, give
\[
 \left\|\frac{\sqrt n(\widehat\psi_d-\psi_d)}
                    {\widehat\sigma_d}\right\|_\infty
 \Rightarrow \|G_d/\sigma_d\|_\infty.
\]
By (T3) and uniform variance consistency, the conditional law of the
bootstrap absolute supremum converges in probability to the same law. Continuity and local strict increase of its distribution function give consistency of the conditional quantile. The probability of the studentized statistic falling below that quantile consequently tends to $1-\alpha$. This event is exactly simultaneous inclusion in \eqref{eq:tate-band}. The quantile and the supremum of the estimated standard deviation are $O_p(1)$, giving the width order.
\end{proof}

\begin{proof}[Proof of Theorem~\ref{thm:tate-sufficient}]
Suppress $d$. All processes below have continuous paths, so suprema
can be restricted to a fixed countable dense set. Boundedness of the
curves, the nuisance fits, and the inverse denominators gives a common finite envelope for oracle and fitted scores. Their paths are uniformly Lipschitz in $t$, because the propensity functions do not depend on $t$. Consequently, for any probability measure $R$ and the class indexed by $t\in T$, an evenly spaced grid gives an $L^2(R)$ covering number at most $C(1+u^{-1})$ at radius $u$. Constants may change below but do not depend on $n$, the fold, or the fitted functions.

First consider the oracle influence functions. The finite entropy
integral and bounded envelope yield the empirical-process central limit theorem. Equivalently, finite-dimensional convergence follows from the ordinary multivariate central limit theorem, and the Lipschitz bound gives a uniform fourth-moment increment bound for the centered empirical process. Tightness then holds in $C(T)$. This proves (T2), with covariance $K(s,t)=E\{\xi(s,O)\xi(t,O)\}$.

For fold $k$, let $h_{k,t}(O)=\rho(t,O;\widehat\eta^{(-k)})-\rho(t,O;\eta)$, and let $P_k$ be the empirical measure on that fold. Conditional on its training data, the validation observations are i.i.d. from $P$, and

\begin{equation}\label{eq:proof-l2}
 \sup_t\|h_{k,t}\|_{P,2}
       \leq C\{p_k+b_{0,k}+b_{1,k}\}=o_p(1).
\end{equation}

This follows by subtracting the two score formulas and using bounded
residuals and denominators. The conditional classes $\{h_{k,t}:t\in T\}$ have the same deterministic bounded-envelope and covering-number bounds just described. The localized empirical-process maximal inequality, with the shrinking $L^2(P)$ radius in \eqref{eq:proof-l2}, therefore gives

\begin{equation}\label{eq:proof-empirical}
 \sup_t|\sqrt{n_k}(P_k-P)h_{k,t}|=o_p(1).
\end{equation}

For clarity, the relevant entropy integral is bounded by
$C\int_0^{c\delta}\sqrt{\log(C/u)}\,du\to0$ as the radius
$\delta\downarrow0$; the bounded-envelope finite-sample remainder
also vanishes. Random radii are handled by first restricting them to
be at most a fixed $\delta$, then taking $n\to\infty$ and
$\delta\downarrow0$. Cross-fitting is used fold by fold; the fitted
functions across different folds are not assumed independent.

Writing $\widehat\pi,\widehat\mu_a$ for the fold fits, direct
conditioning on $X$ gives the exact bias identity
\begin{equation}\label{eq:proof-bias}
 Ph_{k,t}=E\left[(\widehat\pi-\pi)
 \left\{\frac{\widehat\mu_1(t,X)-\mu_1(t,X)}{\widehat\pi}
       +\frac{\widehat\mu_0(t,X)-\mu_0(t,X)}{1-\widehat\pi}
 \right\}\right].
\end{equation}

Hence $\sup_t|Ph_{k,t}|\leq\epsilon^{-1}p_k(b_{0,k}+b_{1,k})
=o_p(n^{-1/2})$. Decomposing the estimator,
\[
 \sqrt n(\widehat\psi-\psi)
 =\frac1{\sqrt n}\sum_i\xi(\cdot,O_i)
  +\sqrt n\sum_k\frac{n_k}{n}\{(P_k-P)h_k+Ph_k\},
\]
proves (T1) by \eqref{eq:proof-empirical}, the bias bound, and fixed $K$.

The same conditional covering argument applied to $h_{k,t}^2$ gives
a uniform law of large numbers. Together with \eqref{eq:proof-l2}, this implies
\[
 \sup_t\frac1n\sum_i
  \{\widehat\xi_i(t)-\xi(t,O_i)\}^2=o_p(1),
\]
where the centering error is controlled by $\|\widehat\psi-\psi\|_\infty=o_p(1)$. Cauchy--Schwarz and the uniform law of large numbers for $\{\xi(s,\cdot)\xi(t,\cdot):(s,t)\in T^2\}$ now yield
\[
 \sup_{s,t}\left|\frac1n\sum_i\widehat\xi_i(s)\widehat\xi_i(t)
                        -K(s,t)\right|=o_p(1).
\]
This proves uniform variance consistency, including consistency of
standard deviations by continuity of the square-root map.

Conditional on the data the proposed multiplier process is Gaussian.
The preceding covariance convergence gives its finite-dimensional
convergence. Uniform Lipschitz bounds for $\widehat\xi_i$ imply
\[
 E_g|\widehat G_n^{\#}(t)-\widehat G_n^{\#}(s)|^4 \leq C|t-s|^4,
\]
and its variance at one fixed point is bounded. These bounds give
conditional tightness in $C(T)$. Covariance convergence in probability, or the usual subsequence characterization thereof, proves (T3). The nondegeneracy and quantile conditions are additional conditions, as stated in the theorem.
\end{proof}

\subsection{Fixed-covariate CTATE inference}
\begin{proof}[Proof of Proposition~\ref{prop:ctate-band}]
The decomposition
\begin{align*}
 \widehat\tau_d(\cdot,x)-\tau_d(\cdot,x)
 &=\sum_i a_{i,n}(x)\varepsilon_{i,d}
   +\left\{\sum_i a_{i,n}(x)\tau_d(\cdot,X_i)-\tau_d(\cdot,x)\right\}\\
 &\quad+\sum_i a_{i,n}(x)
          \{\widehat\rho_{i,d}-\rho_d(\cdot,O_i;\eta)\}
\end{align*}
and (C2)--(C3) give weak convergence of the error divided by $v_n(x)$ to $H_{d,x}$. By (C4), studentization gives the limit
$H_{d,x}/\sigma_{d,x}$ and the conditional multiplier process has the same limit. Its absolute-supremum quantile is consistent by the stated quantile condition. This proves simultaneous coverage. Uniform convergence of $\widehat s_{d,x}/v_n(x)$ and boundedness of the critical value prove the width order.
\end{proof}

\begin{proof}[Proof of Corollary~\ref{cor:ols}]
Write $b_i=b(X_i)$ and suppress $d$. With probability tending to one, $Q_n$ is invertible and its inverse is bounded. Since the basis contains an intercept, $\sum_i a_{i,n}(x)=1$. Algebra gives
\[
 v_n(x)^2=\frac1n b(x)^\top Q_n^{-1}b(x),\qquad
 \max_i|a_{i,n}(x)|=O_p(n^{-1}).
\]
Thus (C1) holds, with $nv_n(x)^2\to\kappa_x=b(x)^\top Q^{-1}b(x)>0$.
Correct specification gives exact reproduction of the conditional mean: $\sum_i a_{i,n}(x)\tau(t,X_i)=\tau(t,x)$.

Let $\varepsilon_i(t)=\rho(t,O_i;\eta)-b_i^\top\beta(t)$.
Its conditional mean given $X_i$ is zero. The bounded basis makes
the foldwise proof of \eqref{eq:proof-empirical} and
\eqref{eq:proof-bias} applicable to each coordinate of $b_i h_{k,t}$. It follows that
\[
 \sup_t\left\|\frac1n\sum_i b_i
      \{\widehat\rho_i(t)-\rho(t,O_i;\eta)\}\right\|
       =o_p(n^{-1/2}).
\]
The complete first-stage part of (C2) follows after multiplication by $b(x)^\top Q_n^{-1}$. Furthermore,
\[
 \widehat\beta(t)-\beta(t)
 =Q_n^{-1}\frac1n\sum_i b_i\varepsilon_i(t)+o_p(n^{-1/2})
\]
uniformly in $t$. Boundedness of $b$ and the oracle curves gives a
functional central limit theorem for the vector process on the right. Since $Q_n^{-1}\to Q^{-1}$,
\[
 \sqrt n\{\widehat\tau(t,x)-\tau(t,x)\}
 =\frac1{\sqrt n}\sum_i\zeta_x(t,O_i)+o_p(1)
\]
in $C(T)$. This verifies (C3), with limiting covariance equal to that of $\zeta_x/\sqrt{\kappa_x}$.

The ordinary residual satisfies
\[
 \widehat r_i(t)-\varepsilon_i(t) =\widehat\rho_i(t)-\rho(t,O_i;\eta) -b_i^\top\{\widehat\beta(t)-\beta(t)\}.
\]
The preceding bounds imply
$\sup_t n^{-1}\sum_i\{\widehat r_i(t)-\varepsilon_i(t)\}^2=o_p(1)$.
Consequently, uniformly in $(s,t)$,
\[
 n\sum_i a_{i,n}(x)^2\widehat r_i(s)\widehat r_i(t)
 \ \longrightarrow\ E\{\zeta_x(s,O)\zeta_x(t,O)\}
\]
in probability. This yields the variance consistency in (C4).
Conditional Gaussian multipliers have exactly this covariance after
root-$n$ scaling. On events where $Q_n^{-1}$ is bounded, their
increment fourth moments are bounded by $C|t-s|^4$: both the fitted
coefficient curve and the residual curves have uniformly bounded
Lipschitz constants there. The same tightness argument as above,
followed by division by the uniformly consistent positive standard
deviation, proves the multiplier part of (C4). The remaining quantile condition is explicitly assumed.
\end{proof}

\subsection{Weighted prediction and estimated weights}

\begin{proof}[Proof of Theorem~\ref{thm:conformal}]
Condition on training data and arm counts. For one arm, the calibration pairs have law $P_a$ and the independent target pair has law $Q_a$ with density proportional to $w_a$ relative to $P_a$.
Conditional on the unordered collection of these $m_a+1$ pairs, the
probability that a given member is the target is proportional to its
weight. This is the weighted-exchangeability argument of
\citet{tibshirani}. The score is a fixed measurable function after conditioning on training data.

Form the weighted empirical distribution of all $m_a+1$ actual scores. The weighted mass at or below its $(1-\alpha_a)$ quantile is at least $1-\alpha_a$. The same is therefore true of the conditional probability that the target score is at or below that quantile. Replacing the target's actual score by $+\infty$ can only increase the quantile. This gives arm coverage at least $1-\alpha_a$, including with ties. With no calibration observations the band is the whole real line, so the claim still holds.

On intersection of the two arm-coverage events, interval subtraction
contains $\phi_d^1(t)-\phi_d^0(t)$ for every $t$. The probability of the complement of the intersection is at most $\alpha_0+\alpha_1$. This union bound requires neither independence
of the potential outcomes nor independence of the arm-coverage events.
\end{proof}

\begin{proof}[Justification of \eqref{eq:tv-conditional}--\eqref{eq:tv-marginal}]
Conditional on training data, $\widehat w_a$ is the exact density
ratio, up to normalization, for the artificial law $\widehat Q_a$.
The preceding weighted-rank argument therefore bounds the arm failure probability under this law by $\alpha_a$. Keeping the calibration law fixed and replacing only the independent target law by $Q_a$ changes the probability of any event by at most
$\operatorname{TV}(Q_a,\widehat Q_a)=\Delta_a$. This follows either
from the product-measure property of total variation or by integrating the event probability, a function taking values in $[0,1]$. Thus the true arm failure probability is at most $\alpha_a+\Delta_a$. The union bound gives the conditional result even though the two tilted arm laws need not share a common covariate distribution. Averaging over training data and counts gives the unconditional result. The normalized $L^1$ formula is the density formula for total variation.
\end{proof}

\begin{proof}[Proof of Proposition~\ref{prop:collapse}]
Put $D=\|\delta_{n+1,d}-\psi_d\|_\infty$ and
$R_n=\|\widehat\psi_d-\psi_d\|_\infty+\|r_n\|_\infty$.
Band inclusion implies $D\leq R_n$. For every $\varepsilon>0$,
\[
 P\{\delta_{n+1,d}\in B_n\}
 \leq P(D\leq\varepsilon)+P(R_n>\varepsilon).
\]
Take a limit superior, then let $\varepsilon\downarrow0$, to obtain
\eqref{eq:collapse}. More precisely, the probability of inclusion
with $D>0$ is bounded by $P(0<D\leq\varepsilon)+P(R_n>\varepsilon)$ and tends to zero. On $D=0$, inclusion is equivalent to $\psi_d\in B_n$. Independence of the new curve and the fitted band gives probability $pP\{\psi_d\in B_n\}$ for this event, proving the last assertion.
\end{proof}

\subsection{Target changes and finite radii}
\begin{proof}[Proof of Theorem~\ref{thm:overlap}]
Bayes' rule gives, on the target support,
\[
 \frac{dQ_{\mathrm{ov},X}}{dP_{X\mid A=a}}(x)
 =\frac{P(A=a)}{E[\pi(X)\{1-\pi(X)\}]}
   \frac{\pi(x)\{1-\pi(x)\}}{e_a(x)}.
\]
The last factor is $1-\pi(x)$ for $a=1$ and $\pi(x)$ for $a=0$.
Consistency and exchangeability identify the conditional response laws on this support. The weighted-rank proof and union bound therefore apply to the overlap target. Regions with deterministic assignment have zero target mass and introduce no unidentified target outcomes.
\end{proof}

\begin{proof}[Proof of Lemma~\ref{lem:finite}]
The total probability mass at finite scores is $U_a/(U_a+u_a(x))$.
A finite quantile exists exactly when this mass is at least
$1-\alpha_a$, which is equivalent to \eqref{eq:finite-condition}.
For $u_a(x)\leq B$, the test mass is at most $B/(U_a+B)$.
The stated lower bound on $U_a$ suffices. Finally, the law of large
numbers gives $U_a/m_a\to E_{P_a}u_a>0$, proving the asymptotic claim.
\end{proof}

\subsection{Silhouette stability and approximation}
The matching, tent perturbation, and power-weight mean-value bounds
below follow the stability argument of \citet[Appendix C.5]{kim-lee}.
Choosing the larger total power weight as denominator and bounding the
largest tent directly gives the diagram-independent constant for $r\geq1$.
At $r=1$ it recovers their constant $3$.
\begin{proof}[Proof of Theorem~\ref{thm:stability}]
Fix a matching, padding each finite diagram with the diagonal points
needed for that matching. Write matched pairs as $(p_i,p_i')$, their
$\ell^\infty$ costs as $c_i$, their persistences as $\ell_i,\ell_i'$
(zero for diagonal points), and their tents as $\lambda_i,\lambda_i'$. Then

\[
 |\ell_i-\ell_i'|\leq2c_i,\qquad
 \|\lambda_i-\lambda_i'\|_\infty\leq c_i.
\]

Put $w_i=\ell_i^r$, $v_i=(\ell_i')^r$, $S=\sum_i w_i$, and
$S'=\sum_i v_i$. If both diagrams are nonempty, interchange them if
necessary so that $S\geq S'>0$. Let $c=\max_i\{\ell_i,\ell_i'\}>0$. Then $S\geq c^r$, and every tent is bounded by $c/2$. Decompose

\begin{align*}
 \Phi_r(D)-\Phi_r(E)
 &=\frac1S\sum_i w_i(\lambda_i-\lambda_i')
   +\frac1S\sum_i(w_i-v_i)\lambda_i'\\
 &\quad+\left(\frac1S-\frac1{S'}\right)\sum_i v_i\lambda_i'.
\end{align*}

The first term has supremum norm at most
$\sum_i(w_i/S)c_i\leq\sum_i c_i$. The remaining two terms together
have norm at most
\[
 \frac{c}{2S}\sum_i|w_i-v_i|
   +\frac{c}{2S}|S-S'|
 \leq\frac cS\sum_i|w_i-v_i|.
\]
For $r\geq1$, the mean-value inequality yields
\[
 \sum_i|w_i-v_i|
 \leq r c^{r-1}\sum_i|\ell_i-\ell_i'|
 \leq2r c^{r-1}\sum_i c_i.
\]
Since $c^r/S\leq1$, the total norm is at most
$(1+2r)\sum_i c_i$. Taking the infimum over matchings proves the
claim. If one diagram is empty, the other silhouette is bounded by
half its largest persistence, which is at most the cost of matching
all its points to the diagonal. If both are empty, the claim is trivial.
\end{proof}

\begin{proof}[Proof of Theorem~\ref{thm:inflation}]
On the approximation event, Theorem~\ref{thm:stability} gives
$\|\phi_{n+1,d}^a-\widehat\phi_{n+1,d}^a\|_\infty
\leq K_r\varepsilon_a$. On intersection with the approximate-curve
coverage event, the exact curve therefore lies in
\eqref{eq:inflation}. The union bound subtracts at most
$\alpha_a+\gamma_a$. Intersecting the two arm events and subtracting
intervals proves the ITTE statement and the half-width formula.
\end{proof}

\subsection{Joint scores and grid extension}
\begin{proof}[Proof of Theorem~\ref{thm:joint}]
The maximum over finitely many continuous-curve scores is a measurable scalar score. The weighted-rank argument applies to it unchanged. An arm score below its quantile is precisely joint inclusion for all $(d,r)\in\mathcal J$ and all $t\in T$. Intersect the two arm events and apply the union bound. The separate-band alternative follows by applying the union bound to the $2|\mathcal J|$ arm events.
\end{proof}

\begin{proof}[Proof of Lemma~\ref{lem:grid}]
For $t\in[t_j,t_{j+1}]$, set $h=t_{j+1}-t_j$ and
$\theta=(t-t_j)/h$. Lipschitz continuity gives
\begin{align*}
 |f(t)-(I_mf)(t)|
 &\leq(1-\theta)|f(t)-f(t_j)|
       +\theta|f(t)-f(t_{j+1})|\\
 &\leq2L\theta(1-\theta)h\leq L\Delta/2.
\end{align*}
On gridwise inclusion, positivity of the interpolation coefficients
gives $I_m\ell\leq I_mf\leq I_mu$. Combining these inequalities
proves the continuous-band inclusion and its probability statement.
\end{proof}

\section{Scope of the guarantees}\label{app:claims}

This register collects the coverage statements and their scope. The TATE
formulation, its AIPW estimation and functional weak convergence, and the
simultaneous-band approach follow \citet{kim-lee};
Sections~\ref{sec:setup}--\ref{sec:confidence} state their role and the
additional sufficient conditions used here. CTATE inference uses
established pseudo-outcome regression, and ITTE prediction combines
the conformal components cited in Section~\ref{sec:conformal}.
The stability result extends the earlier bound for $r>1$ and recovers
its $r=1$ constant.

\begin{center}
\begin{tabular}{p{0.29\textwidth}p{0.63\textwidth}}
\toprule
Claim & Scope and reference\\
\midrule
TATE confidence & Asymptotic, simultaneous over $T$, under (T1)--(T4).
Verified sufficient nuisance and path conditions appear in
Theorem~\ref{thm:tate-sufficient}.\\[2pt]
CTATE confidence & At a prespecified $x$, simultaneous over $T$;
rate $v_n(x)$ and negligible bias are essential. Root-$n$ inference is
verified for the correctly specified fixed-dimensional model in
Corollary~\ref{cor:ols}.\\[2pt]
ITTE prediction & Finite-sample marginal lower coverage bound with true
ratios, independent training, and target-point infinity mass;
Theorem~\ref{thm:conformal}. Not a conditional-at-every-$x$ guarantee.\\[2pt]
Estimated weights & Lower bound reduced by expected total-variation
errors; \eqref{eq:tv-marginal}. Splitting alone does not remove
these errors.\\[2pt]
Overlap target & The true overlap ratios give coverage for the overlap
population. Finite radii still require
\eqref{eq:finite-condition}.\\[2pt]
Shrinking mean band as prediction & Coverage tends to zero if the ITTE
law has no atom at its mean; Proposition~\ref{prop:collapse}.\\[2pt]
Diagram approximation & Forward $W_1$ stability and deterministic
inflation; Theorems~\ref{thm:stability}--\ref{thm:inflation}.
No inverse recovery of diagram distance.\\[2pt]
Several degrees or exponents & One joint arm score or a valid
Bonferroni allocation; Theorem~\ref{thm:joint}.\\[2pt]
Grid implementation & Gridwise inclusion extends to continuous
inclusion with the Lipschitz inflation in Lemma~\ref{lem:grid}.
Numerical approximation of bootstrap quantiles needs its own control.\\
\bottomrule
\end{tabular}
\end{center}

These results distinguish a mathematical coverage guarantee from its
empirical verification and from whether the returned band is
informative. None asserts that the entire individual-effect law is
identified from unpaired observational potential outcomes.

\section{Additional experimental results}
\label{app:experiments}

This appendix supplies the designs and numerical details behind Section~\ref{sec:experiments}. Coverage always refers to the complete finite evaluation grid. Where coverage averages several test curves from one fitted sample, the fitted sample is the replication unit; those test curves must not be treated as independent replications when calculating Monte Carlo uncertainty.

\subsection{Bounded diagram-valued simulation design}
\label{app:bounded-rerun}

The design in Section~\ref{sec:sim-bounded} generates the two nonempty-diagram indicators independently conditional on $X$ and applies the normalized silhouette directly to the resulting finite diagrams. Both bars have persistence two, so equal weights give $\{(2-t)+(t-1)\}/2=1/2$ on $[1.1,1.9]$ for every $r\ge1$. The empty diagram gives zero. All targets and covariates are bounded, strong overlap holds, and the conditional means are linear in $(1,X)$.

The nuisance regression uses five Fourier functions on $50$ filtration points and a linear covariate design. Its coefficients are constrained to $[-2,2]$, which contains the true coefficients in its interior. This supplies deterministic bounded and Lipschitz fitted paths even for nearly singular finite training designs. Known treatment probabilities are used in the AIPW score, with two folds randomized independently of the observations. The second-stage OLS design is $(1,X_1,X_2,X_3)$, and the fixed CTATE profiles are $(-1,0,1)$, $(0,0,0)$, and $(1,-1,0)$. Confidence calibration approximates the raw Gaussian supremum quantile with $999$ multiplier draws at nominal level $0.90$. The constant truth belongs to the Fourier span, so this response projection has no target bias in this design.

Prediction uses independent equal-size training and calibration splits, the constant scale one, and arm error $0.025$ each. At each of $2{,}000$ fitted samples per sample size, $200$ new paired potential outcomes are generated separately under the full, overlap, and matching covariate laws. Rejection sampling implements each declared tilt, and the known propensity gives the exact arm likelihood ratios. The composition event and both arm events are retained. A comparator that scores the paired individual contrast directly is available only because both simulated potential outcomes are known; it is not implementable from the clinical trial's observed outcomes.

The target covariate law is proportional to $h(x)P_X(dx)$, with $h=1$ for the full population, $h=\pi(1-\pi)$ for overlap, and $h=\min\{\pi,1-\pi\}$ for matching. Writing $e_1=\pi$ and $e_0=1-\pi$, the arm ratios are proportional to $h/e_a$, as in Section~\ref{sec:extensions}; their common normalizing constants cancel in the weighted quantile.

At $n=200$, full-population arm coverage is $0.9828$ and $0.9829$, their joint coverage is $0.9663$, and composed effect coverage is $0.99885$. The direct-contrast comparator covers $0.9502$. These values illustrate conservatism without attributing a fixed excess percentage to one algebraic step. Monte Carlo standard errors use the fitted-sample replication as their unit. There are no failed fits in the $6{,}000$ requested replications. The empirical zero unbounded share at $n=500$ and $1{,}000$ is not a universal finite-radius statement.

\subsection{Functional design and complete TATE results}
\label{app:levelA-tate}

The two Gaussian subgroup means are
\[
 m_0=(1,0.6,-0.7,2.2,-1),\qquad m_1=(0.4,-0.4,-0.6,3.3,3),
\]
with covariance $0.5I_5$. Each generated sample contains $\lfloor n/2\rfloor$ units from the first subgroup and the remaining units from the second. This is fixed-composition sampling, rather than iid sampling from a mixture. Assignment follows
\[
 e(x)=\operatorname{expit}\{x^\top b+0.5x_2x_3-0.7x_1x_3\},\qquad
 b=(-0.5,-0.1,0.6,0.1,0.1).
\]
With $B(t)=(1,\cos2\pi t,\sin2\pi t,\cos4\pi t,\sin4\pi t)^\top$, the potential curves have the form
\[
 \phi^a_{i,d}(t)=B(t)^\top\Gamma_{ad}(1,X_i^\top)^\top
 +0.3\sum_{k=0}^4 2^{-k}Z^a_{idk}B_k(t),
\]
where the $Z$ variables are independent standard normal draws. The fixed coefficient matrices $\Gamma_{ad}\in\mathbb{R}^{5\times6}$ are generated once from independent standard normal draws with row $k$ scaled by $2^{-k}$ ($k=0,\dots,4$) using NumPy generator \texttt{default\_rng(0)}, sampled sequentially in order $(a,d)\in\{(0,0),(0,1),(1,0),(1,1)\}$. For degree zero, the explicit values rounded to three decimal places are
\[
\Gamma_{00}\approx\begin{pmatrix}
0.126 & -0.132 & 0.640 & 0.105 & -0.536 & 0.362\\
0.652 & 0.474 & -0.352 & -0.633 & -0.312 & 0.021\\
-0.581 & -0.055 & -0.312 & -0.183 & -0.136 & -0.079\\
0.052 & 0.130 & -0.016 & 0.171 & -0.083 & 0.044\\
0.056 & 0.006 & -0.046 & -0.058 & -0.029 & 0.014
\end{pmatrix},
\]
and
\[
\Gamma_{10}\approx\begin{pmatrix}
-0.436 & -1.170 & 1.739 & -0.496 & 0.329 & -0.259\\
0.792 & 0.660 & 0.317 & -1.102 & 0.026 & 0.342\\
0.251 & -0.155 & 0.456 & -0.330 & -0.165 & 0.234\\
0.006 & 0.250 & 0.024 & -0.079 & -0.047 & -0.136\\
-0.080 & 0.039 & 0.036 & 0.081 & -0.047 & 0.106
\end{pmatrix}.
\]
The conditional mean is linear in $x$, and the marginal mean uses $E[X]=(m_0+m_1)/2$. These signed functional outcomes bypass the diagram map and need not themselves be realizable silhouettes.

The confidence study uses $50$ grid points on $[0,1]$, five Fourier basis functions, two treatment-stratified folds, logistic propensity regression with the correct interaction features, and $1{,}500$ multiplier draws. The Liebl--Reimherr comparator uses four subintervals and the Gaussian critical function. It is an implementation comparison; no local-fairness or smoothness assumption is verified solely by its coverage. Table~\ref{tab:appA-tate} uses replication-weighted aggregation of the replications. The interval-wise testing procedure is excluded because its native inferential object is a p-value function, and the recorded coordinatewise bands are not an established inversion of its interval tests.

\begin{table}[ht]
\centering\small
\caption{Complete functional TATE confidence diagnostics. Each coverage uses $200$ replications; parentheses give Monte Carlo standard errors. Width is the grid-averaged full width.}
\label{tab:appA-tate}\label{tab:sim-tate}
\begin{tabular}{@{}llrrrr@{}}
\toprule
& & \multicolumn{2}{c}{$d=0$}&\multicolumn{2}{c}{$d=1$}\\
\cmidrule(lr){3-4}\cmidrule(lr){5-6}
Procedure&$n$&Coverage&Width&Coverage&Width\\
\midrule
Multiplier&50&0.885 (0.023)&2.3380&0.855 (0.025)&2.5470\\
&100&0.930 (0.018)&0.9979&0.920 (0.019)&1.1644\\
&200&0.935 (0.017)&0.6221&0.945 (0.016)&0.7617\\
&500&0.950 (0.015)&0.3826&0.945 (0.016)&0.4740\\
&1000&0.940 (0.017)&0.2687&0.935 (0.017)&0.3332\\
\midrule
Liebl--Reimherr&50&0.955 (0.015)&3.2275&0.930 (0.018)&3.3324\\
&100&0.980 (0.010)&1.2035&0.960 (0.014)&1.3554\\
&200&0.970 (0.012)&0.7441&0.980 (0.010)&0.8633\\
&500&0.965 (0.013)&0.4569&0.965 (0.013)&0.5349\\
&1000&0.975 (0.011)&0.3208&0.965 (0.013)&0.3762\\
\midrule
Pointwise normal&50&0.790 (0.029)&2.2123&0.725 (0.032)&2.2644\\
&100&0.780 (0.029)&0.7998&0.765 (0.030)&0.8925\\
&200&0.840 (0.026)&0.4915&0.825 (0.027)&0.5681\\
&500&0.835 (0.026)&0.3014&0.785 (0.029)&0.3517\\
&1000&0.865 (0.024)&0.2115&0.775 (0.030)&0.2474\\
\bottomrule
\end{tabular}
\end{table}

\subsection{Iid functional approximation and component studies}
\label{app:functional-iid}

The target-matched functional studies use independent Bernoulli$(1/2)$ mixture memberships with the same means, covariance and Fourier coefficient construction as above. At $n=500$, treatment follows the interaction logit multiplied by $\kappa\in\{1,4\}$ and truncated to $[0.01,0.99]$ as part of the generating mechanism. Mean estimation uses this known propensity, two folds randomized independently of the observations, five Fourier functions, and $999$ raw Gaussian multiplier draws. The three profiles are $m_0$, $(m_0+m_1)/2$, and $m_1$, fixed before fitted-sample generation. Each profile is evaluated separately at level $0.90$.

The base curves have analytic TATE and CTATE references. The stepped design adds $\mathbf{1}\{X_1>0.7\}\sin(\pi t)$ to each treatment component. Its CTATE reference adds this term at the specified profile; its TATE reference adds $P(X_1>0.7)\sin(\pi t)$, with the probability calculated from the known Gaussian mixture. Numerical reference errors are zero. This perturbation deliberately violates the working linear covariate model and fixed periodic Fourier span. Correct propensity preserves the population AIPW mean identity, but it does not eliminate the second-stage approximation requirement for a fixed-profile confidence band.

The three-component study generates three Fourier components, indexed by $d=0,1,2$. Its fixed coefficient draw changes the first two conditional means as well, so this is not a nested extension holding their outcomes fixed. These signed curves bypass persistence, making the extra component a statistical multiplicity experiment rather than a degree-two homology calculation. Prediction uses the common calibration indices and maximum score over all components and grid points. Section~\ref{sec:sim-functional} summarizes the mean calculations, and Table~\ref{tab:remaining-degrees} gives the joint prediction comparison.

\subsection{Fixed-profile CTATE diagnostics}
\label{app:ctate-simulation}

The conditional-mean study evaluates $m_0$, $(m_0+m_1)/2$, and $m_1$ at nominal $0.90$, with the same functional design, $200$ replications, $50$ grid points, five Fourier basis functions, and $1{,}500$ multiplier draws. The second-stage linear model and Fourier projection match the truth. These are profile-wise confidence calculations, with no simultaneous guarantee over all three profiles. Table~\ref{tab:ctate-simulation} reports the multiplier results. Monte Carlo standard errors measure fitted-sample variability; they do not measure model-approximation error outside this design.

\begin{table}[ht]
\centering\small
\caption{Fixed-profile CTATE gridwise coverage and width in the correctly specified functional design. Parentheses give Monte Carlo standard errors from $200$ replications.}
\label{tab:ctate-simulation}
\begin{tabular}{@{}llrrrr@{}}
\toprule
& &\multicolumn{2}{c}{$d=0$}&\multicolumn{2}{c}{$d=1$}\\
\cmidrule(lr){3-4}\cmidrule(lr){5-6}
Profile&$n$&Coverage&Width&Coverage&Width\\
\midrule
$m_0$&200&0.870 (0.024)&0.3615&0.860 (0.025)&0.3579\\
&500&0.895 (0.022)&0.1960&0.920 (0.019)&0.1952\\
&1000&0.915 (0.020)&0.1347&0.900 (0.021)&0.1338\\
\midrule
$(m_0+m_1)/2$&200&0.885 (0.023)&0.2559&0.845 (0.026)&0.2541\\
&500&0.915 (0.020)&0.1407&0.895 (0.022)&0.1395\\
&1000&0.905 (0.021)&0.0952&0.890 (0.022)&0.0951\\
\midrule
$m_1$&200&0.870 (0.024)&0.3427&0.890 (0.022)&0.3428\\
&500&0.905 (0.021)&0.1927&0.900 (0.021)&0.1907\\
&1000&0.920 (0.019)&0.1293&0.890 (0.022)&0.1295\\
\bottomrule
\end{tabular}
\end{table}

\subsection{Point-cloud design and inferential failures}
\label{app:levelB}\label{app:regime}

The point-cloud generator uses the first three coordinates $X=(X_1,X_2,X_3)^\top$ from the Gaussian subgroups, with subgroup means $m_0=(1.0, 0.6, -0.7)^\top$, $m_1=(0.4, -0.4, -0.6)^\top$, covariance $0.5I_3$, and overall mean $E[X]=(0.7, 0.1, -0.65)^\top$. For the fixed unit direction $w=(0.1888, -0.1984, 0.9618)^\top$, giving mean projection $E[X]^\top w = -0.5128$, the arm-specific loop count is
\[
 L^a(x)=\operatorname{clip}\bigl(\operatorname{round}\{2+2a+1.2(x-E[X])^\top w\},0,6\bigr).
\]
Each loop contains $40$ points on a radius-one circle with independent angular positions and isotropic Gaussian coordinate noise of standard deviation $0.08$. Circle centers are spaced by three radius units and placed on a $3\times3$ grid at coordinates $(3(k\bmod 3), 3\lfloor k/3\rfloor)$ for $k=0,\dots,L^a(x)-1$. Six background points are sampled independently from the uniform distribution on the bounding square $[-9,9]^2$. Alpha-complex persistence and the power-weighted silhouette with $r=3$ produce two homology degrees on $45$ grid points in $[0,1]$. Topology values in this stress-testing design follow the $\sqrt{2}$-scaled silhouette convention, $\sqrt{2}\phi_d$, implemented by \href{https://gudhi.inria.fr/python/3.7.0/_modules/gudhi/representations/vector_methods.html#Silhouette}{GUDHI\textquotesingle s \texttt{Silhouette.transform}}. \citet[Appendix D.1]{kim-lee} report using GUDHI for their experiments; their theoretical Equation~(2) defines the unscaled silhouette. This deterministic scaling leaves inclusion events invariant but scales numerical widths and constants in forward stability inequalities by $\sqrt{2}$.

The main stress settings multiply the treatment logit by $\kappa=3$ or $5$, replace the propensity regression by a constant, replace the outcome regression by a constant, or replace both. Other settings change point noise to $0.03$ or $0.22$, loop slope to $0.4$ or $2.6$, or multiply loop radii by $1+0.6|T|$ for standardized Student-$t$ noise. The TATE reference averages $384$ covariate draws with eight independently generated cloud pairs per draw. There are $80$ fitted-sample replications with $600$ multiplier draws each. Table~\ref{tab:sim-levelb} evaluates performance under these stress configurations relative to this numerical reference.

\begin{table}[ht]
\centering\small
\caption{Point-cloud TATE stress tests under severe overlap violation ($\kappa \in \{3,5\}$) and model misspecification, $80$ Monte Carlo replications at nominal level $0.90$. Parentheses report replication-level Monte Carlo standard errors.}
\label{tab:sim-levelb}
\begin{tabular}{@{}lrrr@{}}
\toprule
Scenario & $n$ & Coverage $d=0$ & Coverage $d=1$\\
\midrule
Base & 500 & 0.900 (0.034) & 0.825 (0.042)\\
Weak overlap ($\kappa=3$) & 500 & 0.713 (0.051) & 0.650 (0.053)\\
Very weak overlap ($\kappa=5$) & 500 & 0.375 (0.054) & 0.188 (0.044)\\
Propensity misspecified & 800 & 0.850 (0.040) & 0.613 (0.054)\\
Outcome misspecified & 800 & 0.963 (0.021) & 0.663 (0.053)\\
Both misspecified & 800 & 0.900 (0.034) & 0.063 (0.027)\\
\bottomrule
\end{tabular}
\end{table}

\subsection{Target-matched point-cloud simulation design}
\label{app:topology-rerun}

The target-matched point-cloud design shares the loop geometry and noise parameters of Appendix~\ref{app:levelB}, with $40$ points per loop, six background points, $r=3$, and homology degrees zero and one. Silhouettes are evaluated on $50$ filtration points in standard tent units. Covariates are i.i.d.\ draws from the equal Gaussian mixture. Treatment assignment follows the interaction logistic probability with logit scale $\kappa\in\{1,4\}$, truncated to $[0.01,0.99]$ as part of the generating law. The generating probability, including this truncation, is known to the estimator.

Each setting has $200$ fitted-sample replications at $n=500$. Mean bands use two independent folds, five Fourier functions, a linear covariate regression, and $999$ raw Gaussian multiplier draws at level $0.90$. The marginal reference uses $10{,}000$ independent paired clouds; each of three fixed covariate profiles uses $3{,}000$ independent pairs. The three fixed profiles, evaluated separately, are $x^{(0)}=(0.7032, 0.1043, -0.6482)^\top$, $x^{(1)}=(0.4958, -1.0190, -2.0448)^\top$, and $x^{(2)}=(0.8327, 1.6403, -1.4177)^\top$. Their largest pointwise Monte Carlo standard errors are $0.00132$ and $0.00237$, respectively. These errors are retained but not incorporated into the confidence calibration, so the reported mean coverage is relative to approximate reference curves. The same reference arrays are used at both logit scales because the potential-outcome law is unchanged.

Prediction uses a half-sample training split, a scale fixed at one, true arm ratios, and $200$ new paired outcomes per target per fitted sample. The full, overlap, and matching tilts are those specified in Appendix~\ref{app:bounded-rerun}. Table~\ref{tab:topology-rerun} gives the joint-score calculation over both degrees, with arm errors $0.025$. The composed bands cover every evaluated individual contrast, although both arm events together occur less often. Empirical zero failures do not establish zero failure probability. Width first averages finite predictions within each fitted sample and then averages those sample means; it is not a pooled average weighted by the number of finite predictions. Different target rows concern different populations and cannot establish same-population efficiency.

\begin{table}[ht]
\centering\small
\caption{Target-matched point-cloud prediction, $200$ replications per logit scale, nominal level $0.95$. All events cover both degrees on the evaluation grid. Widths use plain-tent units. Parentheses give Monte Carlo standard errors across fitted samples.}
\label{tab:topology-rerun}
\begin{tabular}{@{}rlrrrr@{}}
\toprule
$\kappa$ & Target & Effect coverage & Arm intersection & Finite width & Unbounded share\\
\midrule
1 & Full & 1.0000 & 0.9618 (0.0014) & 2.1523 & 0.0285\\
1 & Overlap & 1.0000 & 0.9599 (0.0016) & 2.1431 & 0.0000\\
1 & Matching & 1.0000 & 0.9598 (0.0016) & 2.1385 & 0.0000\\
4 & Full & 1.0000 & 0.9662 (0.0015) & 2.1536 & 0.5787\\
4 & Overlap & 1.0000 & 0.9700 (0.0014) & 2.1906 & 0.4425\\
4 & Matching & 1.0000 & 0.9704 (0.0015) & 2.1921 & 0.3330\\
\bottomrule
\end{tabular}
\end{table}

The direct paired-contrast comparator, available only in simulation, has coverage $0.9509$--$0.9548$ across these targets and settings. For TATE, coverage is $0.930$ and $0.895$ in degrees zero and one at scale one, and $0.835$ and $0.845$ at scale four. Across the three profiles and two degrees, CTATE coverage ranges from zero to $0.665$ at scale one and from $0.295$ to $0.760$ at scale four. The linear second stage and Fourier projection have non-negligible approximation error for this rounded-loop design. The low profile coverage is therefore consistent with the bias requirement in Section~\ref{subsec:ctate}; it is not evidence against a theorem that assumes this bias is negligible.

\subsection{Weighting and quantile diagnostics}
\label{app:stabilization}

The iid weighting study crosses three prediction-weight fits with six weight maps in the functional design at $\kappa=1,4$ and the $r=3$ point-cloud design at $\kappa=4$. Every setting has $n=500$, $200$ fitted samples, an independent half-sample training split, a score scale fixed at one, and $200$ new paired outcomes per declared target. The exact-weight fits use the true generating propensity, including its truncation. Learned fits use the interaction features; constant fits use only an intercept. An interaction-logistic regression cannot reproduce the truncated generating propensity exactly. Neither fitted alternative is treated as an exact density ratio or as an observed-data double-robust prediction procedure.

The full, overlap and matching tilts follow Appendix~\ref{app:bounded-rerun}. The fourth population uses $h=\sqrt{\pi(1-\pi)}$. New covariates are drawn using the true generating tilt in each case; substituting the fitted propensity in the calibration weights does not change that evaluation population. Estimated-ratio validity is therefore subject to the discrepancy bound of Section~\ref{subsec:estimated}. Different population rows do not establish efficiency for a common estimand.

\begin{figure}[ht]
\centering
\includegraphics[width=\textwidth]{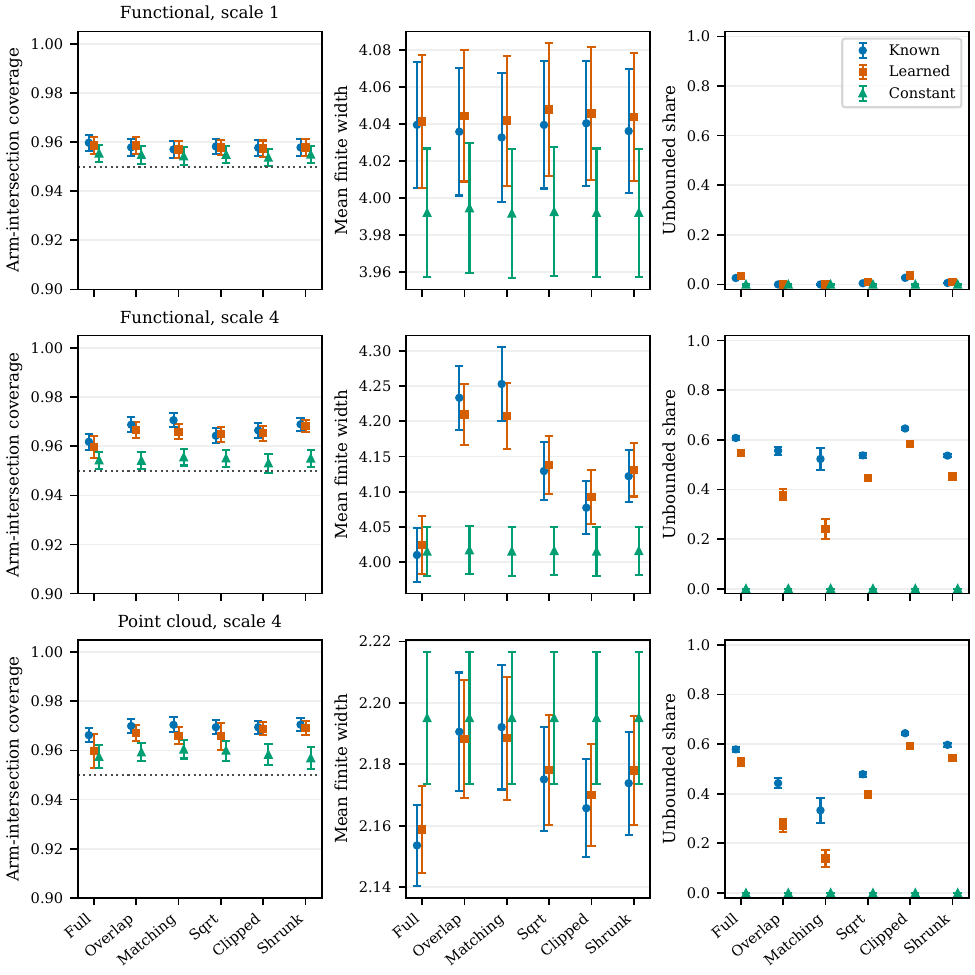}
\caption{Weight and target sensitivities for iid functional and point-cloud designs. Error bars are two Monte Carlo standard errors across $200$ fitted samples. The horizontal arm-intersection reference is $0.95$. Width averages only finite target predictions, first within fitted samples and then across samples with a finite value; unbounded share is reported separately. Full, overlap, matching and square-root rows concern different populations. Clipped and shrunk ratios are evaluated against the full target without an exact full-population guarantee.}
\label{fig:remaining-weights}
\end{figure}

The remaining maps approximate full-population ratios. Clipping replaces the denominator probability by its truncation to $[0.1,0.9]$. Shrinkage replaces an inverse-arm weight $w$ by $w/(1+w/10)$. These maps need not define the same covariate law for both arms; they are evaluated against fresh full-population paired outcomes as sensitivity procedures, without exact full-target claims. Table~\ref{tab:remaining-weights} reports coverage together with unbounded share, and Figure~\ref{fig:remaining-weights} adds finite width and arm-intersection uncertainty. Joint scores cover both components and every evaluated grid point, with arm errors $0.025$.

\begin{table}[ht]
\centering\footnotesize\setlength{\tabcolsep}{3pt}
\caption{Target-specific prediction sensitivities. Each cell reports composed coverage (fitted-sample Monte Carlo standard error), followed by unbounded share. There are $200$ fitted samples and $200$ independent target draws per sample. The clipped and shrunk rows evaluate the full population with approximate arm ratios. Other target rows evaluate their own true covariate tilts. Zero observed failures or unbounded bands do not imply zero probability. Finite width and arm-intersection uncertainty appear in Figure~\ref{fig:remaining-weights}.}
\label{tab:remaining-weights}
\begin{tabular}{@{}lrrr@{}}
\toprule
Target or ratio & Known & Learned & Constant\\
\midrule
\multicolumn{4}{@{}l@{}}{\textit{Functional, $\kappa=1$}}\\
Full & 0.9997 (0.0001) / 0.0261 & 0.9997 (0.0001) / 0.0342 & 0.9996 (0.0001) / 0.0000\\
Overlap & 0.9997 (0.0001) / 0.0000 & 0.9998 (0.0001) / 0.0000 & 0.9997 (0.0001) / 0.0000\\
Matching & 0.9997 (0.0001) / 0.0000 & 0.9997 (0.0001) / 0.0000 & 0.9997 (0.0001) / 0.0000\\
Square-root & 0.9997 (0.0001) / 0.0047 & 0.9997 (0.0001) / 0.0090 & 0.9996 (0.0001) / 0.0000\\
Full, clipped & 0.9995 (0.0001) / 0.0267 & 0.9996 (0.0001) / 0.0366 & 0.9995 (0.0001) / 0.0000\\
Full, shrunk & 0.9996 (0.0001) / 0.0062 & 0.9996 (0.0001) / 0.0100 & 0.9995 (0.0001) / 0.0000\\
\addlinespace
\multicolumn{4}{@{}l@{}}{\textit{Functional, $\kappa=4$}}\\
Full & 0.9999 (0.0001) / 0.6077 & 0.9999 (0.000050) / 0.5458 & 0.9997 (0.0001) / 0.0000\\
Overlap & 0.9999 (0.0001) / 0.5563 & 0.9998 (0.0001) / 0.3783 & 0.9996 (0.0001) / 0.0000\\
Matching & 1.0000 (0.000035) / 0.5229 & 0.9998 (0.0001) / 0.2400 & 0.9995 (0.0001) / 0.0000\\
Square-root & 0.9998 (0.0001) / 0.5374 & 0.9998 (0.0001) / 0.4468 & 0.9997 (0.0001) / 0.0000\\
Full, clipped & 0.9999 (0.000043) / 0.6458 & 0.9999 (0.0001) / 0.5842 & 0.9996 (0.0001) / 0.0000\\
Full, shrunk & 0.9999 (0.000050) / 0.5364 & 0.9999 (0.0001) / 0.4519 & 0.9996 (0.0001) / 0.0000\\
\addlinespace
\multicolumn{4}{@{}l@{}}{\textit{Point cloud, $\kappa=4$}}\\
Full & 1.0000 (0.0000) / 0.5787 & 1.0000 (0.0000) / 0.5273 & 1.0000 (0.0000) / 0.0000\\
Overlap & 1.0000 (0.0000) / 0.4425 & 1.0000 (0.0000) / 0.2717 & 1.0000 (0.0000) / 0.0000\\
Matching & 1.0000 (0.0000) / 0.3330 & 1.0000 (0.0000) / 0.1383 & 1.0000 (0.0000) / 0.0000\\
Square-root & 1.0000 (0.0000) / 0.4786 & 1.0000 (0.0000) / 0.3967 & 1.0000 (0.0000) / 0.0000\\
Full, clipped & 1.0000 (0.0000) / 0.6438 & 1.0000 (0.0000) / 0.5933 & 1.0000 (0.0000) / 0.0000\\
Full, shrunk & 1.0000 (0.0000) / 0.5970 & 1.0000 (0.0000) / 0.5450 & 1.0000 (0.0000) / 0.0000\\
\bottomrule
\end{tabular}
\end{table}

A finite weighted quantile requires adequate calibration mass relative to the new-point atom, as in Lemma~\ref{lem:finite}. Bounded individual weights alone do not ensure this condition. Finite width first averages finite predictions within each fitted sample and then averages the fitted-sample means with a finite value; it is reported alongside the separate unbounded frequency. Weighting calculations evaluating learned overlap weights and calibration-estimated modulation on full populations are treated as sensitivity diagnostics rather than exact-target guarantees.

\subsection{Degrees, exponents, and the forward stability map}
\label{app:degree-r}\label{app:figures}

The three-component functional design uses iid covariates at $n=500$ and logit scale one, known arm ratios and an independently fixed score scale. Joint calibration takes the maximum over all three components; Bonferroni uses arm error $0.025/3$ for each component. The \v{S}id\'ak comparator uses $1-(1-0.025)^{1/3}$ and requires additional dependence conditions for a joint guarantee. A direct paired-contrast score supplies a simulation-only comparison. Table~\ref{tab:remaining-degrees} evaluates each scheme under separate full, overlap and matching target draws.

\begin{table}[ht]
\centering\footnotesize\setlength{\tabcolsep}{3pt}
\caption{Joint prediction across three generated functional components, $n=500$, $200$ fitted samples and nominal level $0.95$. Parentheses give fitted-sample Monte Carlo standard errors. The direct contrast uses unavailable paired outcomes as a simulation comparator. The \v{S}id\'ak rule is a comparator without a general joint guarantee. $R_f$ counts fitted samples with any finite prediction; their finite-prediction means receive equal weight. A width based on one fitted sample has unavailable Monte Carlo standard error, shown as NA. No width exists when $R_f=0$.}
\label{tab:remaining-degrees}
\begin{tabular}{@{}llrrrrr@{}}
\toprule
Target & Scheme & Effect & Arms & Finite width & $R_f$ & Unbounded\\
\midrule
Full & Joint maximum & 1.0000 (0.000035) & 0.9586 (0.0016) & 4.212 (0.015) & 200 & 0.02643 (0.00096)\\
Full & Bonferroni & 1.0000 (0.000025) & 0.9751 (0.0010) & 4.280 (0.015) & 190 & 0.92158 (0.00358)\\
Full & \v{S}id\'ak comparator & 1.0000 (0.000025) & 0.9748 (0.0011) & 4.280 (0.015) & 196 & 0.90760 (0.00363)\\
Full & Direct contrast & 0.9503 (0.0014) & NA & 2.653 (0.007) & 200 & 0.00000 (0.00000)\\
Overlap & Joint maximum & 0.9998 (0.0001) & 0.9584 (0.0016) & 4.206 (0.015) & 200 & 0.00000 (0.00000)\\
Overlap & Bonferroni & 1.0000 (0.0000) & 0.9794 (0.0009) & 4.442 (NA) & 1 & 0.99995 (0.00005)\\
Overlap & \v{S}id\'ak comparator & 1.0000 (0.0000) & 0.9788 (0.0009) & 4.442 (NA) & 1 & 0.99990 (0.00010)\\
Overlap & Direct contrast & 0.9516 (0.0014) & NA & 2.650 (0.007) & 200 & 0.00000 (0.00000)\\
Matching & Joint maximum & 0.9998 (0.0001) & 0.9584 (0.0015) & 4.202 (0.015) & 200 & 0.00000 (0.00000)\\
Matching & Bonferroni & 1.0000 (0.0000) & 0.9836 (0.0007) & NA & 0 & 1.00000 (0.00000)\\
Matching & \v{S}id\'ak comparator & 1.0000 (0.0000) & 0.9835 (0.0007) & NA & 0 & 1.00000 (0.00000)\\
Matching & Direct contrast & 0.9547 (0.0014) & NA & 2.651 (0.007) & 200 & 0.00000 (0.00000)\\
\bottomrule
\end{tabular}
\end{table}

The exponent study holds the point-cloud generator, $n=500$, two degrees, $40$ points per loop and logit scale one fixed while changing the prespecified exponent to $r=1,2,3,4,5$. Every exponent has its own independently evaluated marginal and three-profile mean reference, using $10{,}000$ and $3{,}000$ paired clouds, respectively. All curves and widths are in plain-tent units. The $r=3$ calculation is the primary study of Appendix~\ref{app:topology-rerun}; the other exponents are outcome sensitivities, rather than estimators of the same functional target.

\begin{table}[ht]
\centering\footnotesize\setlength{\tabcolsep}{3pt}
\caption{Exponent sensitivity in plain-tent units at $n=500$ and logit scale one, $200$ fitted samples per exponent. Mean bands target $0.90$ and full-population joint prediction targets $0.95$. Parentheses give fitted-sample Monte Carlo standard errors. CTATE ranges span six separately evaluated profile-degree pairs. Mean coverage is relative to Monte Carlo reference curves whose uncertainty is not included in band calibration. Each exponent defines a different outcome.}
\label{tab:remaining-exponents}
\begin{tabular}{@{}rrrrrrr@{}}
\toprule
$r$ & TATE $d=0$ & TATE $d=1$ & CTATE range & ITTE & Finite width & Unbounded\\
\midrule
1 & 0.920 (0.019) & 0.895 (0.022) & 0.000--0.435 & 1.0000 (0.0000) & 1.757 (0.007) & 0.0285 (0.0009)\\
2 & 0.925 (0.019) & 0.895 (0.022) & 0.000--0.310 & 1.0000 (0.0000) & 2.170 (0.004) & 0.0285 (0.0009)\\
3 & 0.930 (0.018) & 0.895 (0.022) & 0.000--0.665 & 1.0000 (0.0000) & 2.152 (0.006) & 0.0285 (0.0009)\\
4 & 0.930 (0.018) & 0.905 (0.021) & 0.045--0.805 & 1.0000 (0.0000) & 2.119 (0.006) & 0.0285 (0.0009)\\
5 & 0.930 (0.018) & 0.890 (0.022) & 0.075--0.835 & 1.0000 (0.0000) & 2.100 (0.006) & 0.0285 (0.0009)\\
\bottomrule
\end{tabular}
\end{table}

The point-cloud and ORBIT population-mean calculations record pointwise Monte Carlo reference errors. Across these calculations their maximum standard errors range from $1.6\times10^{-5}$ to $3.0\times10^{-3}$; the largest ratio of a pointwise reference standard error to the grid-average band halfwidth is $0.245$. The denominator is an average width, not the smallest pointwise coverage margin. These ratios diagnose reference precision and do not correct coverage or validate the inferential assumptions. Reference uncertainty is not incorporated into confidence calibration.

All five exponents lie within Theorem~\ref{thm:stability}, with forward constants $1+2r$. The theorem concerns diagram $W_1$ error; a small difference between silhouettes supplies no upper bound on the diagram error. Numerical perturbation checks remain implementation diagnostics rather than proofs or inverse guarantees. The prediction guarantee is a lower bound, so near-one composed inclusion does not contradict a $0.95$ target. Armwise events, their intersection and the direct-contrast comparator are reported without assigning a fixed fraction of conservatism to interval composition.

\subsection{ORBIT dynamical-system benchmark}
\label{app:orbit}\label{sec:real-orbit}\label{sec:real-honesty}\label{app:orbit-single}

ORBIT arises from the linked twist map dynamical system used by \citet{adams}, with its corrected sequential update \citep{adams-erratum}. For $h_s(u)=su(1-u)$, the map is
\[
 F_s(x,y)=(x+h_s(y),y)\bmod1,\qquad
 G_s(x,y)=(x,y+h_s(x))\bmod1,\qquad T_s=G_s\circ F_s.
\]
Thus the second coordinate uses the updated first coordinate. For each synthetic unit, an unordered pair is sampled uniformly without replacement from $s\in\{3.5,4,4.1\}$, and one cloud is generated at each parameter using independent initial coordinates. The larger-parameter cloud is assigned to potential arm one with probability $0.7$; the potential-arm ordering is reversed with probability $0.3$. This ordering is separate from realized treatment assignment. Alpha-complex persistence is converted to radii, essential classes are removed, and degree-zero and degree-one silhouettes use $r=3$ on $[0,0.2]$. The covariates and treatment mask are subsequently generated independently of the clouds, following the constructed contrast of \citet{kim-lee}.

The repeated-sampling benchmark generates $1{,}000$ points per cloud and $200$ independent fitted samples of size $n=300$. An independent sample of $10{,}000$ paired outcomes fixes a population mean reference before those fits. Since clouds and covariates are independent, every conditional-mean reference equals that population curve. This differs from comparing a fitted band with the paired contrasts retained in its own sample.

Mean fits use two independently randomized folds, the known interaction propensity truncated to $[0.01,0.99]$ as part of the generating law, five Fourier functions and $999$ raw Gaussian multiplier draws. Three profiles from the independent reference covariate sample, identically matching the profile coordinates $x^{(0)}, x^{(1)}, x^{(2)}$ defined in Appendix~\ref{app:sars-rerun}, are fixed before fitting. Their confidence calculations remain separate. Prediction fixes the scale at one on an independent training split, uses true arm ratios, and evaluates $200$ new paired outcomes under each full, overlap and matching target per fitted sample. Table~\ref{tab:remaining-orbit} reports gridwise population-reference confidence and target-matched individual prediction in plain-tent units.

\begin{table}[ht]
\centering\footnotesize\setlength{\tabcolsep}{3pt}
\caption{ORBIT with $1{,}000$ points per cloud, $n=300$ and $200$ independent fitted samples. TATE gridwise confidence targets $0.90$ and joint individual prediction targets $0.95$. Parentheses give fitted-sample Monte Carlo standard errors. Width is full grid-average width in plain-tent units, restricted to finite bands for prediction. Mean coverage uses a fixed independent Monte Carlo population reference.}
\label{tab:remaining-orbit}
\begin{tabular}{@{}llrrr@{}}
\toprule
Object or target & Degree & Coverage & Width & Unbounded share\\
\midrule
TATE & 0 & 0.910 (0.020) & 0.001968 (0.000019) & NA\\
TATE & 1 & 0.900 (0.021) & 0.009010 (0.000057) & NA\\
ITTE, Full & Joint & 0.9999 (0.000050) & 0.20944 (0.00129) & 0.1420 (0.0026)\\
ITTE, Overlap & Joint & 1.0000 (0.000035) & 0.21077 (0.00132) & 0.0238 (0.0021)\\
ITTE, Matching & Joint & 0.9998 (0.0001) & 0.21097 (0.00134) & 0.0000 (0.0000)\\
\bottomrule
\end{tabular}
\end{table}

\begin{figure}[ht]
\centering
\includegraphics[width=\textwidth]{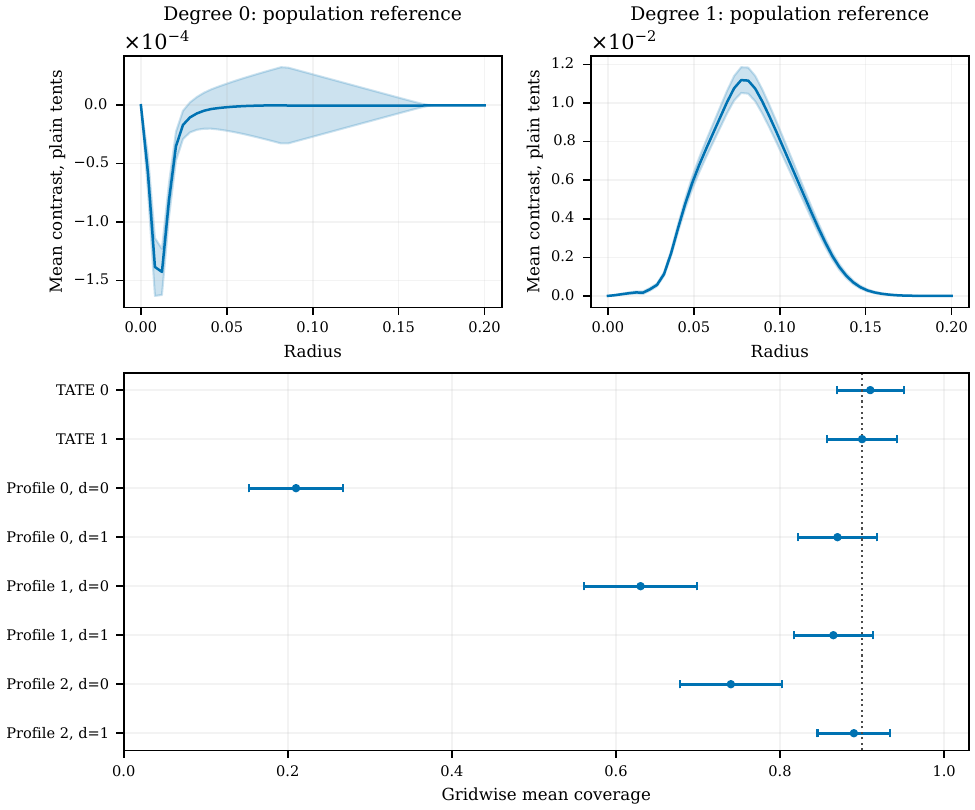}
\caption{ORBIT population-reference study. The first two panels show the independently simulated mean contrast with a pointwise two-standard-error reference envelope, in plain-tent units. The last panel reports fitted-sample gridwise mean coverage with two-Monte-Carlo-standard-error bars and the nominal $0.90$ line. Reference uncertainty is not included in the confidence bands; the envelope is not a simultaneous confidence region. Profiles are evaluated separately.}
\label{fig:orbit-population}
\end{figure}

Across the three separate profiles and two degrees, CTATE coverage ranges from $0.210$ to $0.890$. The largest pointwise reference standard error is $3.4\times10^{-4}$. It is retained but not incorporated into confidence calibration, so these frequencies concern an approximate population reference. They do not establish continuum coverage or joint coverage across profiles. A mean constant in covariates can still lie outside the response Fourier span; low profile coverage does not demonstrate covariate heterogeneity. Figure~\ref{fig:orbit-population} shows the reference and the mean-coverage calculation across profiles.

\begin{figure}[ht]
\centering
\includegraphics[width=0.85\textwidth]{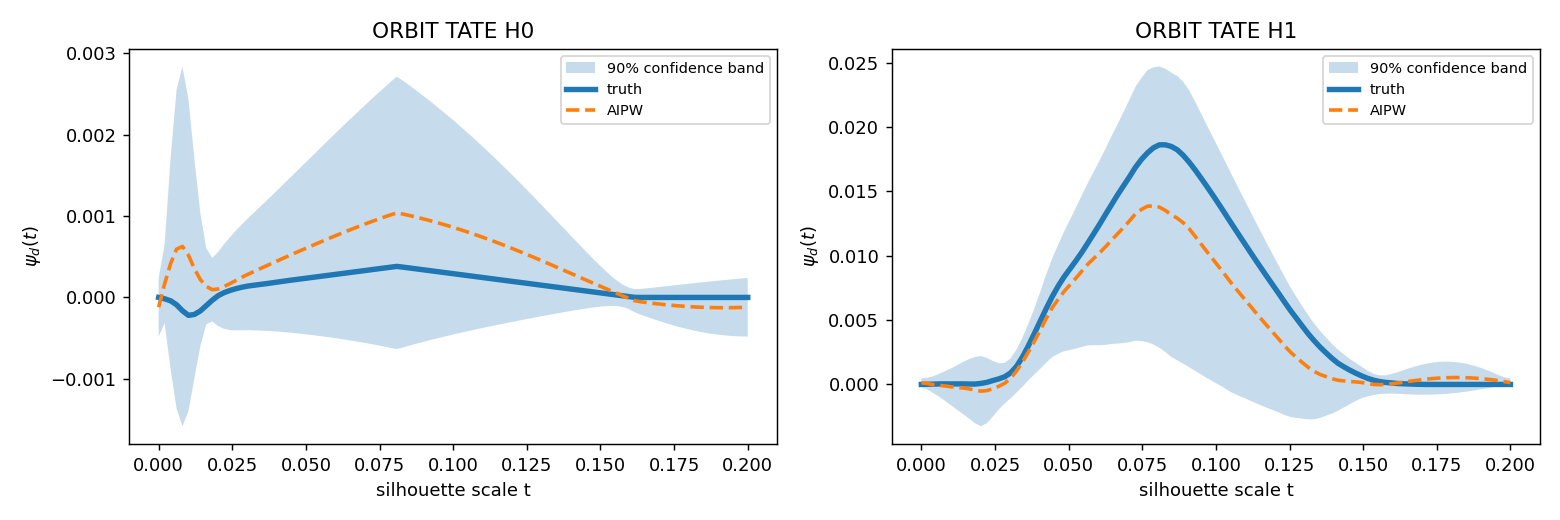}
\caption{Single sample realization of the ORBIT benchmark at $n=300$, showing a nominal $90\%$ gridwise multiplier band and the corresponding sample-specific paired-contrast reference under the $\sqrt{2}\phi_d$ convention. This illustration does not estimate population-mean coverage across fitted samples.}
\label{fig:orbit}
\end{figure}

To visualize an individual sample realization alongside the repeated-sampling population study, Figure~\ref{fig:orbit} illustrates one fitted-sample multiplier band compared with the sample-specific paired-contrast mean at $n=300$. Its gridwise inclusion event is distinct from repeated-sampling coverage of a fixed population mean. Curves in Figure~\ref{fig:orbit} use the rescaled silhouette convention $\sqrt{2}\phi_d$, whereas the population-reference study above uses unscaled tent units.

\subsection{CT-slice benchmark design (\texorpdfstring{$r=1$}{r=1})}
\label{app:sars-rerun}

Following the synthetic slice-pool construction of \citet[Section 6 and Appendix D.2]{kim-lee}, all $1{,}252$ available COVID-19 and $1{,}229$ non-COVID-19 images are processed at their original dimensions, using their first channel divided by $255$, with no resizing, masking, or cropping. The non-COVID-19 count matches the copy reported by \citet[Appendix D.1]{kim-lee}, but is one smaller than the $1{,}230$ non-COVID-19 images reported in the dataset paper \citep[version 2]{soares}; the reason for the discrepancy is undocumented. Finite degree-zero lower-star cubical bars yield plain power-weighted silhouettes at $r=1$ on $50$ points in $[0,1]$. Arm zero resamples COVID-19 slices; arm one independently resamples a mixture containing $75\%$ non-COVID-19 and $25\%$ COVID-19 slices. Both draws are independent of the covariates. Thus the TATE and every CTATE are exactly $0.75$ times the difference between the two empirical pool means. This is an analytic target for the resampling law, without oracle Monte Carlo error.

Covariates are iid draws from the functional design's Gaussian mixture. Treatment follows its logistic rule truncated to $[0.01,0.99]$, with the true probability known to the AIPW and prediction procedures. Two independent folds and five Fourier functions give the mean fit. Confidence calibration uses $999$ raw Gaussian multiplier draws. Prediction uses half the $n=400$ sample for training, a scale fixed at one, and $200$ new paired slices under each declared target per fitted sample. Target tilts follow Appendix~\ref{app:bounded-rerun}. There are $200$ successful fitted-sample replications and no failed fits.

The three profile points, fixed from an independent covariate reference draw before the replications, are approximately
\[
\begin{aligned}
x^{(0)}&=(0.6946,0.0933,-0.6438,2.7483,1.0135),\\
x^{(1)}&=(0.4958,-1.0190,-2.0448,3.9217,3.0105),\\
x^{(2)}&=(0.4651,1.4650,1.0655,2.5499,-1.9537).
\end{aligned}
\]
Table~\ref{tab:sars-rerun} reports mean confidence and joint-score individual prediction. With one homology degree, joint-score and Bonferroni calibration coincide. Mean coverage is tested against the unprojected empirical-pool curve, so a fixed Fourier approximation can still introduce bias even though the true covariate effect is constant. The profile bands are separate, with no simultaneous guarantee across all three points. The prediction width averages finite predictions within each fitted sample and then averages across fitted samples. The comparison between full and tilted targets concerns different populations. Slice resampling and generated treatment assignments do not confer patient-level causal interpretation.

\begin{table}[ht]
\centering\small
\caption{Exponent-one CT-slice resampling study, $200$ replications. Mean bands target level $0.90$ and prediction targets $0.95$. Parentheses give Monte Carlo standard errors over fitted samples. Widths are full grid averages in plain-tent units, restricted to finite bands for prediction.}
\label{tab:sars-rerun}
\begin{tabular}{@{}llrrr@{}}
\toprule
Object & Profile or target & Coverage & Width & Unbounded share\\
\midrule
TATE & Full & 0.920 (0.019) & 0.00234 & \\
CTATE & $x^{(0)}$ & 0.700 (0.032) & 0.00220 & \\
CTATE & $x^{(1)}$ & 0.860 (0.025) & 0.00626 & \\
CTATE & $x^{(2)}$ & 0.895 (0.022) & 0.00772 & \\
ITTE & Full & 0.999925 (0.000043) & 0.07319 & 0.051225\\
ITTE & Overlap & 0.999875 (0.000055) & 0.07336 & 0.000025\\
ITTE & Matching & 0.999950 (0.000035) & 0.07381 & 0.000000\\
\bottomrule
\end{tabular}
\end{table}

\subsection{CT-slice boundary sensitivity (\texorpdfstring{$r=0.1$}{r=0.1})}
\label{app:sars-diagnostic}

Following \citet[Section 6 and Appendix D.2]{kim-lee}, the CT benchmark loads each image in its available resolution, uses its first channel divided by $255$, applies no resizing, lung masking, or cropping, and computes degree-zero lower-star cubical persistence on the resulting intensity field. Infinite classes are removed. Its $r=0.1$ silhouettes use $[0,1]$ with the rescaled $\sqrt{2}\phi_0$ normalization. This exponent falls outside the $r\ge1$ stability regime of Theorem~\ref{thm:stability}.

Arm zero is sampled from the COVID-19 slice pool; arm one is a mixture of $75\%$ non-COVID-19 and $25\%$ COVID-19 slices. The images are paired arbitrarily and covariates are generated independently. The population conditional mean of this constructed contrast is constant in $x$. The dotted curve in the top-left panel of Figure~\ref{fig:sars} is its sample-specific reference. The three profile estimates in the bottom panel differ from this sample-specific reference; this illustration does not estimate conditional bias or coverage. The separate $r=1$ repeated-sampling study in Appendix~\ref{app:sars-rerun} evaluates profile coverage against the empirical-pool target.

At $n=400$, integrated absolute mean discrepancy is $6.2\times10^{-5}$ and integrated confidence width is $4.9\times10^{-4}$. The prediction object's grid-averaged finite width is $0.0208$. The mean and prediction widths are therefore different summaries even before their different targets are considered. Patient identifiers are unavailable in the image-level data used for this analysis; independently resampling slices defines only the empirical slice-pool benchmark.

\section{PACCE construction and exploratory diagnostics}
\label{app:pacce}

\subsection{Released cohort and geometric representation}
\label{app:pacce-construction}

The supplied release contains $842$ subjects; availability varies across outcome and assessment domains. The assigned chemotherapy backbone is oxaliplatin for $336$ subjects per arm and irinotecan for $85$ per arm. The intended-treatment flag is present for all $842$; actual treatment differs from assignment for nine subjects. This analysis uses assigned treatment. Because the sampling mechanism selecting these $842$ patients from the $1{,}053$ randomized participants in the original trial is not documented in the data repository, generalizing findings to the full intention-to-treat trial population requires transportability assumptions.

The longitudinal geometric outcome summarizes target-lesion dynamics during early treatment. Central target-lesion assessments are available at baseline for $808$ subjects and at the scheduled week-12 visit for $662$ subjects ($661$ having both). To construct a non-trivial two-dimensional point cloud for persistent homology, the geometric eligibility criterion requires that a patient have: 1. at least two distinct assessment visit days within the window through study day $117$, and 2. at least four valid target-lesion diameter measurements across these visits. Among the $842$ supplied subjects, $566$ satisfy these criteria ($288$ control, $278$ panitumumab). The remaining $276$ subjects fail eligibility due to: 1. fewer than two distinct assessment visits ($144$ subjects), 2. fewer than four total measurements ($99$ subjects), or 3. no target-lesion records within the day 0--117 window ($33$ subjects). Because these criteria depend on survival, visit adherence, and radiological re-assessment up to week 12, they constitute post-randomization observation exclusions rather than baseline eligibility conditions.

For each eligible subject, each recorded lesion measurement is mapped to planar coordinates $u=\mathrm{day}/117$ and $z=\log\{\max(\mathrm{diameter},5)\}$, where the horizontal axis represents normalized time and the vertical axis represents log-diameter floored at the $5$~mm imaging resolution threshold. Pooled coordinate means are $(0.286116,3.231917)$ and standard deviations are $(0.413488,0.635882)$. To mitigate degeneracies in Delaunay triangulation during alpha-complex computation, exact duplicate coordinates receive a deterministic $10^{-6}$ jitter; this does not guarantee general position. Alpha-complex filtration values are converted to radii, and essential classes of infinite persistence are discarded. The primary summary is the degree-zero persistence silhouette with order $r=3$ (with degree one evaluated as an exploratory alternative). The $100$-point filtration interval $[0,0.83]$ uses the historically fixed endpoint $0.83$. It was selected treatment-blind before pooled coordinate standardization, by taking the $99$th percentile of the per-subject, per-degree maximum finite death radii in the raw-coordinate diagrams and rounding upward to the next hundredth ($0.8223$ to $0.83$). The same endpoint was retained for the standardized-coordinate analysis. Although coordinate scaling and the filtration endpoint were selected without reference to treatment assignment, they were derived from the same eligible cohort used for inference, defining a data-adaptive representation. This motivates the held-out split-sample sensitivity analysis in Appendix~\ref{app:pacce-pilot}.

The baseline design includes chemotherapy backbone, age, sex, ECOG status, baseline target-lesion count, metastatic-site count, KRAS indicators including missingness, and an LDH indicator. Post-treatment response, progression-free survival, and treatment duration are excluded. The AIPW calculation uses a five-function Fourier outcome model, two treatment-stratified folds, a learned chemotherapy-only propensity model, and $4{,}999$ Rademacher multiplier draws. The confidence band is studentized with a numerical variance floor, whereas the zero-curve test uses the unstudentized absolute supremum. These are working models for the selected observed data. The main calculation uses treatment-stratified folds, whereas Theorem~\ref{thm:tate-sufficient} assumes a partition independent of the observations. We do not claim that this clinical implementation satisfies that sufficient regime; its bands remain exploratory under the working models. Randomization by itself does not provide the selected-cohort likelihood ratio, and a minimization design is not replaced here by an iid assignment assumption.

Holding the $566$ subjects, pooled representation, nuisance specifications, and multiplier method fixed, we also use five random seeds for observation-independent two-fold allocation and multiplier resampling. Their nominal $90\%$ bands include zero at every grid node; the sup-test $p$-values range from $0.787$ to $0.979$, peak absolute estimates from $0.0051$ to $0.0108$, and mean full widths from $0.0405$ to $0.0414$ in rescaled units. At the six age-$61$ profiles, a corresponding independently allocated two-fold fit also gives zero-containing CTATE bands, with sup-test $p$-values $0.363$--$0.956$. These checks address fold allocation only; they do not certify nuisance rates, remove the pooled representation's sample dependence, or identify a post-selection causal effect.

\subsection{Split-sample sensitivity analysis with held-out scale estimation}
\label{app:pacce-pilot}

To assess sensitivity to data-dependent representation choices, a split-sample sensitivity analysis reserves a randomly selected $25\%$ auxiliary sample ($210$ subjects) solely for fitting coordinate scale parameters, strictly excluding them from all subsequent outcome regressions and confidence calibration. Held-out eligible lesion points yield coordinate means $(0.293533,3.197913)$ and standard deviations $(0.405892,0.634060)$. The remaining $632$ subjects contain $419$ eligible patients, $210$ control and $209$ panitumumab. The calculation conditions on the held-out coordinate scaling. The filtration interval $[0,1]$, degree zero, and $r=3$ are held fixed; the interval endpoint does not use the eligible cohort's empirical death quantile. As an observational reanalysis of historical trial data, this investigation provides descriptive sensitivity evidence rather than confirmatory clinical claims.

\begin{figure}[t]
\centering
\includegraphics[width=0.82\textwidth]{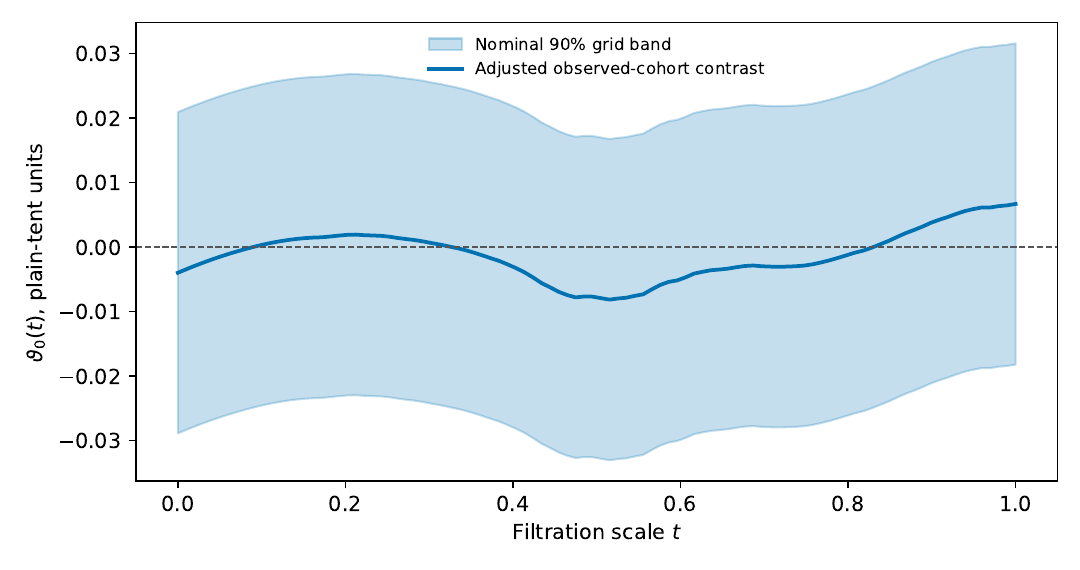}
\caption{PACCE split-sample sensitivity analysis among $419$ eligible analysis subjects with coordinate scales pre-fitted on an independent held-out sample of $210$ subjects (excluded from inference), using standard tent units and a nominal $90\%$ Gaussian unstudentized grid band. Selection after treatment and the working nuisance models limit interpretation to an adjusted observed-cohort contrast.}
\label{fig:pacce-pilot}
\end{figure}

Two data-independent folds fit full-baseline logistic propensity and five-function Fourier outcome regressions. Gaussian unstudentized confidence calibration uses $4{,}999$ draws, avoiding division at zero-variance filtration points. The observed-cohort contrast has peak absolute estimate $0.008139$, mean full width $0.049796$, and no grid point excluding zero. These quantities use plain silhouettes, so they are not directly comparable to the differently selected and rescaled calculation in Figure~\ref{fig:pacce-tate}. Figure~\ref{fig:pacce-pilot} shows the fitted curve and its nominal band. The fitted propensity is a model for treatment conditional on eligibility, not the original randomized probability. No nuisance-rate or post-selection causal identification claim is inferred from this fit.

Changing only the CTATE second-stage Fourier basis to $3$, $5$, or $9$ functions, in the same $419$ eligible analysis subjects, gives profile peak absolute estimates between $0.0127$ and $0.0668$ and mean widths between $0.0915$ and $0.1661$. All six profile bands contain zero over the grid in each version. Exact backbone-and-burden support counts in these $419$ subjects are $75$, $39$, $18$, $16$, $13$, and $3$ in the same profile order as Table~\ref{tab:pacce-ctate}. These are working-model comparisons, not a bias certificate. They do not change the TATE fit, which uses the same first-stage specification in all three versions.

\subsection{The log-summed-diameter comparator}
\label{app:pacce-comparator}

The comparator uses the same $566$ eligible subjects and baseline adjustment design as the topological analysis. Let $S_{i0}$ and $S_{i1}$ be the sums of central target-lesion diameters in the baseline and week-12 windows. Set $z_{ij}=\{\log(\max(S_{ij},5))-3.231917\}/0.635882$ for $j=0,1$. The reference assessment coordinates are $u_0=\{(-6/117)-0.286116\}/0.413488$ and $u_1=\{(81/117)-0.286116\}/0.413488$. On the same $100$-point grid over $[0,0.83]$, the constructed response is
\[
 g_i(t)=\frac12\left\{z_{i0}+
       \frac{t-u_0}{u_1-u_0}(z_{i1}-z_{i0})\right\}.
\]
This is a linearly interpolated log-summed-diameter summary on a standardized assessment coordinate, rather than a diameter measured in millimeters or a topological summary. Its adjusted mean contrast is fitted with the same five-function Fourier outcome regression, chemotherapy-only propensity model, two treatment-stratified folds, and $4{,}999$ multiplier draws. The reported $p=0.175$ tests its zero adjusted mean curve with an absolute-supremum statistic. It shares the observation-selection and working-model qualifications of the topological calculation.

\subsection{Conditional profiles and prediction objects}
\label{app:pacce-profiles}

The profiles query chemotherapy backbone crossed with target-lesion counts $2$, $4$, and $6$, while fixing the remaining covariates, as presented in Table~\ref{tab:pacce-ctate} of the main text. The six profiles fix age at $61$ years, male sex, ECOG zero, two metastatic sites, wild-type KRAS, and normal LDH, while varying chemotherapy backbone and target-lesion count. Age enters the fitted design in years. The second-stage functional DR learner uses four Fourier basis functions. Its nominal $90\%$ studentized bands use $1{,}999$ Rademacher multiplier draws and a numerical variance floor; the separately simulated sup-test $p$-values use $1{,}999$ draws with fixed, distinct seeds. The feature design retains the raw burden count and adds indicators for counts $3$--$4$ and at least $5$, subject to a support rule. These profiles are individual covariate points in that working regression, not averages over the burden bins.

Every fitted band in Table~\ref{tab:pacce-ctate} contains zero over the grid. This is a description of the fitted curves, not proof that all true conditional effects vanish. The largest exact burden matches are not equivalent to the full support for all profile covariates. Neither the peaks' similar locations nor the absence of nominal rejection establishes that the excursions are noise or that heterogeneity is absent. The available simulation evidence documents approximation-bias failures of the conditional-mean fit, so the table is exploratory.

For exploratory individual prediction, we use separate armwise bands rather than the second-stage mean confidence bands. Half the $566$ eligible subjects are reserved for calibration; five-function Fourier outcome regressions and a chemotherapy-only logistic model are fitted on the other half. The scale in every nonconformity score is fixed at $s_a(t,x)=1$ before calibration. Calibration residuals are used only to compute the supremum scores and their weighted quantiles, not to estimate a modulation function. The armwise error budgets are $0.05$ each, giving a nominal $90\%$ composed prediction level under the theorem's assumptions. Because the likelihood ratios are learned, the pooled representation is data-adaptive, and post-treatment selection remains unresolved, these clinical prediction objects are exploratory; the exact-ratio theorem does not certify them, and they do not establish conditional coverage at the queried profiles.

The representation uses $\widetilde\phi_0=\sqrt{2}\Phi_3$ on $[0,0.83]$, where $\Phi_3$ is the normalized silhouette of standard tents. Units are standardized alpha-complex radii, not millimeters or log-diameter. Since alpha-complex births satisfy $b\geq0$, every standard tent satisfies $0\leq\lambda_{b,d}(t)\leq t$ for $t\geq0$; its normalized nonnegative weighted average does too. Therefore, any individual contrast of two such summaries satisfies
\[
 -\sqrt{2}t\leq\delta_i(t)\leq\sqrt{2}t.
\]
We intersect the composed band $[L(t,x),U(t,x)]$ with this deterministic support, obtaining
\[
 \left[\max\{L(t,x),-\sqrt{2}t\},\,
       \min\{U(t,x),\sqrt{2}t\}\right].
\]
This intersection preserves any coverage possessed by the original band because a supported true curve cannot be removed from a band that contains it. It does not repair estimated-ratio or selection limitations. Widths below average the full width equally over the $100$ grid nodes. The complete support envelope alone has mean full width $1.174$ on this grid.

\begin{table}[t]
\centering\small
\caption{Exploratory PACCE prediction widths at age $61$ years in rescaled units $\sqrt{2}\Phi_3$. The score scale is fixed at one before calibration. All six composed bands are finite and contain zero at every grid node before and after intersection with the deterministic support.}
\label{tab:pacce-prediction}
\begin{tabular}{@{}llrr@{}}
\toprule
Backbone&Burden&Unrestricted mean width&Support-intersected mean width\\
\midrule
Oxaliplatin&2&1.730&1.090\\
&4&1.730&1.083\\
&6&1.730&1.086\\
Irinotecan&2&1.639&1.063\\
&4&1.639&1.053\\
&6&1.639&1.056\\
\bottomrule
\end{tabular}
\end{table}

The unrestricted mean full widths are $1.639$--$1.730$, and support intersection reduces them to $1.053$--$1.090$ (Table~\ref{tab:pacce-prediction}). The broad intervals reflect the supremum calibration and conservative composition of two armwise bands; their width is not itself an estimate of individual-effect variability. Zero inclusion does not prove individual equivalence or test an individual causal null. The ITTE, CTATE, and TATE calculations concern the selected topological summary and do not establish agreement with the distinct toxicity and progression-free-survival findings of \citet{hecht2009}. Individual potential-outcome pairs are unavailable in PACCE, so individual-effect coverage cannot be empirically checked. We report aggregate prediction summaries and retain patient-level trajectories within the restricted analysis environment.

\subsection{Controls and selection sensitivity}
\label{app:pacce-controls}

Baseline-only geometry gives a working sup-test $p=0.122$ in $486$ subjects. Two hundred within-backbone treatment-label permutations yield $18$ nominal rejections, a proportion $0.090$ with Monte Carlo standard error $0.020$. These are diagnostics on the selected records. Permuting labels after eligibility selection does not reproduce the original randomization distribution when treatment can affect availability, nor does a single nonsignificant baseline test validate all modeling assumptions.

Preprocessing alternatives change the diameter floor to $3$, $5$, or $10$ mm, include additional visits through day $120$, require six points, use $r=2$ or $4$, or use degree one. Their working test $p$-values range from $0.528$ to $0.969$; the broader day window includes $627$ subjects and the six-point rule includes $471$. A re-fit at the 5 mm floor and nominal cohort gives $p=0.656$, rather than the main calculation's $0.982$, because nuisance and resampling choices also differ. Thus these are joint analysis variants, not isolated changes holding every other element fixed. None supplies evidence of a detected observed-group difference, and they are not an equivalence analysis.

The observation-weight sensitivity tilts a fitted baseline eligibility model by standardized target burden and reweights observed subjects. Such a family can illustrate sensitivity of a weighted observed contrast. It does not specify the missing outcomes and cannot establish robustness to general missing-not-at-random selection, death, or treatment-affected assessment. An effect for a complete latent outcome would additionally require that this outcome be well defined for excluded patients and that the chosen missingness assumptions identify it. We therefore draw no selection-robust causal-null conclusion from this calculation.

\subsection{Data governance and access}
\label{app:pacce-data-access}

The patient-level clinical trial records from the PACCE trial (Amgen study 20040249) were analyzed under a data use agreement with Project Data Sphere. In compliance with the Project Data Sphere Data Use Agreement and patient privacy standards, individual-level clinical records cannot be redistributed, published, or hosted in third-party public repositories. This restriction also extends to individual topological treatment effect (ITTE) trajectory curves and individual patient prediction bands, which cannot be published or included in the public GitHub repository to safeguard participant anonymity. Researchers seeking to reproduce or extend analyses on this clinical cohort must register and submit a formal data request directly through Project Data Sphere (\url{https://data.projectdatasphere.org/}). All aggregation code and sensitivity evaluation scripts that operate on the restricted data are provided in the project repository.

\end{document}